\documentclass[aps, prd, reprint, superscriptaddress, amsmath, amssymb, floatfix]{revtex4-2}

\usepackage{graphicx}
\usepackage{epstopdf}
\usepackage{dcolumn}
\usepackage{bm}
\usepackage[utf8]{inputenc}
\usepackage{color,xcolor}
\usepackage{subcaption}
\usepackage{url}
\usepackage[hidelinks]{hyperref}
\usepackage{booktabs}
\usepackage{multirow}
\usepackage{float}

\begin{document}

\title{Identifiability of \texorpdfstring{$g$}{g} mode Resonances in Eccentric Binary Neutron Stars with Multidetector Observations}

\newcommand{\CQUPhys}{Department of Physics, Chongqing University, Chongqing 401331, P.R. China}
\newcommand{\CQUKeyLab}{Chongqing Key Laboratory for Strongly Coupled Physics, Chongqing University, Chongqing 401331, P.R. China}
\newcommand{\CQUInter}{Institute of Advanced Interdisciplinary Studies, Chongqing University, Chongqing 401331, China}
\newcommand{\XTUInfo}{Department of Electronical Information Science and Technology, Xingtai University, Xingtai 054001, P.R. China}
\newcommand{\SUSTechEarth}{Department of Earth and Sciences, Southern University of Science and Technology, Shenzhen 518055, P.R. China}
\newcommand{\SUSTechPhys}{Department of Physics, Southern University of Science and Technology, Shenzhen 518055, P.R. China}
\newcommand{\CQUMicro}{School of Microelectronics and Communication Engineering, Chongqing University, Chongqing 401331, P.R. China}

\author{Mengfei~Sun}
\affiliation{\CQUPhys}
\affiliation{\CQUKeyLab}

\author{Jie~Wu}
\affiliation{\CQUPhys}
\affiliation{\CQUKeyLab}

\author{Qianning~Hu}
\affiliation{\CQUPhys}
\affiliation{\CQUKeyLab}

\author{Jin~Li}
\email{cqujinli1983@cqu.edu.cn}
\affiliation{\CQUPhys}
\affiliation{\CQUKeyLab}
\affiliation{\CQUInter}

\author{Nan~Yang}
\affiliation{\CQUKeyLab}
\affiliation{\XTUInfo}

\author{Xianghe~Ma}
\affiliation{\CQUPhys}
\affiliation{\CQUKeyLab}

\author{Borui~Wang}
\affiliation{\SUSTechEarth}

\author{Minghui~Zhang}
\affiliation{\SUSTechPhys}

\author{Yuanhong~Zhong}
\email{zhongyh@cqu.edu.cn}
\affiliation{\CQUMicro}

\date{\today}

\begin{abstract}
$g$ mode resonances in eccentric binary neutron star systems are potential probes of internal stratification, superfluidity, composition gradients, and the equation of state. Although such weak dynamical tidal signatures are unlikely to be resolved with current detector sensitivities, third generation observations may make them accessible, in which case identifying the weak resonant phase shift would provide information beyond the bulk adiabatic tidal deformability. We build a four class dataset in an eccentric harmonic framework, containing point particle, adiabatic tide, resonant $g$ mode, and pure noise samples, and use Einstein Telescope (ET) and Cosmic Explorer (CE) detector data to test whether this weak resonant phase signature can be identified from noisy time domain strain. The ET, CE, and ET+CE deep learning models reach accuracies of {$0.650$, $0.809$, and $0.875$}, respectively. On the same simulated samples, the matched filtering method reaches lower accuracies of {$0.534$, $0.703$, and $0.718$}. This result arises from the fact that the resonant correction manifests as a weak phase morphology difference superimposed on the adiabatic tidal background, whereas matched filtering is sensitive only to the overall similarity. Hence, in the presence of weak phase differences, {the neural network employed for the classification task} is better able to learn these local phase and morphology features from the complete time domain strain segment. The results indicate that joint third generation observations improve the identifiability of weak internal mode phase information.
\end{abstract}

\maketitle

\section{Introduction}
\label{sec:introduction}

{Binary neutron star (BNS) mergers provide a gravitational wave (GW) probe of dense nuclear matter, whose equation of state (EoS) governs neutron star structure and macroscopic observables \citep{LattimerPrakash2016DenseMatter}. GW170817 demonstrated that tidal deformability can constrain neutron star radii and the EoS \citep{ligo2017gw170817,Abbott2018RadiiEOS,De2018GW170817Radii,Annala2018EOS,gwtc3,Annala2022MultimessengerEOS}. Because matter above nuclear saturation density remains poorly constrained, waveform signatures beyond bulk tidal deformability may provide complementary probes of internal structure. Static and adiabatic tides characterize the global stellar response through the tidal Love number and dimensionless tidal deformability \citep{Hinderer2008Love,DamourNagar2009Tidal,Hinderer2010TidalEOS,YagiYunes2013ILoveQ}. More generally, dynamical tides describe how the star responds to the orbital tidal field across frequencies. They include a nonresonant mode response, dominated by the fundamental $f$ mode and approaching the adiabatic limit at low frequencies, as well as resonant excitation when a tidal harmonic approaches a stellar eigenfrequency \citep{Hinderer2016DynamicTides,Steinhoff2016DynamicTides,Pratten2022DynamicTidesEOS}. The $f$ mode is mainly determined by bulk stellar properties such as mass and radius, whereas the $g$ mode supported by buoyancy is more sensitive to the density dependence of the lepton fraction, superfluidity, crustal structure, and density discontinuities induced by strong first order phase transitions \citep{KantorGusakov2014Gmodes,PassamontiAnderssonHo2016Buoyancy,Constantinou2021GmodesCrossover,Jaikumar2021HybridGmodes,Counsell2025InterfaceModes,passamonti2021crust}. Identifying such modes would therefore probe neutron star microphysics beyond static tidal parameters.}

The main observational challenge for the $g$ mode is that its coupling to the external tidal field is weak. Compared with the dominant adiabatic tidal response, a mode resonance usually does not produce a strong amplitude variation in the waveform, but instead appears mainly as a weak phase shift. In realistic noisy data, this weak phase perturbation can be masked by the dominant tidal phase, uncertainties in the mass parameters, and noise fluctuations \citep{Lai1994ResonantTides,ReiseneggerGoldreich1994Modes,KokkotasSchafer1995TidalResonant}.

{
The detectability of tidally excited mode resonances has been assessed with progressively more realistic methods. An early Fisher matrix study found that second generation detectors would require phase shifts of order unity to several radians for resonance frequencies in the tens of hertz, suggesting limited prospects for weakly coupled modes \citep{BalachandranFlanagan2007Detectability}. Ho and Andersson subsequently argued, using phase uncertainties inferred from GW170817 and projected A+ and CE sensitivities, that criteria based on phase shifts of a few radians are overly conservative and that future observations could constrain the energy transfer and coupling of $g$ mode resonances \citep{HoAndersson2023DynamicalTides}. More recent mismatch calculations weighted by detector noise showed that observability is governed by the time advance induced by the resonance and the dephasing accumulated across the detector band, and that estimates at a single frequency can overstate the sensitivity \citep{RevillaPena2026Detectability}. A fully Bayesian population study for ET found detectable resonances in a substantial fraction of the loudest simulated events, with sensitivity to phase shifts reaching about $0.03$ rad for favorable sources, and showed that neglecting resonances can bias the inference of tidal deformability \citep{Gittins2026DetectingResonances}. These studies establish tidal resonances as an observational target, but focus primarily on quasicircular binaries.
}

In eccentric orbits, the tidal driving contains multiple orbital harmonics, which modifies the accumulation of the $g$ mode resonant phase shift. Higher orbital harmonics can accumulate larger phase shifts at an earlier inspiral stage and bring these shifts into the more sensitive frequency bands of the detector. Eccentric binaries can also undergo multiple epicyclic resonances, further increasing the total phase shift \citep{VickLai2019EccentricTides,Takatsy2024EccentricDynamicTides,takatsy2026gmode}. Thus, eccentricity is not only an additional complication in waveform modeling, but also a possible mechanism for enhancing otherwise weak internal mode information.

{Eccentric BNS mergers are expected to be rare \citep{DietrichHindererSamajdar2021BNSReview}.} Isolated field binaries are usually close to circular by the time they enter the ground based detector band, whereas dynamical interactions in dense clusters or hierarchical systems may allow binaries to retain measurable eccentricity \citep{Peters1963,Peters1964,Gold2012EccentricBNS,East2012DynamicalCapture,Paschalidis2015OneArm,VickLai2019EccentricTides}. {Recent studies have also investigated the possibility of measurable eccentricity in mergers between neutron stars and black holes \citep{Morras2026EccentricNSBH,Clarke2026GW200105}.} Searches for eccentric BNS signals and reanalyses of existing events have not yet produced a confirmed detection, but eccentricity measurements have become part of current data analysis practice. If residual eccentricity is ignored, tidal deformability and EoS inference may suffer from systematic biases \citep{nitz2019eccbns,lenon2020ecc,Lenon2021CEeccentric,DuttaRoySaini2024UnmodeledEccentricity,KacanjaSoniNitz2025EccentricitySignatures}. In this context, detecting weak $g$ mode signatures in eccentric BNS signals is relevant to eccentric orbit formation, dynamical tide modeling, and constraints on neutron star internal structure.

{Third generation detectors such as the Einstein Telescope (ET) and Cosmic Explorer (CE) will improve strain sensitivity and extend the observable band to lower frequencies, providing access to eccentric, multiharmonic, and weak dynamical tide signals \citep{Punturo2010ET,Hild2011ET,maggiore2020et,Reitze2019CosmicExplorer,evans2021ce,Hall2022CosmicExplorer}. Combining detectors can increase the signal to noise ratio and provide complementary information, motivating a direct test of whether an ET+CE network improves $g$ mode identifiability. Matched filtering and Bayesian inference remain standard GW analysis tools, but searches involving eccentricity, dynamical tides, multiple harmonics, and long third generation signals are computationally demanding \citep{Allen2012FindChirp,Veitch2015LALInference,Ashton2019Bilby}. Deep learning offers an efficient complement and has been applied to detection, classification, parameter estimation, waveform reconstruction, and denoising \citep{george2017deeplearn,Sun2024DeepLearningDECIGO,Gabbard2018DeepNetworks,Sun2025ConditionalAutoencoderBNS,Cuoco2020MLReview,Sun2023DECIGOIMBBH,ChuaVallisneri2020Posteriors,Sun2025LensedDMHalos,Dax2021NPE}.}

{Building on recent studies of eccentric $g$ mode resonances \citep{VickLai2019EccentricTides,Takatsy2024EccentricDynamicTides,takatsy2026gmode}, we ask whether a deep learning classifier can identify resonant phase shifts against an adiabatic tidal background in noisy third generation data while distinguishing point particle (PP), adiabatic tide (AT), resonant $g$ mode (AT+$g$), and pure noise (PN) samples. We construct a data set with four classes for ET, CE, and ET+CE, compare a 1D ResNet with matched filtering on the same test samples, and analyze performance across mass, eccentricity, mode frequency, and tidal parameters. The waveform construction maps resonance effects on the orbital evolution to the corresponding radiative harmonics, allowing us to quantify whether joint observations improve the identifiability of weak $g$ mode signatures.}

\section{Theoretical Framework}
\label{sec:theory}
This section summarizes the eccentric BNS $g$ mode resonance phase prescription adopted in this work and specifies how it is used for waveform construction, dataset generation, and identification analysis.
{Throughout the analytical expressions, we use geometrized units with $G=c=1$.}

\subsection{Eccentric orbit harmonics and \texorpdfstring{$g$}{g} mode resonance}
\label{subsec:eccentric_harmonics}

Following Refs.~\citep{Lai1994ResonantTides,ReiseneggerGoldreich1994Modes,KokkotasSchafer1995TidalResonant,takatsy2026gmode}, we adopt an eccentric orbit harmonic framework to describe the $g$ mode resonant phase. In this construction, the resonant phase shift is first computed at the orbital frequency level and then mapped to the corresponding radiative harmonics. The calculation uses a Newtonian tidal orbital coupling model, retains only the quadrupolar tidal degrees of freedom, and labels the modes by $\alpha=\{n,l=2,m\}$, where $n$ denotes the radial mode order rather than a radiative harmonic index. Spin is neglected and the orbit is restricted to the orbital plane. The effective Lagrangian of the primary star can be written as {\citep{FlanaganHinderer2008TidalLove,Lai1994ResonantTides,ReiseneggerGoldreich1994Modes,KokkotasSchafer1995TidalResonant,takatsy2026gmode}}
\begin{align}
L={}&\left[\frac{1}{2}\mu \dot{r}^{2}
+\frac{1}{2}\mu r^{2}\dot{\phi}^{2}
+\frac{m_{\rm tot}\mu}{r}\right]
-\sum_{\alpha}\frac{1}{2}Q_{ij}^{\alpha}E_{ij} \notag\\
&+\sum_{\alpha}\frac{1}{4\lambda_1^\alpha \omega_\alpha^2}
\left(\dot{Q}_{ij}^{\alpha}\dot{Q}_{ij}^{\alpha}
-\omega_\alpha^{2}Q_{ij}^{\alpha}Q_{ij}^{\alpha}\right),
\end{align}
{
Here $m_1$ is the mass of the tidally deformed primary star and $m_2$ is the companion mass, $m_{\rm tot}=m_1+m_2$, and $\mu=m_1m_2/m_{\rm tot}$. The relative separation is $r=|\boldsymbol{r}|=|\boldsymbol{x}_1-\boldsymbol{x}_2|$, and $\phi$ is the true longitude of $\boldsymbol{r}$ in the orbital plane, measured from a fixed $x$ axis. The quantity $Q_{ij}^{\alpha}$ is the symmetric and trace free dynamical mass quadrupole tensor associated with mode $\alpha$, while $E_{ij}$ is the external tidal tensor generated by the companion. The indices $i,j\in\{1,2,3\}$ are Cartesian spatial indices, and repeated spatial indices are summed. The explicit sum over $\alpha$ runs over the retained quadrupolar stellar modes. The quantities $\lambda_1^\alpha$ and $\omega_\alpha$ are the tidal deformability and eigenangular frequency of mode $\alpha$ in star 1, respectively, and overdots denote time derivatives.
}
Following the notation commonly used in GW tidal modeling, we use
\begin{align}
k_\alpha=\frac{2\pi}{2l+1}\frac{\tilde{Q}_\alpha^{2}}{\tilde{\omega}_\alpha^{2}},\qquad
k_\alpha=\frac{3}{2}\frac{\lambda_1^\alpha}{R^{5}},
\end{align}
{
Here $k_\alpha$ is the contribution of mode $\alpha$ to the quadrupolar tidal Love number, $\tilde{Q}_\alpha$ and $\tilde{\omega}_\alpha$ are the dimensionless mode overlap and eigenfrequency in this normalization, and $R$ is the stellar radius.
}
We also use the dimensionless mode tidal parameter $\Lambda_1^\alpha=\lambda_1^\alpha/m_1^5$. For the first order $g$ mode considered here, we denote the corresponding coupling by $\Lambda_g$. {Through Eq.~(2), $\Lambda_g$ is determined by the dimensionless mode overlap, mode eigenfrequency, and stellar compactness, making it sensitive to the neutron star equation of state and internal composition. Microphysical calculations for this $g$ mode with the highest frequency, characterized by a single radial node, therefore span $10^{-2}\lesssim\Lambda_g\lesssim1$ \citep{takatsy2026gmode}.} In the simulations below, we focus on a weak coupling portion of this range, taking $\Lambda_g$ from $0.03$ to $0.30$ to examine how the resonant phase amplitude affects identification performance.

The external tidal tensor is
\begin{align}
E_{ij}=-m_2\,\partial_i\partial_j\left(\frac{1}{r}\right).
\end{align}
{
Here $\partial_i\equiv\partial/\partial x^i$ denotes a Cartesian spatial derivative.
}
{
The symmetric and trace free tensors $Q_{ij}^{\alpha}$ and $E_{ij}$ each have five independent quadrupolar components and can be expanded in a rank 2 tensor harmonic basis as \citep{Takatsy2024EccentricDynamicTides,takatsy2026gmode}
\begin{align*}
Q_{ij}^{\alpha}=\sum_{m=-2}^{2}Q_m^{\alpha}\mathcal{Y}_{ij}^{2m},\qquad
E_{ij}=\sum_{m=-2}^{2}E_m\mathcal{Y}_{ij}^{2m}.
\end{align*}
Here $\mathcal{Y}_{ij}^{2m}$ denotes a rank 2 symmetric and trace free spherical harmonic basis tensor, normalized through $Y_{2m}(\theta,\varphi)=\mathcal{Y}_{ij}^{2m}n^i n^j$, where $n^i$ are the Cartesian components of the unit radial vector. The quantities $Q_m^{\alpha}$ and $E_m$ are the corresponding expansion coefficients, with $m=-2,-1,0,1,2$.

Varying the conservative Lagrangian in Eq.~(1) with respect to $Q_{ij}^{\alpha}$ and projecting onto $\mathcal{Y}_{ij}^{2m}$ yields
\begin{align*}
\ddot{Q}_m^\alpha+\omega_\alpha^{2}Q_m^\alpha
=-\lambda_1^\alpha \omega_\alpha^{2}E_m .
\end{align*}
To describe dissipation phenomenologically, we then add a linear damping term, so that the driven mode amplitude obeys
}
\begin{align}
\ddot{Q}_m^\alpha+2\gamma_\alpha \dot{Q}_m^\alpha+\omega_\alpha^{2}Q_m^\alpha
=-\lambda_1^\alpha \omega_\alpha^{2}E_m .
\end{align}
{
Here $\gamma_\alpha$ is the linear damping rate of mode $\alpha$, accounting for dissipative effects such as microscopic viscosity and gravitational wave backreaction from the tidal deformation. For $\gamma_\alpha=0$, Eq.~(4) reduces to the conservative equation derived from Eq.~(1). For nonzero $\gamma_\alpha$, Eq.~(4) is a phenomenological dissipative extension and does not follow directly from the conservative Lagrangian.
}
For $g$ modes, viscous damping and radiation reaction are typically much weaker than the orbital radiation driving, so we take $\gamma_\alpha\simeq 0$ at leading order. The key feature of an eccentric orbit is that the driving term naturally contains multiple orbital harmonics. Its Fourier expansion can be written as {\citep{Takatsy2024EccentricDynamicTides,takatsy2026gmode}}
\begin{align}
-\lambda_1^\alpha \omega_\alpha^{2}E_m(t)
=C_m\lambda_1^\alpha \omega_\alpha^{2}\frac{m_2}{a^{3}}
\sum_{\ell=-\infty}^{\infty}X_{-\ell}^{-3,-m}e^{-i\ell\Phi},
\end{align}
{
Here $a$ is the semimajor axis, $\Phi$ is the mean anomaly, $\ell\in\mathbb{Z}$ is the orbital harmonic index, $C_m$ is the geometric coefficient for the $m$ component, and $X_{\ell}^{k,m}(e)$ is the Hansen coefficient.
}
The resonance condition
\begin{align}
\ell_r \Omega=\mathrm{sgn}(m)\,\omega_\alpha
\end{align}
{
contains the mean orbital angular frequency $\Omega\equiv\dot{\Phi}$. Here $\ell_r$ is the integer orbital harmonic satisfying the resonance condition, and a subscript $r$ denotes evaluation at resonance.
}
This condition shows that the same eccentric system can trigger mode resonances at multiple harmonics. In the following formulas we take the positive resonant harmonics $\ell_r>0$, corresponding to the $m=2$ component; the $m=-2$ contribution has the same amplitude and is included by symmetry. Thus, in an eccentric waveform, the $g$ mode contributes phase perturbations distributed across the harmonic structure.

Within the stationary phase approximation, the corresponding single resonance mode amplitude and exchanged energy are approximately {\citep{Lai1994ResonantTides,VickLai2019EccentricTides,takatsy2026gmode}}
\begin{align}
|q_m^\alpha|\approx
\frac{C_m\lambda_1^\alpha \omega_\alpha m_2}{2a^{3}}
X_{\ell_r}^{-3,|m|}\sqrt{\frac{2\pi}{|\ell_r|\dot{\Omega}_r}},
\end{align}
\begin{align}
\Delta E\approx
\frac{9\pi \lambda_1^\alpha \omega_\alpha^{2} m_2^{2}}{8\ell_r \dot{\Omega}_r a^{6}}
\left(X_{\ell_r}^{-3,2}\right)^2 .
\end{align}
{
Here $q_m^\alpha$ is the complex amplitude of the $m$ component of mode $\alpha$, $\Delta E$ is the energy transferred from the orbit to the mode, and $\dot{\Omega}_r=(d\Omega/dt)_r$ is the derivative of the orbital angular frequency evaluated at resonance.
}
The associated single resonance phase shift is related to the energy perturbation by
\begin{align}
\Delta\Phi\approx-\frac{3\Omega^{2}\Delta E}{2\dot{\Omega}|E_{\rm orb}|}.
\end{align}
{Here $E_{\rm orb}$ is the orbital binding energy.}
Using the eccentricity dependent Peters factor {\citep{Peters1964,takatsy2026gmode}}
\begin{align}
\mathcal{F}(e)=
\left(1+\frac{73}{24}e^2+\frac{37}{96}e^4\right)(1-e^2)^{-7/2},
\end{align}
and the mass ratio $q=m_2/m_1$, this can be written as
\begin{align}
\Delta\Phi_1\approx
-\frac{75\pi}{8192}\frac{\ell_r\Lambda_1^\alpha}{q(q+1)}
\left[\frac{X_{\ell_r}^{-3,2}}{\mathcal{F}(e_r)}\right]^2 .
\end{align}
For the two neutron stars, we compute the phase shifts separately using their masses, mass ratio, and tidal parameters, and then sum the two contributions. In the equal mass limit, this reduces to the direct sum of the two stellar contributions \citep{takatsy2026gmode}. This expression shows that the visibility of the resonance effect is controlled not only by the coupling strength $\Lambda_g$, but also by the harmonic weights determined by eccentricity, the orbital evolution rate at resonance, and the mass ratio structure.

{Following Refs.~\citep{VickLai2019EccentricTides,Takatsy2024EccentricDynamicTides,takatsy2026gmode}, we retain only harmonic resonances satisfying the coherence condition}
\begin{align}
\delta N_r=\frac{\Omega_r}{2\pi}\delta t_r\gtrsim 1,\qquad
\delta t_r=\sqrt{\frac{2\pi}{\ell_r\dot{\Omega}_r}}.
\end{align}
{
Here $\delta t_r$ is the characteristic time spent crossing the $\ell_r$th resonance, and $\delta N_r$ is the corresponding number of orbital cycles accumulated during this interval. Following Ref.~\citep{takatsy2026gmode}, we retain only resonances with $\delta N_r\gtrsim 1$, for which the excitation remains coherent for at least approximately one orbit and the prescription adopted here for the resonance amplitude is applicable. This condition concerns model applicability rather than a general detectability threshold. Ho and Andersson~\citep{HoAndersson2023DynamicalTides} and Gittins et al.~\citep{Gittins2026DetectingResonances} address sensitivity to smaller phase shifts in quasicircular binaries, rather than the eccentric multiharmonic coherence condition used here.
}
Summing all included coherent resonance terms gives the total orbital frequency level phase correction,
\begin{align}
\Delta\Phi_{\rm tot}=\sum_{\delta N_r\mathrel{\gtrsim}1}\Delta\Phi_1(\ell_r,e_r).
\end{align}

\subsection{Resonant phase shift and multiharmonic phase correction}
\label{subsec:phase_injection}

Because the phase correction is applied to frequency domain waveforms, we use its frequency domain representation. Under the stationary phase approximation,
\begin{align}
\Delta\Psi(F)=2\pi F\Delta t(F)-\Delta\Phi(F),
\end{align}
where {$F=\Omega/(2\pi)$ is the orbital frequency and} $\Delta t(F)$ is the time delay for the perturbed system to reach orbital frequency $F$. For a single resonance, $\Delta t_1=\Delta\Phi_1/\Omega_r$. {Adopting the phase convention in which the correction induced by the resonance vanishes above each resonance frequency \citep{FlanaganRacine2007Rossby,takatsy2026gmode}, the total frequency domain correction to the orbital phase is}
\begin{align}
\Delta\Psi_{\rm tot}(F)=
\sum_{\delta N_r\mathrel{\gtrsim}1}\left(1-\frac{F}{F_r}\right)\Delta\Phi_1\,\Theta(F_r-F).
\end{align}
Here {$F_r=\Omega_r/(2\pi)$ is} the orbital frequency at resonance, and $\Theta$ is the Heaviside step function.
{
{We use the orbital frequency $F$ to organize the eccentric harmonic expansion. The emitted GW frequency is intrinsically tied to the orbital motion: $f=2F$ for the dominant harmonic of a quasi-circular binary, and $f=\ell F$ for the $\ell$th eccentric harmonic. A resonance changes the orbital motion and the associated radiation together.} An eccentric waveform contains multiple radiative harmonics, with the $\ell$th harmonic contributing at the gravitational wave frequency $f=\ell F$ in the detector frame. Thus, at fixed $f$, the $\ell$th harmonic samples the orbital evolution at $F=f/\ell$. We define $\Delta\psi_\ell(f)$ as the additive correction induced by the resonance to the Fourier phase $\psi_\ell(f)$ of the $\ell$th harmonic, such that $\psi_\ell(f)\rightarrow\psi_\ell(f)+\Delta\psi_\ell(f)$. The required mapping is \citep{takatsy2026gmode}
}
\begin{align}
\Delta\psi_\ell(f)=\ell\,\Delta\Psi\!\left(\frac{f}{\ell}\right).
\end{align}
{
The factor $\ell$ arises because the phase of the $\ell$th radiative harmonic is $\ell$ times the orbital phase. Consequently, different harmonics at the same detector frequency sample different epochs of the inspiral.
}

\section{Detector Configuration and Noise Modeling}
\label{sec:detector}

This section describes the detector responses of ET and CE, the noise modeling procedure, and the construction of noisy strain data at the detector level.

\subsection{Detector response functions}
\label{subsec:detector_response}

Given the source frame polarization waveforms, the signal must be projected onto a specified detector configuration. For each detector $I\in\{\mathrm{ET},\mathrm{CE}\}$, the frequency domain signal response is
\begin{align}
\tilde h_I^{\rm det}(f)=F^I_+(\alpha,\delta,\psi)\,\tilde h_+(f)
+F^I_\times(\alpha,\delta,\psi)\,\tilde h_\times(f),
\end{align}
{
Here $\tilde h_+(f)$ and $\tilde h_\times(f)$ are the plus and cross polarization strains of the incident gravitational wave in the Fourier domain before projection onto detector $I$, and a tilde denotes a Fourier transform.
}
The quantities $F^I_+$ and $F^I_\times$ are antenna pattern functions, and $(\alpha,\delta,\psi)$ denote right ascension, declination, and polarization angle. Sky position, inclination, polarization angle, and coalescence phase are randomized so that the data marginalize over source geometry and do not encode class labels through a fixed geometric configuration.

The ET and CE response functions and target sensitivity curves are based on the ET design sensitivity study \citep{Hild2011ET} and the CE next generation ground based detector design \citep{Hall2022CosmicExplorer}. {Here $S_{n,I}(f)$ denotes the noise power spectral density of detector $I$ defined for positive frequencies.} In the simulations, the corresponding $F^I_+$, $F^I_\times$, and $S_{n,I}(f)$ enter the detector projection and colored noise generation.

\subsection{Noise modeling and strain construction}
\label{subsec:noise_input}

For each detector, a colored Gaussian noise realization $n_I(t)$ is generated with PyCBC from the one sided detector power spectral density (PSD) $S_{n,I}(f)$ \citep{alex_nitz_2024_10473621}. The ET and CE PSDs are taken from their corresponding design sensitivity curves \citep{Hild2011ET,Hall2022CosmicExplorer}, so that the simulated noise reflects the instrumental noise properties in each frequency band. After the projected detector response is transformed to the time domain, the noisy detector strain is written as $d_I(t)=h_I^{\rm det}(t)+n_I(t)$.

The noisy time domain strain from each detector is used as one input channel. Using the training samples, we compute a normalization at each time sample for each detector channel,
\begin{align}
x_I(t)=\frac{d_I(t)-\mu_I(t)}{\sigma_I(t)+\varepsilon},
\end{align}
where $\mu_I(t)$ and $\sigma_I(t)$ are computed from the stacked training samples at the corresponding time sample, and $\varepsilon$ is a small constant preventing division by zero. The dual detector input consists of ET and CE time domain sequences of length $4096$, while the single detector experiments use a single channel of the same length, ensuring comparability across configurations.

\section{Dataset Construction and Simulation Setup}
\label{sec:dataset}

This section presents the parameter space, waveform construction procedure, and four class sample construction used in the simulations.

\subsection{Parameter space and waveform generation pipeline}
\label{subsec:parameter_space}

The synthetic BNS data are specified by the intrinsic physical parameter vector
\begin{align}
\boldsymbol{\theta}=
\left(m_1,m_2,\Lambda_{\rm tide,1},\Lambda_{\rm tide,2},
e_{10{\rm Hz}},f_g,\Lambda_g\right).
\end{align}
The parameter ranges are listed in Table~\ref{tab:parameter_space}, and the dataset and input settings are summarized in Table~\ref{tab:dataset_setup}. We also record the redshift $z$ and the corresponding luminosity distance $D_L$, and randomly sample sky position, inclination, polarization angle, and coalescence phase to construct the detector projected noisy strain data.

\begin{figure}[t]
\centering
\includegraphics[width=0.88\linewidth]{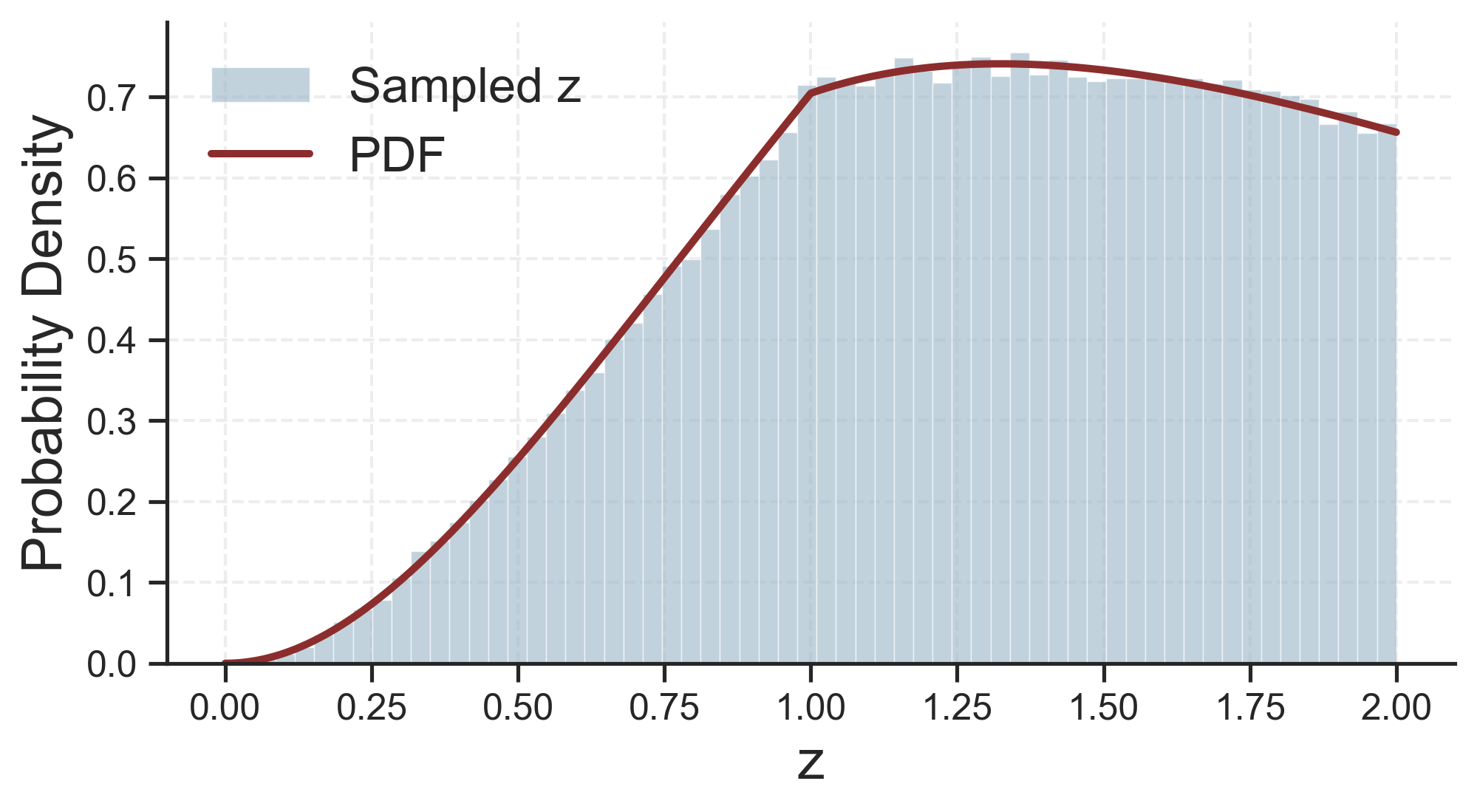}
\caption{BNS redshift distribution. The histogram shows the sampled redshifts, and the solid curve shows the normalized probability density corresponding to Eq.~\eqref{eq:redshift_pdf}.}
\label{fig:z_distribution}
\end{figure}

The BNS redshift distribution is sampled from the probability density \citep{Zhao2011ETCosmology,Sun2024DeepLearningDECIGO}
\begin{align}
\rho(z) \sim \frac{4 \pi d_{C}^{2}(z) R(z)}{H(z)(1+z)} ,
\label{eq:redshift_pdf}
\end{align}
where the comoving distance is
\begin{align}
d_{C}(z)=\int_{0}^{z}\frac{1}{H\left(z^{\prime}\right)}d z^{\prime}.
\label{eq:comoving_distance}
\end{align}
The luminosity distance is $D_L=d_C(1+z)$. The range $z\in[0,2]$ is chosen to cover cosmological BNS events expected to be accessible to third generation detector networks while keeping the simulated classification set within the redshift range considered in this work. The source evolution rate is \citep{Schneider2001LowFrequency,CutlerHolz2009Cosmology,CaiYang2017ETCosmology}
\begin{align}
R(z)=
\left\{
\begin{array}{cc}
1+2z, & z\leq 1, \\
\frac{3}{4}(5-z), & 1<z<5, \\
0, & z\geq 5 .
\end{array}
\right.
\label{eq:source_evolution}
\end{align}
Figure~\ref{fig:z_distribution} compares the sampled redshift distribution with the normalized theoretical probability density from Eq.~\eqref{eq:redshift_pdf}, showing that the sampling is consistent with the target distribution.

\begin{table}[t]
\centering
\caption{Physical parameter space. {$m_1>m_2$ and $\Lambda_{\rm tide,1}<\Lambda_{\rm tide,2}$.}}
\label{tab:parameter_space}
\footnotesize
\setlength{\tabcolsep}{1pt}
\begin{tabular}{lll}
\hline\hline
Parameter & Symbol & Range/value \\
\midrule
Component mass & $m_1$ & $\mathcal{U}(1.2,2.0)\,M_\odot$ \\
Component mass & $m_2$ & $\mathcal{U}(1.2,2.0)\,M_\odot$ \\
Tidal parameter & $\Lambda_{\rm tide,1}$ & $\mathcal{U}(100,1000)$ \\
Tidal parameter & $\Lambda_{\rm tide,2}$ & $\mathcal{U}(100,1000)$ \\
Eccentricity at $10\,\mathrm{Hz}$ & $e_{10{\rm Hz}}$ & $\mathcal{U}(0,0.5)$ \\
$g$ mode frequency & $f_g$ & from $50$ to $300\,\mathrm{Hz}$, step $25\,\mathrm{Hz}$ \\
$g$ mode coupling & $\Lambda_g$ & $0.03$, $0.05$, $0.07$, $0.10$ \\
 & & $0.15$, $0.20$, $0.30$ \\
Redshift & $z$ & Eq.~\eqref{eq:redshift_pdf}, $z\in[0,2]$ \\
Luminosity distance & $D_L$ & from $z$ (Eq.~\eqref{eq:redshift_pdf}) \\
Right ascension & $\alpha$ & $\mathcal{U}(0,2\pi)$ \\
Declination & $\delta$ & $\sin\delta\sim\mathcal{U}(-1,1)$ \\
Inclination & $\iota$ & $\cos\iota\sim\mathcal{U}(-1,1)$ \\
Polarization & $\psi$ & $\mathcal{U}(0,\pi)$ \\
Phase & $\phi_c$ & $\mathcal{U}(0,2\pi)$ \\
\hline\hline
\end{tabular}
\end{table}

\begin{table}[b]
\centering
\caption{Dataset composition and input settings.}
\label{tab:dataset_setup}
\footnotesize
\begin{tabular}{ll}
\hline\hline
Item & Setting \\
\midrule
Classes & PP, AT, AT+$g$, PN \\
Samples per class & $5\times10^4$ \\
Total samples per task & $2\times10^5$ \\
Detector configurations & ET, CE, ET+CE \\
Sampling rate & $2048\,\mathrm{Hz}$ \\
Time window & $2\,\mathrm{s}$ \\
Radiation harmonics & $\ell=1$--$40$ \\
\hline\hline
\end{tabular}
\end{table}

The waveform generation pipeline proceeds as follows. First, {we generate eccentric BNS waveforms in the time domain using the TEOBResumS-Dal\'i waveform model \citep{albanesi2025effective}, with the component masses, orbital eccentricity, and tidal deformabilities as input parameters, yielding the two polarizations $h_+(t)$ and $h_\times(t)$.} The waveform is generated from an initial GW frequency of $20\,\mathrm{Hz}$, and the network input is the final $2\,\mathrm{s}$ premerger segment of the generated waveform. With a sampling rate of $2048\,\mathrm{Hz}$, each input contains 4096 time samples. Here $e_{10\mathrm{Hz}}$ labels the binary eccentricity at the $10\,\mathrm{Hz}$ reference frequency.

{Bayesian analysis of third generation BNS signals down to approximately $5\,\mathrm{Hz}$ has already been demonstrated \citep{Smith2021ThirdGenerationInference}. The $20\,\mathrm{Hz}$ starting frequency and final $2\,\mathrm{s}$ input window adopted here therefore define the scope of this classification study. The classifier does not directly use the earlier inspiral data or the resonance information available from those data. Our results describe the separability of the simulated classes within this window and do not establish the sensitivity of a full inspiral analysis extending to the low frequency limit of ET or CE.}

The waveform is then transformed to the frequency domain, where the resonance conditions, Hansen coefficients, coherence indicators, and single resonance phase shifts are computed for each harmonic. Coherent terms are summed to obtain the orbital frequency phase correction $\Delta\Psi(F)$. Finally, using the harmonic mapping described above, this correction is mapped harmonic by harmonic back to radiative frequency to construct the frequency domain waveform containing the $g$ mode resonance. The corrected frequency domain waveform is inverse transformed to the time domain before detector noise is added, so the machine learning input is the noisy time domain strain after the frequency domain resonant phase correction has been applied. {This sequence of transforming from the time domain to the frequency domain and back is used because the prescription for eccentric resonances gives a Fourier phase correction separately for each harmonic, $\Delta\psi_\ell(f)=\ell\Delta\Psi(f/\ell)$, whereas TEOBResumS-Dal\'i provides the baseline eccentric tidal waveform in the time domain. Simply applying a common phase factor to the total time domain polarizations after the harmonics have been combined would not reproduce this harmonic by harmonic mapping.}

\begin{figure}[t]
\centering
\includegraphics[width=\linewidth]{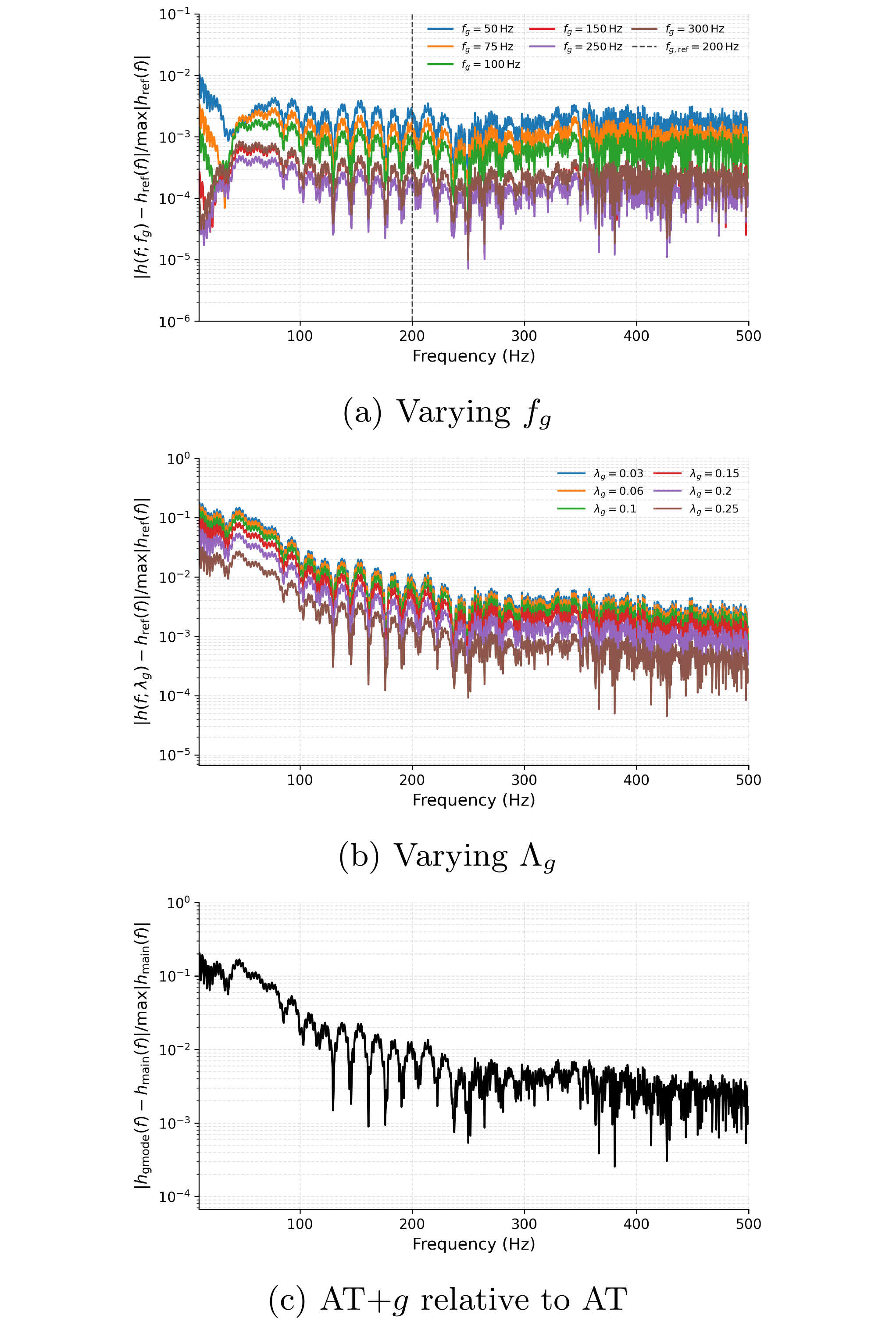}
{%
\phantomsubcaption\label{fig:fg_clean_residual}%
\phantomsubcaption\label{fig:lambdag_clean_residual}%
\phantomsubcaption\label{fig:reference_gmode_perturbation}%
}
\caption{Normalized frequency domain residuals under noiseless conditions. Panels (a) and (b) show the waveform with a resonant $g$ mode phase correction relative to the reference waveform when $f_g$ and $\Lambda_g$ are varied, respectively. Panel (c) shows the normalized frequency domain difference between the AT+$g$ and AT waveforms for the reference event.}
\label{fig:clean_residuals}
\end{figure}

\begin{figure}[t]
\centering
\includegraphics[width=0.88\linewidth]{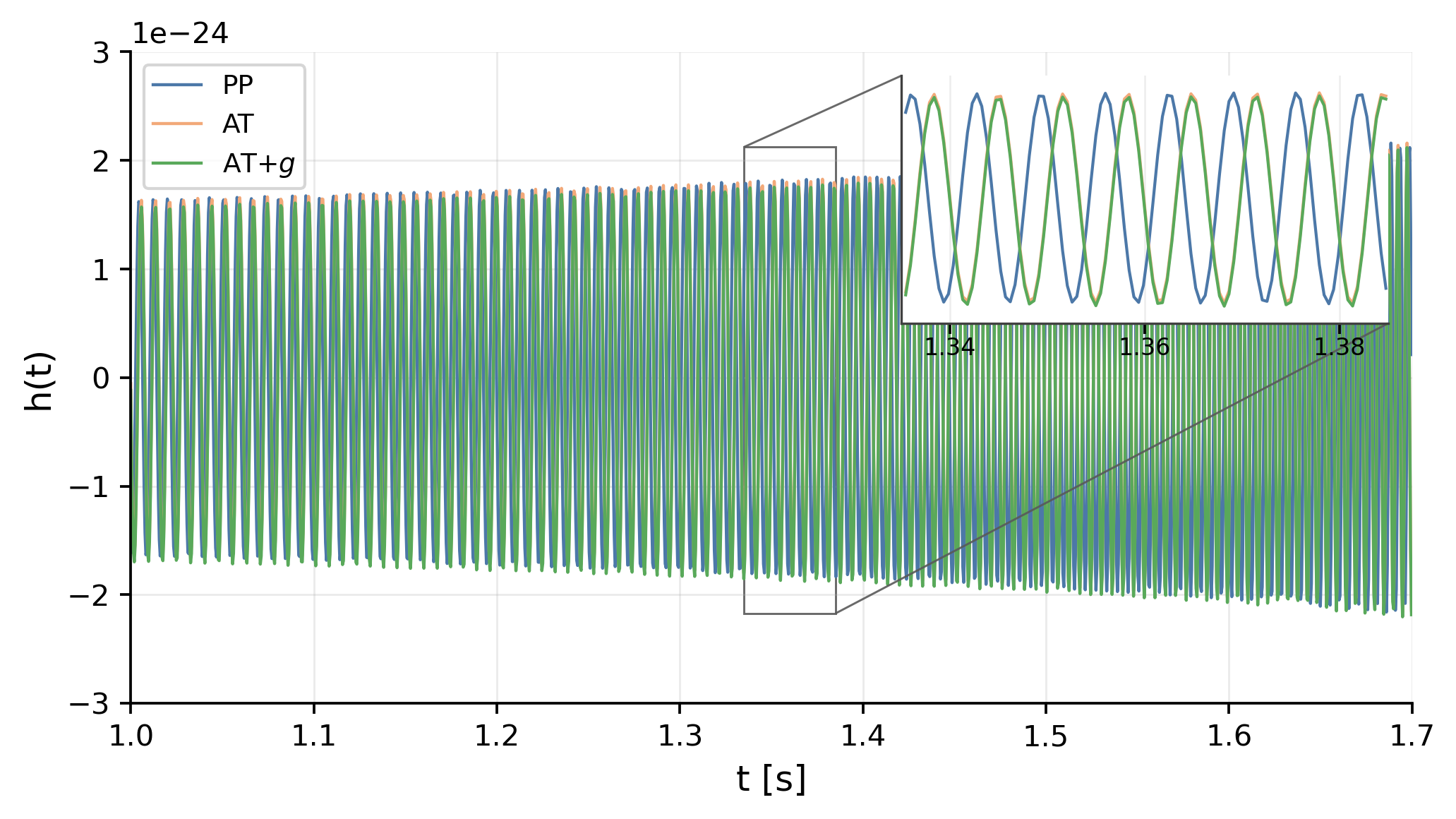}
\caption{Time domain reference waveforms for the same sample used in Fig.~\ref{fig:clean_residuals}. The curves show how the frequency domain resonant phase correction is transformed back to the time domain before detector noise is added and before the strain is supplied to the classifier.}
\label{fig:time_domain_reference}
\end{figure}

We select one sample from the noiseless dataset as a reference event to illustrate how the resonant phase correction affects the waveform morphology. The component masses of this event are $m_1=1.786\,M_\odot$ and $m_2=1.214\,M_\odot$, the tidal deformabilities are {$\Lambda_{\rm tide,1}=763.94$ and $\Lambda_{\rm tide,2}=925.99$}, and the eccentricity at $10\,\mathrm{Hz}$ is $e_{10\mathrm{Hz}}=0.253$. The $g$ mode parameters used for the resonant phase correction are $f_g=200\,\mathrm{Hz}$ and $\Lambda_g=0.30$, with redshift $z=0.366$ and luminosity distance $D_L=1953.61\,\mathrm{Mpc}$. We then vary $f_g$ and $\Lambda_g$ separately, while keeping the other parameters fixed, and compute normalized residuals of the noiseless frequency domain waveform relative to the reference waveform. Figure~\ref{fig:fg_clean_residual} shows that changing $f_g$ shifts and redistributes the main frequency domain contribution of the resonant perturbation. Figure~\ref{fig:lambdag_clean_residual} shows that increasing $\Lambda_g$ increases the relative difference between the corrected waveform and the reference waveform for the same event. Figure~\ref{fig:reference_gmode_perturbation} further compares the normalized difference between the AT+$g$ and AT waveforms for the same reference event, showing that the resonant $g$ mode contribution mainly appears as a weak frequency domain perturbation from low frequencies to several hundred Hz. Figure~\ref{fig:time_domain_reference} shows the corresponding time domain waveforms used as network inputs before detector noise is added. Together these figures connect the frequency domain phase correction to the time domain strain supplied to the classifier.

\subsection{Sample classes and identification task}
\label{subsec:classes}

We define four sample classes according to their physical content. {The PP class is generated using TEOBResumS-Dal\'i with the tidal deformabilities set to zero, $\Lambda_{\rm tide,1}=\Lambda_{\rm tide,2}=0$, and therefore represents eccentric point particle waveforms without tidal corrections. The AT class is generated with nonzero $\Lambda_{\rm tide,1}$ and $\Lambda_{\rm tide,2}$, and includes the adiabatic tidal effects modeled in TEOBResumS-Dal\'i, but does not include the additional resonant $g$ mode phase correction. The AT+$g$ class further incorporates the multiharmonic resonant $g$ mode phase correction on top of the AT waveform.} The PN class contains only colored noise matched to the detector PSD and does not include a GW signal. With these definitions, the identification problem tests whether noisy data can separate pure noise, nontidal signals, adiabatic tidal backgrounds, and weak resonant phase shifts. The AT+$g$ class therefore represents a resonant phase perturbation embedded in an adiabatic tidal background. In addition to the four class assignments, we store physical parameters such as $e_{10{\rm Hz}}$, $\Lambda_g$, and $f_g$, which are later used to analyze how identification performance varies with these parameters.

\begin{figure*}[t]
\centering
\includegraphics[width=\textwidth]{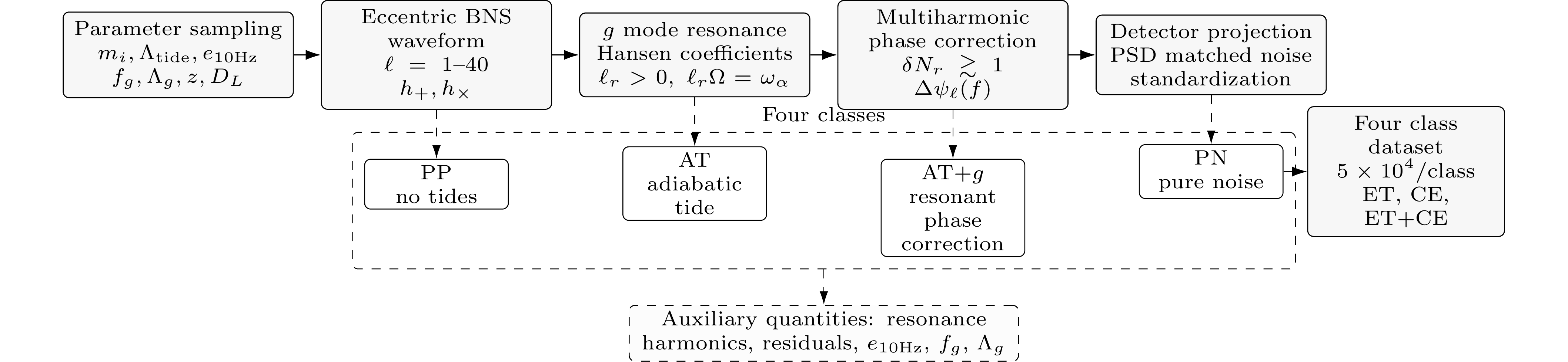}
\caption{Data generation pipeline. The procedure starts from physical parameters, source geometry, and distance information to generate an eccentric BNS reference waveform. Coherent $g$ mode resonances are then selected in the eccentric orbit harmonic framework and represented as resonant phase corrections. The physical classes are projected onto ET/CE detectors, PSD matched colored Gaussian noise is added, and standardized PP, AT, AT+$g$, and PN samples are obtained. Auxiliary quantities are stored for the conditional analysis.}
\label{fig:pipeline}
\end{figure*}

Figure~\ref{fig:pipeline} summarizes the full pipeline from parameter sampling, waveform generation, and resonant phase correction to detector projection, noise addition, and sample construction.

\section{Matched Filtering and Deep Learning}
\label{sec:network}

This section describes the matched filtering method, the deep learning classifier, and the training and evaluation procedure used to test the identifiability of weak resonant phase shifts in noisy time domain strain.

\subsection{Matched filtering}
\label{subsec:matched_filtering}

We use the standard noise weighted inner product for detector $I$ \citep{Allen2012FindChirp},
\begin{align}
(a|b)_I
=
4\,{\rm Re}\int_{f_{\min}}^{f_{\max}}
\frac{\tilde{a}_I(f)\tilde{b}_I^*(f)}
{S_{n,I}(f)}\,df ,
\end{align}
where $S_{n,I}(f)$ is the one sided detector PSD, and the integral is evaluated over the analysis frequency band used for the waveform comparison. For a detector configuration $q$, with detector set $\mathcal{I}_q$, the network inner product is
\begin{align}
(a|b)_q
=
\sum_{I\in\mathcal{I}_q}(a_I|b_I)_I .
\end{align}
Let $c\in\{\mathrm{PP},\mathrm{AT},\mathrm{AT}+g\}$ denote the three signal containing classes. The template bank for class $c$ is
\begin{align}
\mathcal{B}_c^{(q)}
=\left\{h_{c,k}^{(q)}(\theta_k)\right\}_{k=1}^{N_c},
\qquad
c\in\{\mathrm{PP},\mathrm{AT},\mathrm{AT}+g\},
\end{align}
where $N_c$ is the number of templates for class $c$. The templates are noiseless waveforms drawn from the same simulated database, excluding the fixed evaluation samples, and $\theta_k$ denotes the source parameters of the $k$th template. For a noisy strain segment $d^{(q)}$, the matched filter (MF) output for template $h_{c,k}^{(q)}$ is the noise weighted correlation normalized by the template norm,
\begin{align}
\rho_{{\rm MF},c,k}^{(q)}(d)
=
\frac{
\left(d^{(q)}|h_{c,k}^{(q)}\right)_q
}{
\sqrt{
\left(h_{c,k}^{(q)}|h_{c,k}^{(q)}\right)_q
}
}.
\end{align}
For the four class assignment we use the corresponding normalized match,
\begin{align}
M_{c,k}^{(q)}
=
\frac{
\left(d^{(q)}|h_{c,k}^{(q)}\right)_q
}{
\sqrt{
\left(d^{(q)}|d^{(q)}\right)_q
\left(h_{c,k}^{(q)}|h_{c,k}^{(q)}\right)_q
}
}.
\end{align}
The normalization by $\sqrt{(d^{(q)}|d^{(q)})_q}$ does not change the ordering of templates for a fixed data segment, but makes the statistic a dimensionless match. The comparison is performed on the same final premerger time window used throughout the analysis, so the statistic measures waveform consistency within this fixed segment rather than carrying out a full search over arrival time. For each physical signal family, we retain the largest match over its template bank,
\begin{align}
\rho_c^{(q)}(d)
&=
\max_{h_{c,k}^{(q)}\in\mathcal{B}_c^{(q)}} M_{c,k}^{(q)},\\
\rho_{\max}^{(q)}(d)
&=
\max_{c\in\{\mathrm{PP},\mathrm{AT},\mathrm{AT}+g\}}
\rho_c^{(q)}(d).
\end{align}

The PN class has no signal template. We therefore determine a threshold $\tau^{(q)}$ from a separate subset disjoint from both the template bank and the evaluation set by maximizing the macro averaged $F_1$ on that subset. The four class prediction rule is
\begin{align}
\hat{y}(d^{(q)})
=
\begin{cases}
\mathrm{PN}, & \rho_{\max}^{(q)}(d) < \tau^{(q)},\\[3pt]
\displaystyle \arg\max_{c\in\{\mathrm{PP},\mathrm{AT},\mathrm{AT}+g\}}
\rho_c^{(q)}(d), & \rho_{\max}^{(q)}(d) \ge \tau^{(q)}.
\end{cases}
\end{align}
Thus the procedure first tests whether the strain is consistent with any signal template and then assigns the most strongly matched signal class. We report the matched filtering results on the same fixed evaluation samples used for the neural network models.

\subsection{Network architecture}
\label{subsec:network_architecture}

\begin{figure*}[t]
\centering
\includegraphics[width=\textwidth]{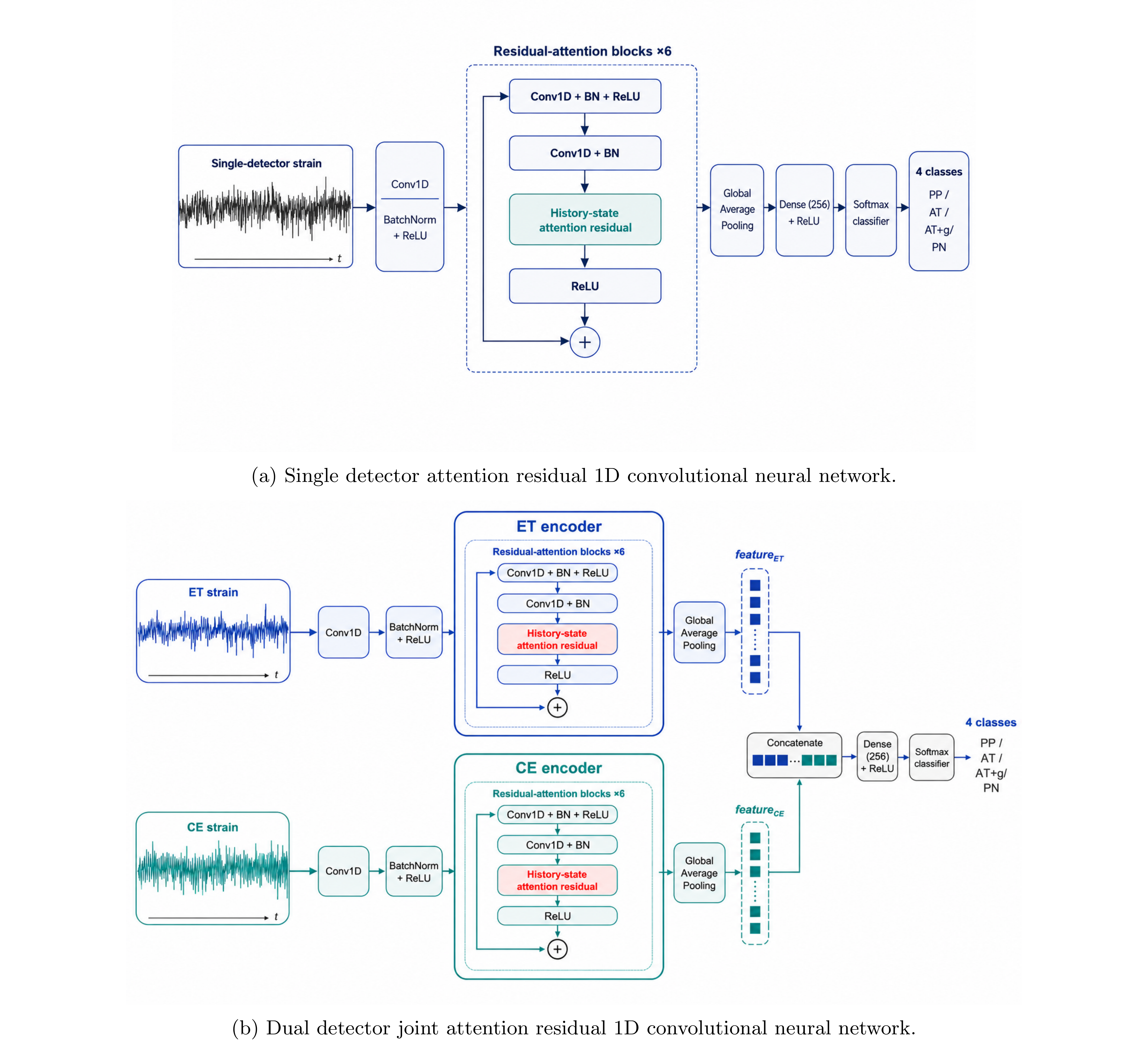}
{%
\phantomsubcaption\label{fig:arch_single}%
\phantomsubcaption\label{fig:arch_dual}%
}
\caption{{Network architecture used in this work. Panel (a) shows the single input version, whose backbone consists of an initial convolutional feature extractor, historical state attention residual blocks, global average pooling, and a fully connected output layer. Panel (b) shows the dual input version, where the two input sequences are encoded by isomorphic branches and then jointly classified after feature vector concatenation.}}
\label{fig:arch}
\end{figure*}

{Using the four sample classes defined above, we train classifiers to distinguish PP, AT, AT+$g$, and PN samples from noisy time domain strain. The models take either one input time series or two input time series processed by separate encoder branches. All three models employ 1D convolutional feature extraction and historical state attention residual blocks. Convolutional networks learn local features through shared kernels \citep{Goodfellow2016DeepLearning}, while residual architectures retain information from earlier representation levels \citep{He2016DeepResidualLearning}. Here the attention weights aggregate states across network depth, following the attention residual construction \citep{Kimi2026AttentionResiduals}. This mechanism motivates retaining several representation levels while processing small waveform differences. Figure~\ref{fig:arch} shows the original single input and dual input models.}

{For the single input model, the input time series first passes through an initial convolution, Batch Normalization, and ReLU activation layer to produce the initial feature $\mathbf{H}^{(0)}$. The network then stacks six historical state attention residual blocks. In block $l$, two convolutional layers first produce the current feature $\tilde{\mathbf{H}}^{(l)}$. All previous historical states and the current feature are then jointly passed into the attention residual module for weighted fusion,}
\begin{align}
\tilde{\mathbf{H}}^{(l)} &= \mathcal{R}^{(l)}\big(\mathbf{H}^{(l-1)}\big),\\
\mathbf{H}^{(l)} &= \phi\!\left(\sum_{i=0}^{l}\alpha_i^{(l)}\,\mathbf{V}_i^{(l)}\right).
\end{align}
Here $\mathcal{R}^{(l)}$ denotes the local feature transformation formed by two 1D convolutional layers, $\mathbf{V}_i^{(l)}\in\{\mathbf{H}^{(0)},\mathbf{H}^{(1)},\ldots,\mathbf{H}^{(l-1)},\tilde{\mathbf{H}}^{(l)}\}$ denotes the historical states and current feature participating in fusion, $\alpha_i^{(l)}$ is the attention weight, and $\phi(\cdot)$ is the pointwise ReLU activation. Each state is first summarized by global average pooling and then passed through RMSNorm, a shared linear mapping, and softmax to obtain $\alpha_i^{(l)}$. The encoder output is then passed through global average pooling and a hidden fully connected layer, followed by a softmax output over the four classes.

{In the dual input mode, the two input time series are processed by parallel encoder branches. The two branches use isomorphic historical state attention residual encoders, but retain their detector specific feature statistics before fusion, allowing the model to adapt to different detector noise spectra and response functions. Each branch outputs a global feature vector, and the two vectors are concatenated in feature space and passed through a fully connected layer to produce the joint prediction. The near-zero AT Recall of the ET-only classifier describes its final four-class decisions; it rounds to $0.000$ at the precision used in Table~\ref{tab:overall_metrics_singlecol}. In the joint model, the classification head combines continuous features from both detectors and can learn a decision boundary different from those of the single-detector classifiers. This feature fusion provides a plausible explanation for the improved AT recognition reported in Sec.~\ref{subsec:overall_performance}.}

\begin{table}[t]
\centering
\caption{{Encoder and classification head specifications. All convolutional layers use stride 1, same padding, and He normal initialization. BN denotes Batch Normalization.}}
\label{tab:architecture_specs}
\footnotesize
\setlength{\tabcolsep}{3pt}
\resizebox{\columnwidth}{!}{\begin{tabular}{ll}
\hline\hline
Specification & Setting \\
\hline
Input per encoder & $4096\times1$ \\
Stem & Conv(128, 2), BN, ReLU \\
Blocks per encoder & 6 single input; 6 dual input \\
Each block & Conv(128, 2), BN, ReLU; Conv(128, 2), BN \\
State aggregation & Historical attention, then ReLU \\
Encoder output & GAP, 128 features \\
Dual encoder fusion & Concatenate, 256 features \\
Classification head & Dense(256), ReLU; Dense(4), softmax \\
\hline\hline
\end{tabular}}
\end{table}

{In Table~\ref{tab:architecture_specs}, Conv($C,k$) denotes a convolution with $C$ filters and kernel size $k$, and GAP denotes global average pooling. Batch Normalization uses momentum $0.99$ and $\epsilon=10^{-3}$. The attention key dimension is 32, the key projection has no bias, and RMSNorm uses $\epsilon=10^{-6}$. Each block has its own learned query and key projection, with weights shared across the historical states entering that block.}

{To examine the conventional alternatives, we train three supplementary models: a single channel convolutional neural network (CNN) without attention for ET, a two channel CNN without attention for ET+CE, and a two channel single encoder historical state attention model for ET+CE. All encoders contain six blocks. The single channel CNN retains 128 filters. The two channel controls use respectively 189 and 185 filters, with the same kernel, stride, normalization, pooling and classification head settings; the attention dimension remains 32. The two channel controls and the proposed dual encoder each have approximately $9.15\times10^5$ trainable parameters. The single channel CNN and the single input attention model have approximately $4.33\times10^5$ and $4.58\times10^5$ trainable parameters, respectively. These counts are estimated from the architectural specifications. Within each detector configuration, the supplementary models and their corresponding original model use identical training and validation samples, preprocessing parameters, training settings and the same fixed independent evaluation dataset. Only the network architecture is changed in these comparisons.}

\subsection{Training strategy}
\label{subsec:training}

{For the original attention models, the 200,000 development samples are split into training and validation subsets in a $9:1$ ratio, giving 180,000 and 20,000 samples, respectively. The training subset is used for gradient updates, while the validation subset is used only for learning rate adjustment, early stopping and model selection. A separate dataset of 10,000 samples is used exclusively for final evaluation and is excluded from the training and validation subsets. The original models and supplementary controls are evaluated on this same fixed independent dataset after the training configurations and model weights have been fixed.}

\begin{figure}[t]
\centering
\includegraphics[width=0.88\linewidth]{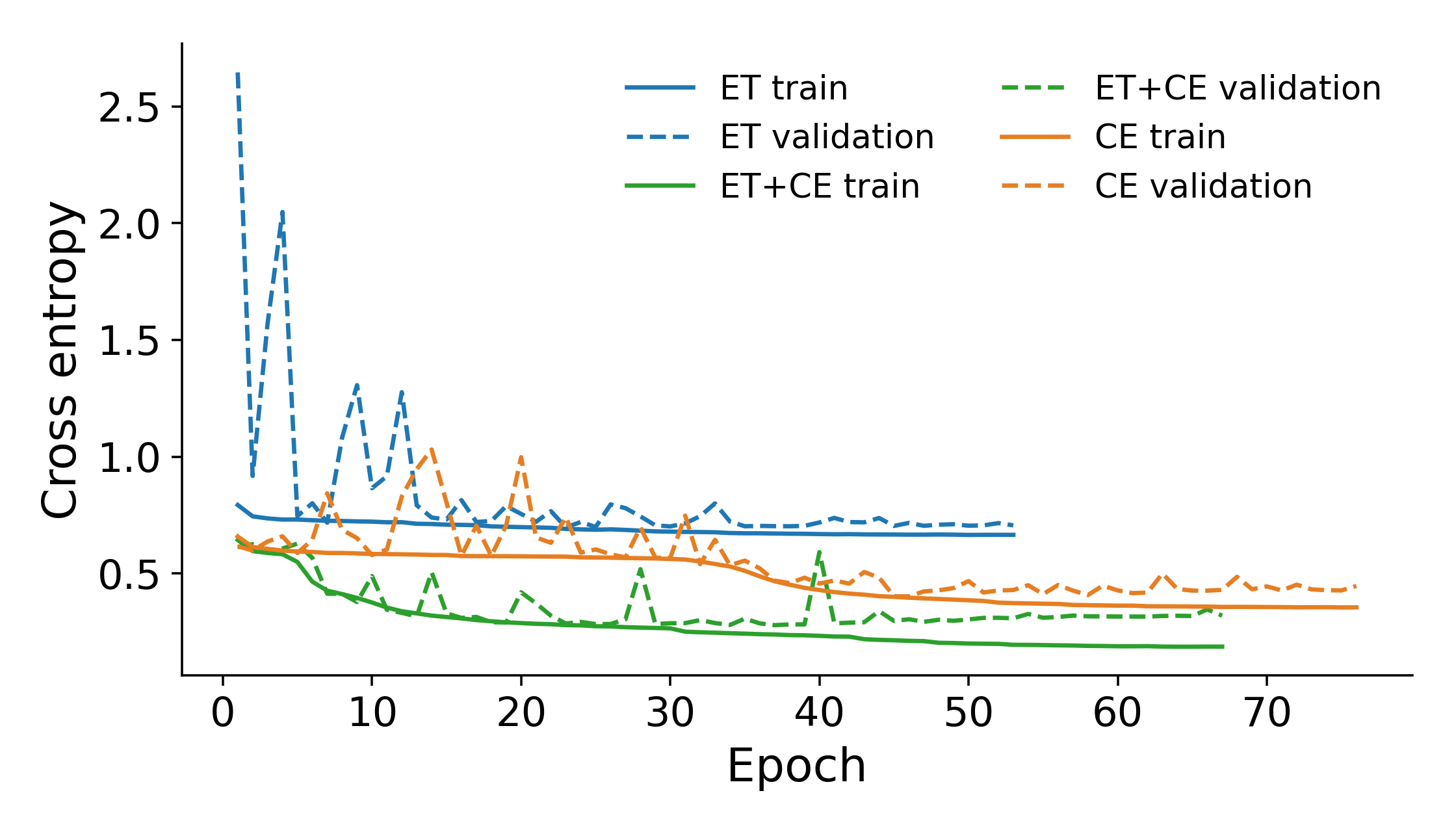}
\caption{{Training (solid) and validation (dashed) cross entropy for the three original attention models. The selected models have the minimum validation loss, at epochs 23 for ET, 46 for CE, and 37 for ET+CE.}}
\label{fig:learning_loss}
\end{figure}

\begin{figure}[t]
\centering
\includegraphics[width=0.88\linewidth]{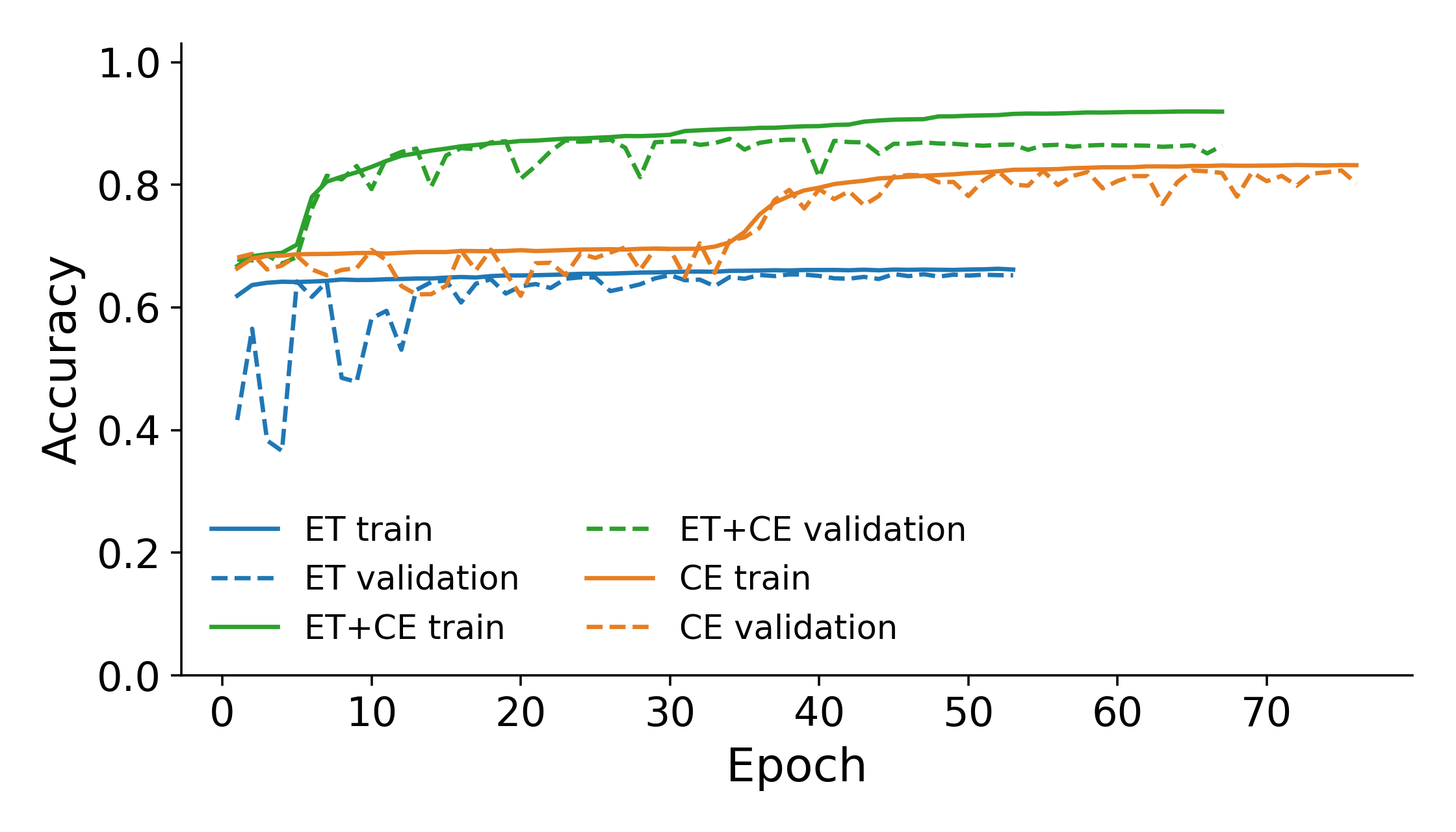}
\caption{{Training (solid) and validation (dashed) accuracy for the same models. The validation subset is not used for gradient updates. Model selection follows validation loss rather than the largest training accuracy.}}
\label{fig:learning_accuracy}
\end{figure}

{The original attention models are trained with Adam using initial learning rate $10^{-4}$, $\beta_1=0.9$, $\beta_2=0.999$, $\epsilon=10^{-7}$, and zero weight decay. The loss is sparse categorical cross entropy. Models with one input sequence use batch size 32 and models with two input sequences use batch size 16, with a maximum of 200 epochs. ReduceLROnPlateau monitors the validation loss, reduces the learning rate by a factor of $0.5$ after five epochs without an improvement of at least $10^{-4}$, and uses a lower limit of $10^{-7}$. Early stopping monitors the validation loss with patience 30 and restores the weights with the minimum validation loss. Figures~\ref{fig:learning_loss} and~\ref{fig:learning_accuracy} show the training and validation loss and accuracy, respectively, with the selected epochs stated in the captions. The training and validation subsets are formed with random state 42, and the model weights are randomly initialized. Training and inference use an NVIDIA A800 80 GB graphics processing unit (GPU).}

{The three supplementary models use the same training and validation samples and preprocessing parameters as the corresponding original models, with the detector-specific training settings stated above. The selected epochs are 9, 14 and 39 for the ET CNN, two channel CNN and single encoder attention model, respectively.}


\section{Results and Analysis}
\label{sec:results}

This section examines the identifiability of the four simulated classes under different detector configurations and then analyzes how the recovery of AT+$g$ depends on physical parameters. In the four class task, PN serves mainly as a noise control, while the AT/AT+$g$ confusion directly probes the separability of an additional resonant phase shift on top of the adiabatic tidal background. The analysis therefore reports both global classification metrics and diagnostics tied to the AT/AT+$g$ boundary.

All neural network models and the matched filtering method are evaluated on the same fixed set of $10^4$ samples. Overall performance is quantified using Accuracy and Average $F_1$. Class level behavior is described by Precision, Recall, and $F_1$ computed from the multiclass confusion matrix, with particular attention to the Recall of AT+$g$ because it measures the recovery of weak resonant phase samples. {For the AT+$g$ class, Recall is the fraction of true AT+$g$ samples that are correctly identified.} Score level separability is evaluated using receiver operating characteristic area under the curve (ROC AUC) and precision recall area under the curve (PR AUC); micro averaged values summarize the four class output, while class level curves are used to examine the separability of individual classes. Using the same evaluation samples allows the overall metrics, class level metrics, conditional statistics, and matched filtering results to be directly compared.

To connect the identification performance with the underlying physical parameters, we further compute conditional statistics in bins of $e_{10\mathrm{Hz}}$, $\Lambda_g$, and $f_g$. These parameters respectively affect the eccentric harmonic weights, the resonant coupling strength, and the frequency range in which the resonant phase perturbation enters the detector sensitive band.

\subsection{Overall Identifiability Across Detector Configurations}
\label{subsec:overall_performance}

{Table~\ref{tab:global_summary} reports global metrics for all methods evaluated on the same independent samples, separately from the class-specific values in Table~\ref{tab:overall_metrics_singlecol}. The ET, CE and ET+CE neural networks give Accuracy $0.650$, $0.809$, and $0.875$ and macro averaged $F_1$ $0.565$, $0.805$, and $0.873$. Matched filtering gives Accuracy $0.534$, $0.703$, and $0.718$. ET+CE gives the highest neural classification performance under this evaluation setup.}

\begin{table}[t]\centering
\caption{{Overall metrics on the independent evaluation dataset.}}
\label{tab:global_summary}
\begin{tabular}{llcc}\hline\hline
Method & Detectors & Accuracy & Macro $F_1$ \\\hline
MF & ET & 0.534 & 0.527 \\
MF & CE & 0.703 & 0.702 \\
MF & ET+CE & 0.718 & 0.717 \\
DL & ET & 0.650 & 0.565 \\
DL & CE & 0.809 & 0.805 \\
DL & ET+CE & 0.875 & 0.873 \\
\hline\hline\end{tabular}\end{table}

\begin{table}[t]\centering\small
\setlength{\tabcolsep}{4pt}\renewcommand{\arraystretch}{0.86}
\caption{{Class-specific metrics on the independent evaluation dataset. Average denotes the macro average. The near-zero ET AT Recall rounds to $0.000$ at three decimal places.}}
\label{tab:overall_metrics_singlecol}
\begin{tabular*}{\columnwidth}{@{\extracolsep{\fill}}llccc@{}}\hline\hline
Method & Class & Prec. & Rec. & $F_1$ \\\hline
\multirow{18}{*}{MF} & \multicolumn{4}{c}{ET} \\
 & PP & 0.805 & 0.543 & 0.649 \\
 & AT & 0.324 & 0.294 & 0.308 \\
 & AT+$g$ & 0.533 & 0.451 & 0.489 \\
 & PN & 0.540 & 0.850 & 0.661 \\
 & Average & 0.551 & 0.534 & 0.527 \\
 & \multicolumn{4}{c}{CE} \\
 & PP & 0.957 & 0.927 & 0.942 \\
 & AT & 0.460 & 0.482 & 0.471 \\
 & AT+$g$ & 0.482 & 0.453 & 0.467 \\
 & PN & 0.911 & 0.951 & 0.930 \\
 & Average & 0.702 & 0.703 & 0.702 \\
 & \multicolumn{4}{c}{ET+CE} \\
 & PP & 0.964 & 0.953 & 0.958 \\
 & AT & 0.471 & 0.495 & 0.483 \\
 & AT+$g$ & 0.489 & 0.452 & 0.470 \\
 & PN & 0.940 & 0.974 & 0.957 \\
 & Average & 0.716 & 0.718 & 0.717 \\
\hline
\multirow{18}{*}{DL} & \multicolumn{4}{c}{ET} \\
 & PP & 0.465 & 0.695 & 0.557 \\
 & AT & 0.500 & 0.000 & 0.001 \\
 & AT+$g$ & 0.664 & 0.957 & 0.784 \\
 & PN & 0.890 & 0.947 & 0.918 \\
 & Average & 0.630 & 0.650 & 0.565 \\
 & \multicolumn{4}{c}{CE} \\
 & PP & 0.720 & 0.765 & 0.742 \\
 & AT & 0.681 & 0.562 & 0.615 \\
 & AT+$g$ & 0.819 & 0.910 & 0.862 \\
 & PN & 0.999 & 1.000 & 1.000 \\
 & Average & 0.805 & 0.809 & 0.805 \\
 & \multicolumn{4}{c}{ET+CE} \\
 & PP & 0.887 & 0.825 & 0.855 \\
 & AT & 0.771 & 0.731 & 0.750 \\
 & AT+$g$ & 0.840 & 0.943 & 0.888 \\
 & PN & 1.000 & 1.000 & 1.000 \\
 & Average & 0.875 & 0.875 & 0.873 \\
\hline
\hline\end{tabular*}\end{table}

{Matched filtering has lower Accuracy and AT+$g$ Recall under all three configurations. Its AT+$g$ Recall values are $0.451$, $0.453$, and $0.452$, compared with $0.957$, $0.910$, and $0.943$ for the neural classifiers.}

{ET+CE gives PN Precision, Recall and $F_1$ of $1.000$. Its PP, AT and AT+$g$ $F_1$ scores are $0.855$, $0.750$, and $0.888$. AT Recall rises from $0.562$ for CE to $0.731$ for ET+CE, indicating more balanced recovery of the signal classes.}

\begin{figure*}[t]
\centering
\includegraphics[width=0.82\textwidth]{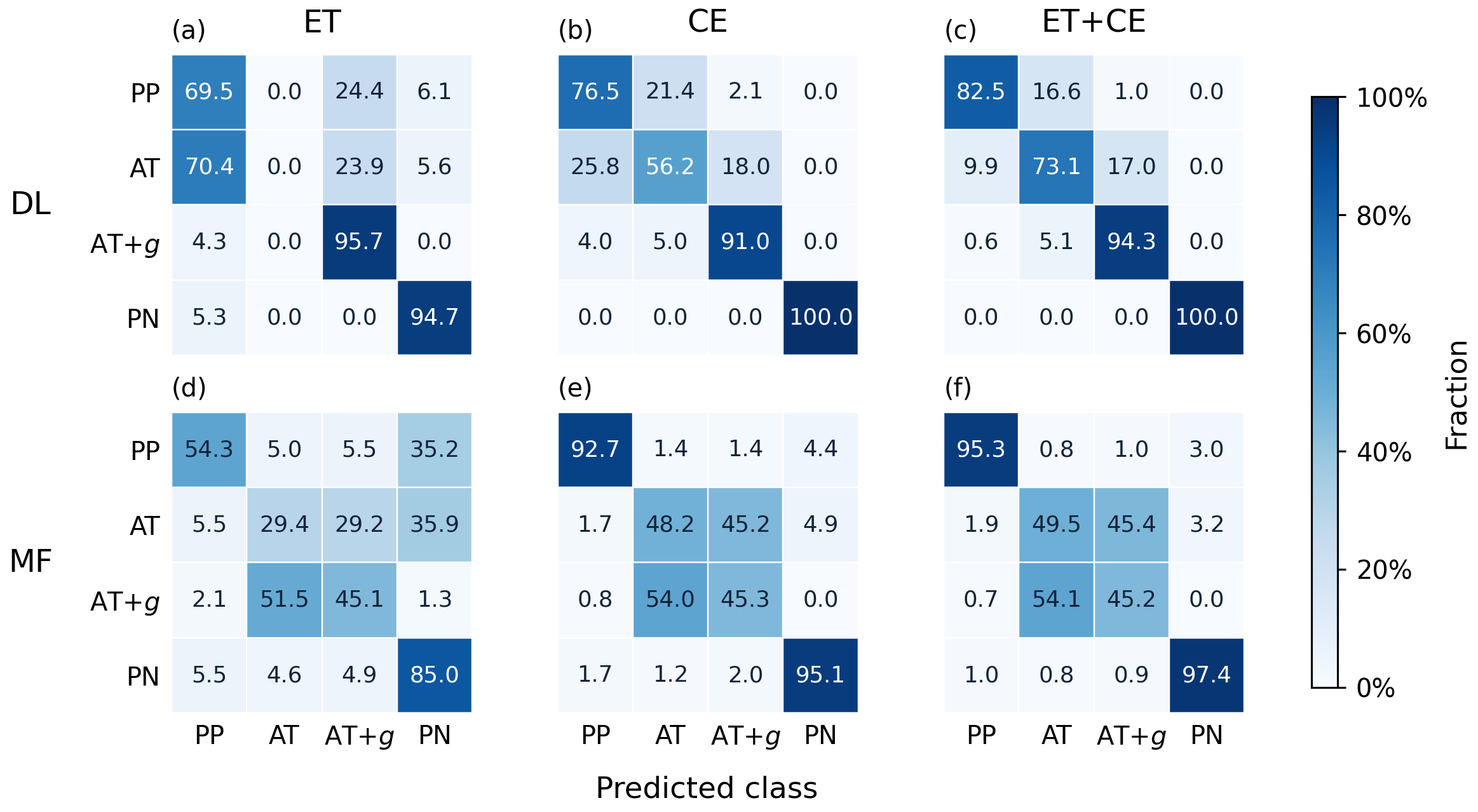}
\caption{{Confusion matrices on the independent evaluation dataset. The columns correspond to ET, CE, and ET+CE. The upper row shows deep learning (DL), and the lower row matched filtering (MF). Within each matrix, rows denote true classes and columns predicted classes. Entries are row-normalized percentages, with a common color scale.}}
\label{fig:cm_panel}
\label{fig:cm_ce}
\label{fig:cm_etce}
\end{figure*}

{Figure~\ref{fig:cm_panel} shows all six confusion matrices. The ET neural model has AT Recall close to zero, displayed as $0.000$ in Table~\ref{tab:overall_metrics_singlecol}. The dominant AT error is AT$\rightarrow$PP, accounting for $70.40\%$ of AT samples; $23.92\%$ are assigned to AT+$g$. AT$\rightarrow$PP confusion decreases from $25.76\%$ for CE to $9.88\%$ for ET+CE. AT$\rightarrow$AT+$g$ also decreases slightly, while the reverse AT+$g\rightarrow$AT rate remains similar for CE and ET+CE.}

{PN is recovered reliably by the neural models, with Recall $0.947$, $1.000$, and $1.000$. Remaining signal errors include PP/AT confusion and confusion between AT and AT+$g$. We use AT+$g$ Recall, PR AUC and the two AT/AT+$g$ error directions as diagnostics below.}

\begin{figure}[h]
\centering
\includegraphics[width=0.88\linewidth]{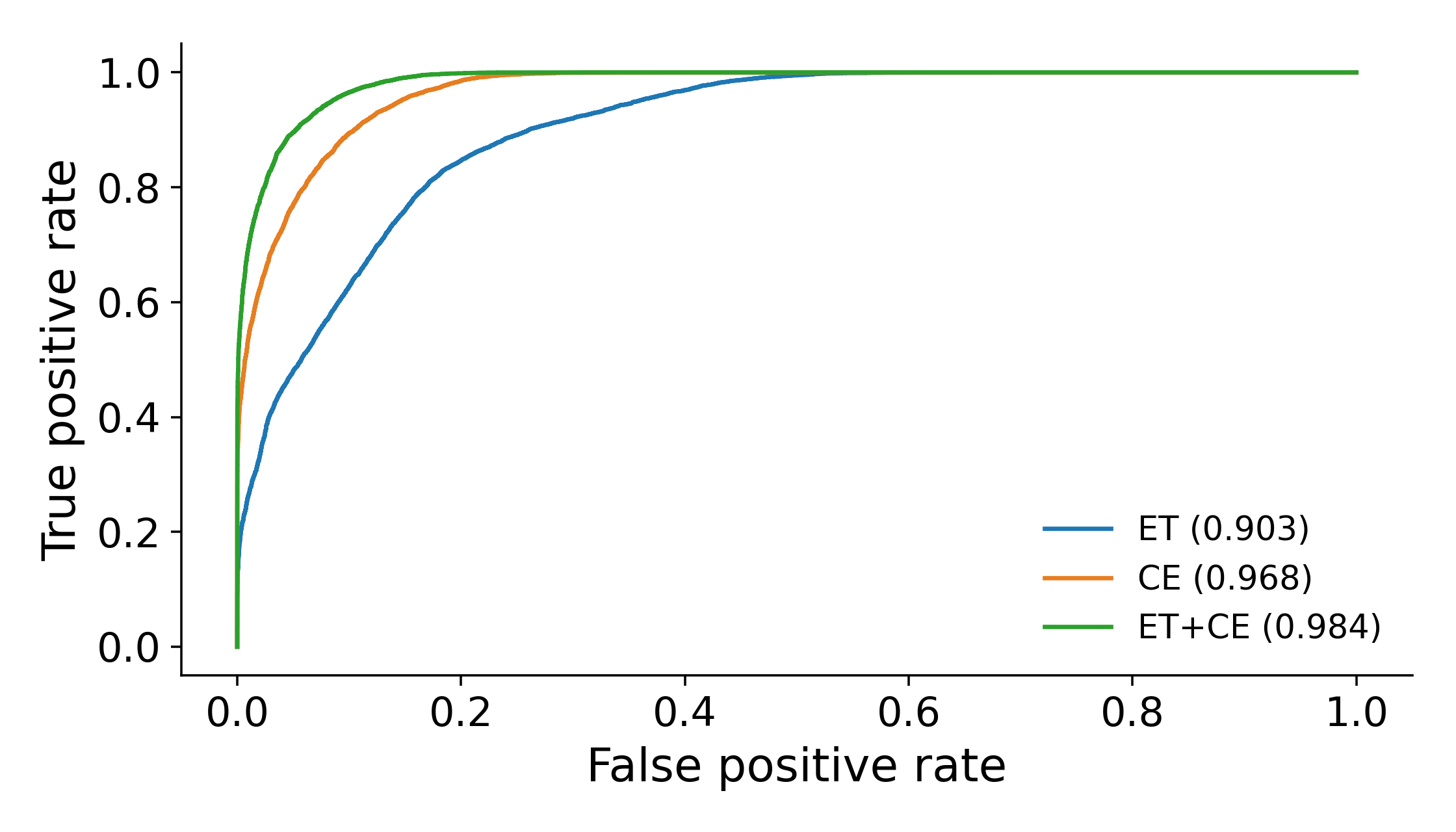}
\caption{{Micro averaged ROC curves for ET, CE, and ET+CE on the independent evaluation dataset.}}
\label{fig:roc_average}
\end{figure}

\begin{figure}[!tb]
\centering
\includegraphics[width=\linewidth]{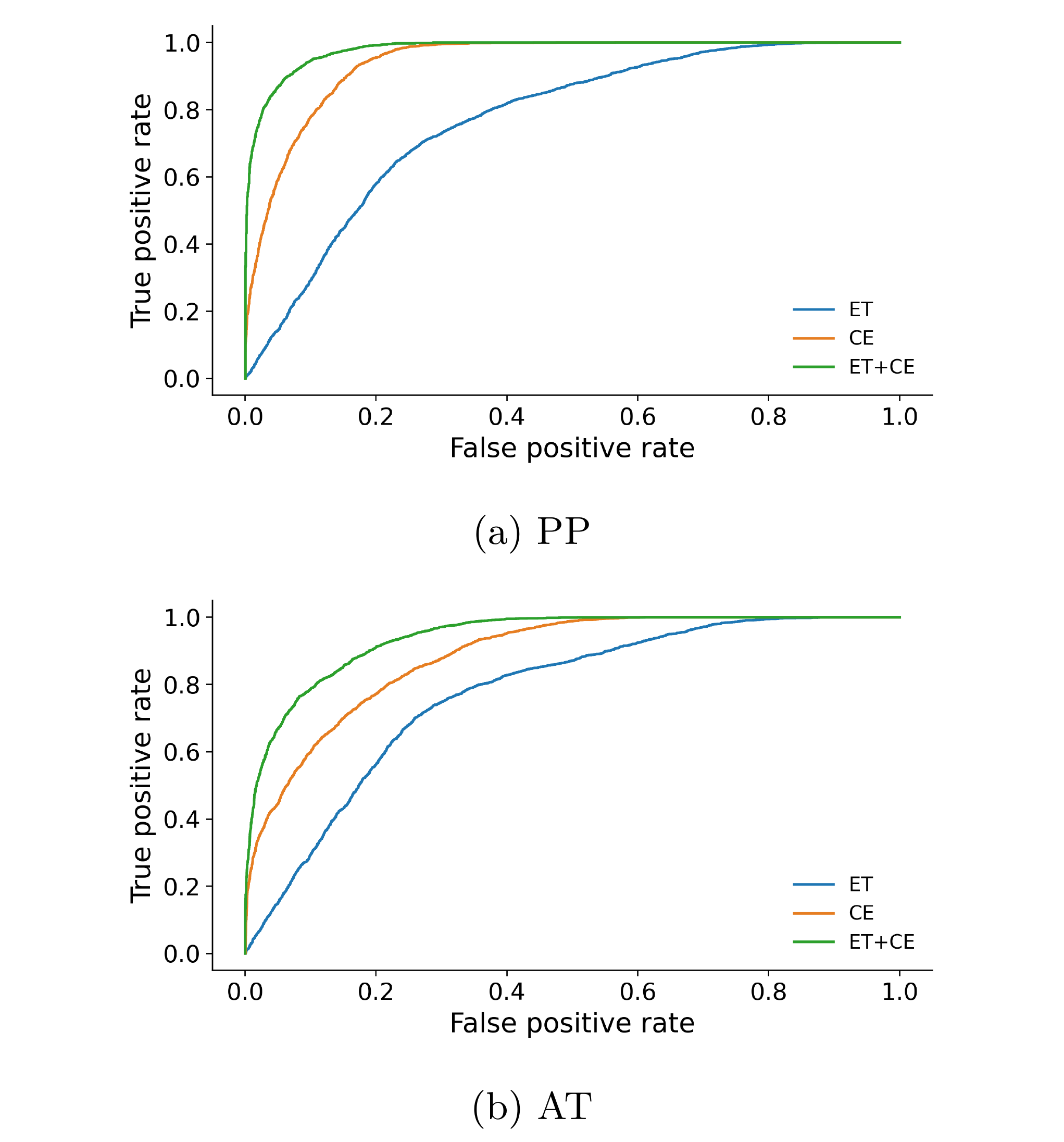}
{%
\phantomsubcaption\label{subfig:roc_pp}%
\phantomsubcaption\label{subfig:roc_at}%
}
\caption{{Class-specific ROC curves for PP (a) and AT (b).}}
\label{fig:roc_classes}
\end{figure}

\begin{figure}[!tb]
\centering
\includegraphics[width=\linewidth]{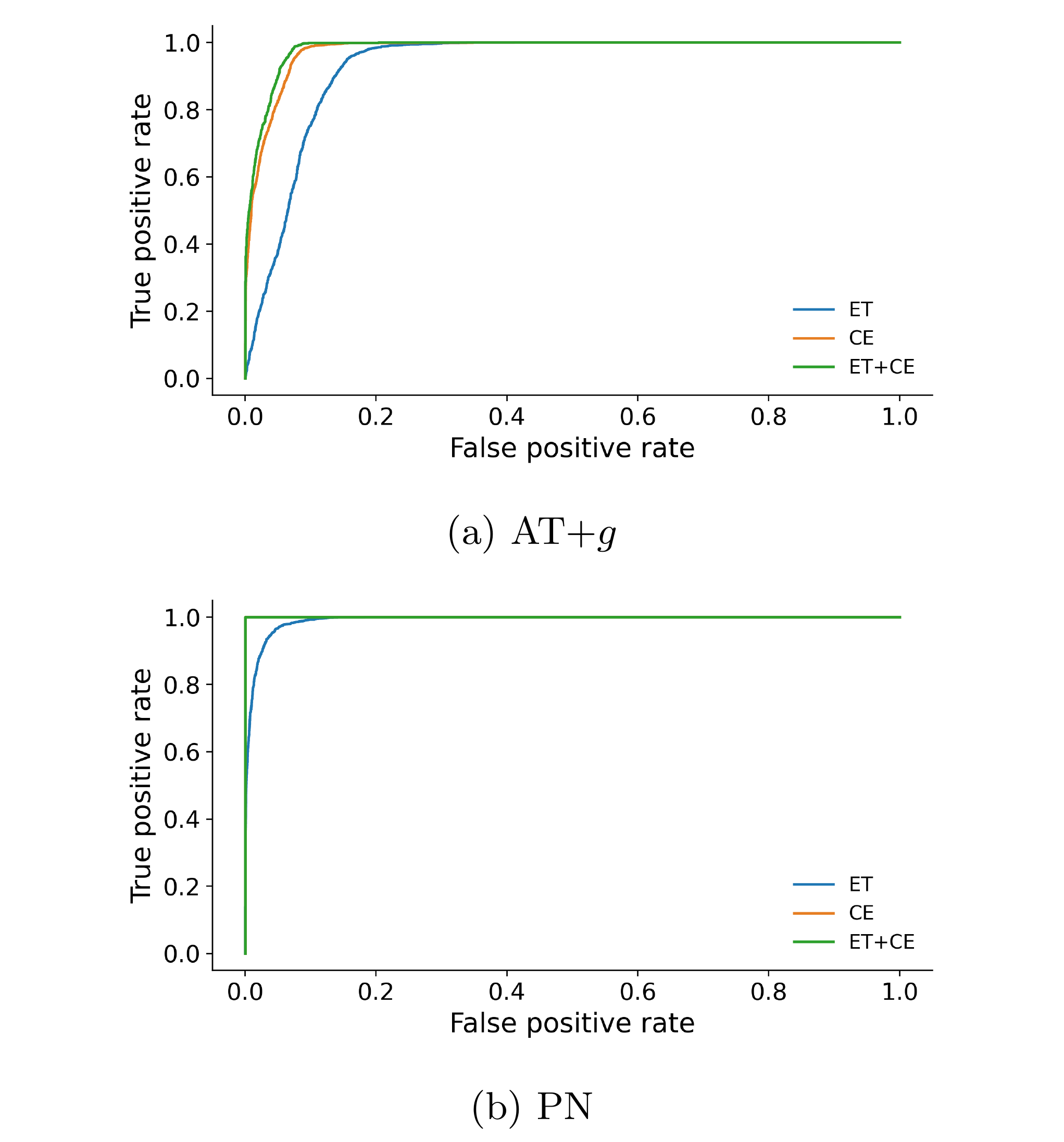}
{%
\phantomsubcaption\label{subfig:roc_gmode}%
\phantomsubcaption\label{subfig:roc_pn}%
}
\caption{{Class-specific ROC curves for AT+$g$ (a) and PN (b).}}
\label{fig:roc_classes_gpn}
\end{figure}

{Figure~\ref{fig:roc_average} shows micro averaged ROC curves, and Figs.~\ref{fig:roc_classes} and~\ref{fig:roc_classes_gpn} show the class-specific curves. ET+CE has micro ROC AUC $0.984$ and micro PR AUC $0.956$, higher than the corresponding ET and CE values. These score-based metrics complement the classification results.}

{The ET model has AT+$g$ Recall $0.957$ but AT Recall close to zero, so resonant-class recovery must be interpreted together with false assignments from PP and AT. CE gives Accuracy $0.809$ and AT+$g$ Recall $0.910$. ET+CE gives Accuracy $0.875$ and AT+$g$ Recall $0.943$, with AT Recall $0.731$.}

This behavior is consistent with the physical nature of the $g$ mode resonance. The $g$ mode mainly leaves distributed perturbations in the frequency domain phase through the multiharmonic structure of eccentric orbits. If the adiabatic tidal phase, mass parameter variation, and noise fluctuations produce similar waveform responses in some regions, AT and AT+$g$ can be confused. These trends motivate the use of AT+$g$ Recall, PR AUC, and the two AT/AT+$g$ misclassification directions as diagnostics of weak phase identifiability in the conditional analysis below.

\begin{table}[t]
\centering
\footnotesize
\setlength{\tabcolsep}{3pt}
\caption{{ET+CE architecture comparison on the same independent evaluation dataset.}}
\label{tab:architecture_comparison}
\begin{tabular*}{\columnwidth}{@{\extracolsep{\fill}}lcc@{}}
\toprule
ET+CE architecture & Accuracy & Macro $F_1$ \\
\midrule
Proposed dual encoder & 0.875 & 0.873 \\
Two channel CNN & 0.8431 & 0.8425 \\
Single encoder attention & 0.8437 & 0.8366 \\
\bottomrule
\end{tabular*}
\end{table}

{For the single detector comparison, the ET CNN gives Accuracy $0.6462$ and macro averaged $F_1$ $0.5622$, close to $0.650$ and $0.565$ for the ET attention model. Table~\ref{tab:architecture_comparison} directly compares the three ET+CE architectures with comparable parameter counts. The proposed dual encoder gives Accuracy $0.875$ and macro averaged $F_1$ $0.873$, higher than $0.8431$ and $0.8425$ for the two channel CNN and $0.8437$ and $0.8366$ for the single encoder attention model. These results support the dual encoder choice for overall four class classification under the common training and evaluation setup. The AT+$g$ Recall values are $0.943$, $0.8856$ and $0.9848$ for the dual encoder, two channel CNN and single encoder attention model, respectively. The single encoder attention model has the highest resonant class Recall in this comparison, while the dual encoder gives the highest overall Accuracy and macro averaged $F_1$.}

\subsection{Dependence on physical conditions}
\label{subsec:physical_dependence}

We evaluate the trained classifiers in bins of $e_{10\mathrm{Hz}}$, $\Lambda_g$, and $f_g$, using AT+$g$ Precision, Recall, PR AUC, and the fraction misclassified as AT. The matched filtering result is used as a global four class reference; the binned analysis is restricted to the neural network outputs because it is intended to diagnose the parameter dependence of AT+$g$ identifiability.

\begin{figure*}[t]
\centering
\includegraphics[width=0.70\textwidth]{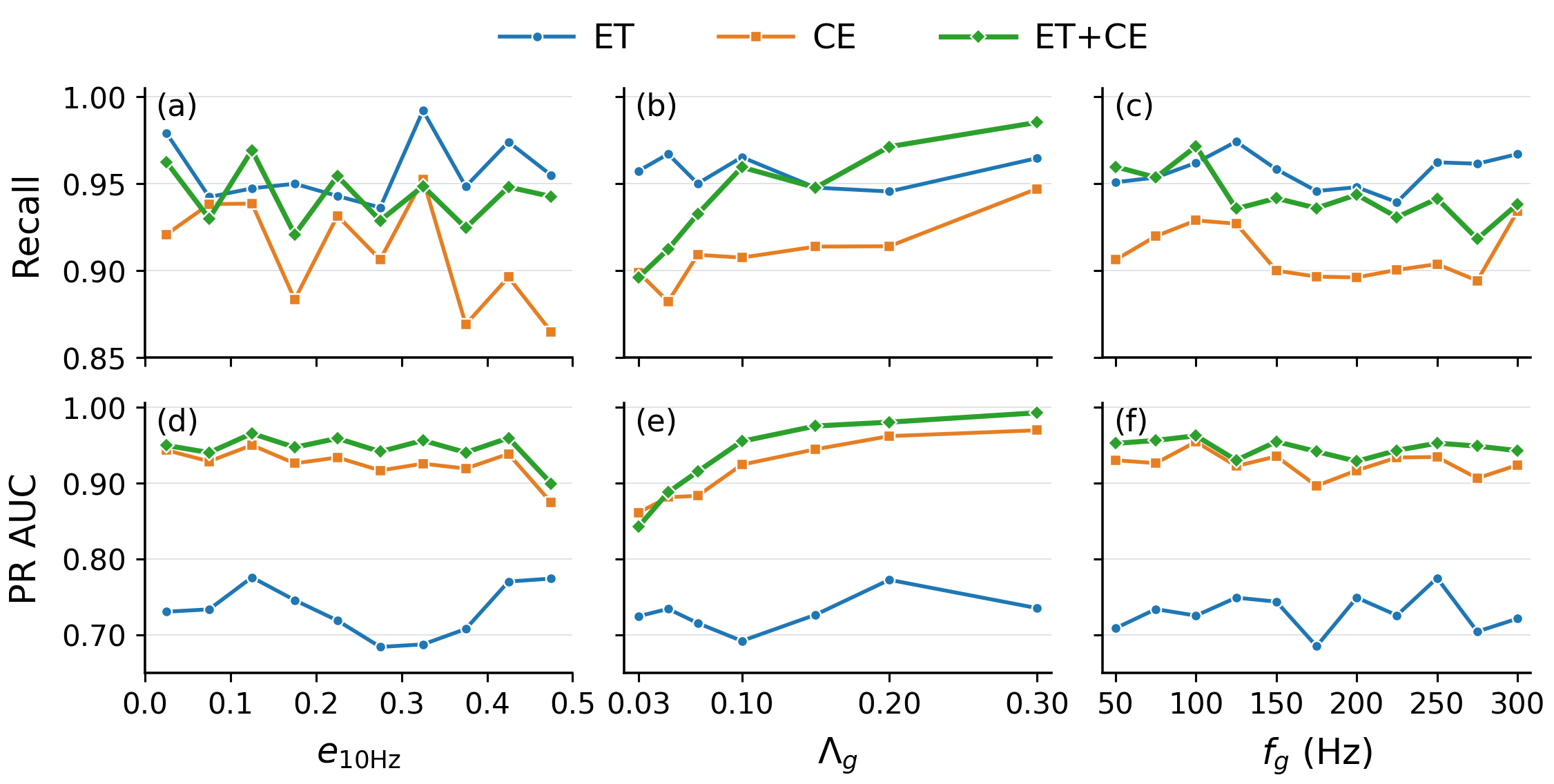}
{\phantomsubcaption\label{subfig:gmode_recall_e10hz}%
\phantomsubcaption\label{subfig:gmode_recall_lambdag}%
\phantomsubcaption\label{subfig:gmode_recall_fghz}%
\phantomsubcaption\label{subfig:gmode_prauc_e10hz}%
\phantomsubcaption\label{subfig:gmode_prauc_lambdag}%
\phantomsubcaption\label{subfig:gmode_prauc_fghz}}
\caption{{AT+$g$ identification on the independent evaluation dataset. The columns show the dependence on $e_{10\mathrm{Hz}}$, $\Lambda_g$, and $f_g$, respectively. The upper row, panels (a), (b), and (c), gives Recall; the lower row, panels (d), (e), and (f), gives PR AUC. The vertical range is common within each row.}}
\label{fig:gmode_binned_all}
\label{fig:gmode_binned_lambdag}
\label{fig:gmode_binned_fg}
\end{figure*}

{Across the eccentricity bins in Fig.~\ref{fig:gmode_binned_all}(a,d), ET+CE AT+$g$ Recall lies between $0.921$ and $0.969$, and PR AUC between $0.899$ and $0.965$. There is no clear monotonic dependence on eccentricity within this parameter range.}

{Figure~\ref{fig:gmode_binned_all}(b,e) shows a stronger overall dependence on coupling strength. At $\Lambda_g=0.03$, ET+CE Recall, PR AUC and AT+$g\rightarrow$AT are $0.896$, $0.843$, and $0.101$. At $\Lambda_g=0.30$, they are $0.985$, $0.992$, and $0.015$. The values fluctuate between adjacent bins, but stronger-coupling samples are generally better separated from the AT background.}

{For the discrete frequencies from $50$ to $300\,\mathrm{Hz}$ in Fig.~\ref{fig:gmode_binned_all}(c,f), ET+CE Recall lies between $0.918$ and $0.972$, and PR AUC between $0.929$ and $0.962$. The frequency dependence is weaker than the overall coupling-strength dependence in this simulation.}

\begin{table}[h]\centering\scriptsize\setlength{\tabcolsep}{3pt}
\caption{{AT+$g$ identification metrics of ET+CE under different physical conditions. $P$ and $R$ denote Precision and Recall.}}
\label{tab:gmode_binned_etce}
\begin{tabular}{cccccc}\hline\hline
Variable & Bin & $P$ & $R$ & PR AUC & AT+$g\rightarrow$AT \\\hline
\multirow{10}{*}{$e_{10\mathrm{Hz}}$} & $[0.00, 0.05]$ & 0.859 & 0.963 & 0.950 & 0.033 \\
 & $[0.05, 0.10]$ & 0.831 & 0.930 & 0.940 & 0.070 \\
 & $[0.10, 0.15]$ & 0.867 & 0.969 & 0.965 & 0.031 \\
 & $[0.15, 0.20]$ & 0.850 & 0.921 & 0.947 & 0.067 \\
 & $[0.20, 0.25]$ & 0.839 & 0.954 & 0.959 & 0.046 \\
 & $[0.25, 0.30]$ & 0.816 & 0.929 & 0.941 & 0.067 \\
 & $[0.30, 0.35]$ & 0.811 & 0.949 & 0.956 & 0.047 \\
 & $[0.35, 0.40]$ & 0.829 & 0.925 & 0.940 & 0.067 \\
 & $[0.40, 0.45]$ & 0.871 & 0.948 & 0.959 & 0.037 \\
 & $[0.45, 0.50]$ & 0.830 & 0.943 & 0.899 & 0.045 \\
\hline
\multirow{7}{*}{$\Lambda_g$} & 0.03 & 0.828 & 0.896 & 0.843 & 0.101 \\
 & 0.05 & 0.835 & 0.912 & 0.888 & 0.079 \\
 & 0.07 & 0.826 & 0.933 & 0.915 & 0.062 \\
 & 0.1 & 0.849 & 0.960 & 0.955 & 0.029 \\
 & 0.15 & 0.838 & 0.948 & 0.975 & 0.039 \\
 & 0.2 & 0.845 & 0.971 & 0.980 & 0.029 \\
 & 0.3 & 0.857 & 0.985 & 0.992 & 0.015 \\
\hline
\multirow{11}{*}{$f_g/\mathrm{Hz}$} & 50 & 0.837 & 0.960 & 0.952 & 0.031 \\
 & 75 & 0.837 & 0.954 & 0.956 & 0.042 \\
 & 100 & 0.865 & 0.972 & 0.962 & 0.024 \\
 & 125 & 0.813 & 0.936 & 0.930 & 0.060 \\
 & 150 & 0.869 & 0.942 & 0.955 & 0.050 \\
 & 175 & 0.823 & 0.936 & 0.941 & 0.059 \\
 & 200 & 0.826 & 0.944 & 0.929 & 0.052 \\
 & 225 & 0.843 & 0.931 & 0.943 & 0.065 \\
 & 250 & 0.862 & 0.941 & 0.953 & 0.050 \\
 & 275 & 0.841 & 0.918 & 0.949 & 0.072 \\
 & 300 & 0.823 & 0.938 & 0.943 & 0.058 \\
\hline
\hline\end{tabular}\end{table}

{Table~\ref{tab:gmode_binned_etce} gives the joint-model values. At $\Lambda_g=0.03$, $0.10$ and $0.30$, the AT+$g\rightarrow$AT fractions are $10.11\%$, $2.89\%$, $1.47\%$, respectively. The coupling strength is a principal control on AT/AT+$g$ separation within the parameter ranges considered.}

\section{Conclusion and Outlook}
\label{sec:conclusion}

This work investigated whether weak resonant phase shifts from eccentric BNS $g$ mode resonances can be identified in noisy third generation detector data. We constructed a four class simulated dataset by applying a multiharmonic resonant phase correction to eccentric BNS waveforms, projecting the resulting signals onto ET and CE detector responses, and adding detector colored Gaussian noise. In this setup, the $g$ mode contribution appears as a small phase perturbation on top of the adiabatic tidal waveform, so the central task is to distinguish AT+$g$ waveforms from the AT background while also separating PP signals and PN samples. Under this fixed four class evaluation, the ET, CE, and ET+CE deep learning models reach Accuracy values of {$0.650$, $0.809$, and $0.875$}, with corresponding Average $F_1$ values of {$0.565$, $0.805$, and $0.873$}. The matched filtering method evaluated on the same samples gives lower Accuracies of {$0.534$, $0.703$, and $0.718$}. The joint ET+CE model also reaches a micro ROC AUC of {$0.984$} and a micro PR AUC of {$0.956$}, and identifies AT+$g$ samples over most of the simulated parameter space. The conditional analysis shows that the remaining AT/AT+$g$ confusion is most strongly controlled by $\Lambda_g$: weak coupling samples are more easily absorbed into the AT background, whereas the dependence on $e_{10\mathrm{Hz}}$ and the discrete $f_g$ values considered here is weaker.

Future work can extend this analysis by broadening the physical parameter space and connecting the identification results more directly to neutron star physics. Including adiabatic tidal parameters, inclination, and source population information in the conditional analysis would help clarify how the AT+$g$ separability varies across different physical and observational conditions. The framework can also be extended to broader waveform and detector configurations to examine how the identified AT/AT+$g$ separation trends depend on waveform modeling and detector response. A natural next step is to develop a Bayesian inference framework for eccentric BNS $g$ mode resonances, using the present identification results to guide parameter space selection and likelihood construction. Such an extension would allow the recovered weak resonant phase signatures to be translated into posterior constraints on mode parameters and neutron star internal physics.

\section*{Acknowledgements}

This work was supported by the National Key Research and Development Program of China (Grant No. 2023YFC2206702), the Fundamental Research Funds for the Central Universities Project (Grant No. 2024IAIS-ZD009), the National Natural Science Foundation of China (Grant Nos. 12575072, 12547101 and 125B2102), and the Natural Science Foundation of Chongqing (Grant No. CSTB2023NSCQ-MSX0103).

\clearpage
%


\begin{thebibliography}{73}%
\makeatletter
\providecommand \@ifxundefined [1]{%
 \@ifx{#1\undefined}
}%
\providecommand \@ifnum [1]{%
 \ifnum #1\expandafter \@firstoftwo
 \else \expandafter \@secondoftwo
 \fi
}%
\providecommand \@ifx [1]{%
 \ifx #1\expandafter \@firstoftwo
 \else \expandafter \@secondoftwo
 \fi
}%
\providecommand \natexlab [1]{#1}%
\providecommand \enquote  [1]{``#1''}%
\providecommand \bibnamefont  [1]{#1}%
\providecommand \bibfnamefont [1]{#1}%
\providecommand \citenamefont [1]{#1}%
\providecommand \href@noop [0]{\@secondoftwo}%
\providecommand \href [0]{\begingroup \@sanitize@url \@href}%
\providecommand \@href[1]{\@@startlink{#1}\@@href}%
\providecommand \@@href[1]{\endgroup#1\@@endlink}%
\providecommand \@sanitize@url [0]{\catcode `\\12\catcode `\$12\catcode
  `\&12\catcode `\#12\catcode `\^12\catcode `\_12\catcode `\%12\relax}%
\providecommand \@@startlink[1]{}%
\providecommand \@@endlink[0]{}%
\providecommand \url  [0]{\begingroup\@sanitize@url \@url }%
\providecommand \@url [1]{\endgroup\@href {#1}{\urlprefix }}%
\providecommand \urlprefix  [0]{URL }%
\providecommand \Eprint [0]{\href }%
\providecommand \doibase [0]{https://doi.org/}%
\providecommand \selectlanguage [0]{\@gobble}%
\providecommand \bibinfo  [0]{\@secondoftwo}%
\providecommand \bibfield  [0]{\@secondoftwo}%
\providecommand \translation [1]{[#1]}%
\providecommand \BibitemOpen [0]{}%
\providecommand \bibitemStop [0]{}%
\providecommand \bibitemNoStop [0]{.\EOS\space}%
\providecommand \EOS [0]{\spacefactor3000\relax}%
\providecommand \BibitemShut  [1]{\csname bibitem#1\endcsname}%
\let\auto@bib@innerbib\@empty
\bibitem [{\citenamefont {Lattimer}\ and\ \citenamefont
  {Prakash}(2016)}]{LattimerPrakash2016DenseMatter}%
  \BibitemOpen
  \bibfield  {author} {\bibinfo {author} {\bibfnamefont {J.~M.}\ \bibnamefont
  {Lattimer}}\ and\ \bibinfo {author} {\bibfnamefont {M.}~\bibnamefont
  {Prakash}},\ }\bibfield  {title} {\bibinfo {title} {The equation of state of
  hot, dense matter and neutron stars},\ }\href
  {https://doi.org/10.1016/j.physrep.2015.12.005} {\bibfield  {journal}
  {\bibinfo  {journal} {Physics Reports}\ }\textbf {\bibinfo {volume} {621}},\
  \bibinfo {pages} {127} (\bibinfo {year} {2016})},\ \Eprint
  {https://arxiv.org/abs/1512.07820} {arXiv:1512.07820 [astro-ph.SR]}
  \BibitemShut {NoStop}%
\bibitem [{\citenamefont {{LIGO Scientific Collaboration}}\ and\ \citenamefont
  {{Virgo Collaboration}}(2017)}]{ligo2017gw170817}%
  \BibitemOpen
  \bibfield  {author} {\bibinfo {author} {\bibnamefont {{LIGO Scientific
  Collaboration}}}\ and\ \bibinfo {author} {\bibnamefont {{Virgo
  Collaboration}}},\ }\bibfield  {title} {\bibinfo {title} {Gw170817:
  Observation of gravitational waves from a binary neutron star inspiral},\
  }\href {https://doi.org/10.1103/PhysRevLett.119.161101} {\bibfield  {journal}
  {\bibinfo  {journal} {Physical Review Letters}\ }\textbf {\bibinfo {volume}
  {119}},\ \bibinfo {pages} {161101} (\bibinfo {year} {2017})},\ \Eprint
  {https://arxiv.org/abs/1710.05832} {arXiv:1710.05832 [gr-qc]} \BibitemShut
  {NoStop}%
\bibitem [{\citenamefont {{LIGO Scientific Collaboration}}\ and\ \citenamefont
  {{Virgo Collaboration}}(2018)}]{Abbott2018RadiiEOS}%
  \BibitemOpen
  \bibfield  {author} {\bibinfo {author} {\bibnamefont {{LIGO Scientific
  Collaboration}}}\ and\ \bibinfo {author} {\bibnamefont {{Virgo
  Collaboration}}},\ }\bibfield  {title} {\bibinfo {title} {{GW170817}:
  Measurements of neutron star radii and equation of state},\ }\href
  {https://doi.org/10.1103/PhysRevLett.121.161101} {\bibfield  {journal}
  {\bibinfo  {journal} {Physical Review Letters}\ }\textbf {\bibinfo {volume}
  {121}},\ \bibinfo {pages} {161101} (\bibinfo {year} {2018})}\BibitemShut
  {NoStop}%
\bibitem [{\citenamefont {De}\ \emph {et~al.}(2018)\citenamefont {De},
  \citenamefont {Finstad}, \citenamefont {Lattimer}, \citenamefont {Brown},
  \citenamefont {Berger},\ and\ \citenamefont {Biwer}}]{De2018GW170817Radii}%
  \BibitemOpen
  \bibfield  {author} {\bibinfo {author} {\bibfnamefont {S.}~\bibnamefont
  {De}}, \bibinfo {author} {\bibfnamefont {D.}~\bibnamefont {Finstad}},
  \bibinfo {author} {\bibfnamefont {J.~M.}\ \bibnamefont {Lattimer}}, \bibinfo
  {author} {\bibfnamefont {D.~A.}\ \bibnamefont {Brown}}, \bibinfo {author}
  {\bibfnamefont {E.}~\bibnamefont {Berger}},\ and\ \bibinfo {author}
  {\bibfnamefont {C.~M.}\ \bibnamefont {Biwer}},\ }\bibfield  {title} {\bibinfo
  {title} {Tidal deformabilities and radii of neutron stars from the
  observation of {GW170817}},\ }\href
  {https://doi.org/10.1103/PhysRevLett.121.091102} {\bibfield  {journal}
  {\bibinfo  {journal} {Physical Review Letters}\ }\textbf {\bibinfo {volume}
  {121}},\ \bibinfo {pages} {091102} (\bibinfo {year} {2018})}\BibitemShut
  {NoStop}%
\bibitem [{\citenamefont {Annala}\ \emph {et~al.}(2018)\citenamefont {Annala},
  \citenamefont {Gorda}, \citenamefont {Kurkela},\ and\ \citenamefont
  {Vuorinen}}]{Annala2018EOS}%
  \BibitemOpen
  \bibfield  {author} {\bibinfo {author} {\bibfnamefont {E.}~\bibnamefont
  {Annala}}, \bibinfo {author} {\bibfnamefont {T.}~\bibnamefont {Gorda}},
  \bibinfo {author} {\bibfnamefont {A.}~\bibnamefont {Kurkela}},\ and\ \bibinfo
  {author} {\bibfnamefont {A.}~\bibnamefont {Vuorinen}},\ }\bibfield  {title}
  {\bibinfo {title} {Gravitational-wave constraints on the neutron-star-matter
  equation of state},\ }\href {https://doi.org/10.1103/PhysRevLett.120.172703}
  {\bibfield  {journal} {\bibinfo  {journal} {Physical Review Letters}\
  }\textbf {\bibinfo {volume} {120}},\ \bibinfo {pages} {172703} (\bibinfo
  {year} {2018})}\BibitemShut {NoStop}%
\bibitem [{\citenamefont {{LIGO Scientific Collaboration}}\ \emph
  {et~al.}(2021)\citenamefont {{LIGO Scientific Collaboration}}, \citenamefont
  {{Virgo Collaboration}},\ and\ \citenamefont {{KAGRA
  Collaboration}}}]{gwtc3}%
  \BibitemOpen
  \bibfield  {author} {\bibinfo {author} {\bibnamefont {{LIGO Scientific
  Collaboration}}}, \bibinfo {author} {\bibnamefont {{Virgo Collaboration}}},\
  and\ \bibinfo {author} {\bibnamefont {{KAGRA Collaboration}}},\ }\bibfield
  {title} {\bibinfo {title} {Gwtc-3: Compact binary coalescences observed by
  ligo and virgo during the second part of the third observing run},\
  }\bibfield  {journal} {\bibinfo  {journal} {arXiv preprint arXiv:2111.03606}\
  }\href {https://doi.org/10.48550/arXiv.2111.03606}
  {10.48550/arXiv.2111.03606} (\bibinfo {year} {2021}),\ \Eprint
  {https://arxiv.org/abs/2111.03606} {arXiv:2111.03606 [gr-qc]} \BibitemShut
  {NoStop}%
\bibitem [{\citenamefont {Annala}\ \emph {et~al.}(2022)\citenamefont {Annala},
  \citenamefont {Gorda}, \citenamefont {Katerini}, \citenamefont {Kurkela},
  \citenamefont {N{\"a}ttil{\"a}}, \citenamefont {Paschalidis},\ and\
  \citenamefont {Vuorinen}}]{Annala2022MultimessengerEOS}%
  \BibitemOpen
  \bibfield  {author} {\bibinfo {author} {\bibfnamefont {E.}~\bibnamefont
  {Annala}}, \bibinfo {author} {\bibfnamefont {T.}~\bibnamefont {Gorda}},
  \bibinfo {author} {\bibfnamefont {E.}~\bibnamefont {Katerini}}, \bibinfo
  {author} {\bibfnamefont {A.}~\bibnamefont {Kurkela}}, \bibinfo {author}
  {\bibfnamefont {J.}~\bibnamefont {N{\"a}ttil{\"a}}}, \bibinfo {author}
  {\bibfnamefont {V.}~\bibnamefont {Paschalidis}},\ and\ \bibinfo {author}
  {\bibfnamefont {A.}~\bibnamefont {Vuorinen}},\ }\bibfield  {title} {\bibinfo
  {title} {Multimessenger constraints for ultradense matter},\ }\href
  {https://doi.org/10.1103/PhysRevX.12.011058} {\bibfield  {journal} {\bibinfo
  {journal} {Physical Review X}\ }\textbf {\bibinfo {volume} {12}},\ \bibinfo
  {pages} {011058} (\bibinfo {year} {2022})}\BibitemShut {NoStop}%
\bibitem [{\citenamefont {Hinderer}(2008)}]{Hinderer2008Love}%
  \BibitemOpen
  \bibfield  {author} {\bibinfo {author} {\bibfnamefont {T.}~\bibnamefont
  {Hinderer}},\ }\bibfield  {title} {\bibinfo {title} {Tidal love numbers of
  neutron stars},\ }\href {https://doi.org/10.1086/533487} {\bibfield
  {journal} {\bibinfo  {journal} {The Astrophysical Journal}\ }\textbf
  {\bibinfo {volume} {677}},\ \bibinfo {pages} {1216} (\bibinfo {year}
  {2008})}\BibitemShut {NoStop}%
\bibitem [{\citenamefont {Damour}\ and\ \citenamefont
  {Nagar}(2009)}]{DamourNagar2009Tidal}%
  \BibitemOpen
  \bibfield  {author} {\bibinfo {author} {\bibfnamefont {T.}~\bibnamefont
  {Damour}}\ and\ \bibinfo {author} {\bibfnamefont {A.}~\bibnamefont {Nagar}},\
  }\bibfield  {title} {\bibinfo {title} {Relativistic tidal properties of
  neutron stars},\ }\href {https://doi.org/10.1103/PhysRevD.80.084035}
  {\bibfield  {journal} {\bibinfo  {journal} {Physical Review D}\ }\textbf
  {\bibinfo {volume} {80}},\ \bibinfo {pages} {084035} (\bibinfo {year}
  {2009})}\BibitemShut {NoStop}%
\bibitem [{\citenamefont {Hinderer}\ \emph {et~al.}(2010)\citenamefont
  {Hinderer}, \citenamefont {Lackey}, \citenamefont {Lang},\ and\ \citenamefont
  {Read}}]{Hinderer2010TidalEOS}%
  \BibitemOpen
  \bibfield  {author} {\bibinfo {author} {\bibfnamefont {T.}~\bibnamefont
  {Hinderer}}, \bibinfo {author} {\bibfnamefont {B.~D.}\ \bibnamefont
  {Lackey}}, \bibinfo {author} {\bibfnamefont {R.~N.}\ \bibnamefont {Lang}},\
  and\ \bibinfo {author} {\bibfnamefont {J.~S.}\ \bibnamefont {Read}},\
  }\bibfield  {title} {\bibinfo {title} {Tidal deformability of neutron stars
  with realistic equations of state and their gravitational wave signatures in
  binary inspiral},\ }\href {https://doi.org/10.1103/PhysRevD.81.123016}
  {\bibfield  {journal} {\bibinfo  {journal} {Physical Review D}\ }\textbf
  {\bibinfo {volume} {81}},\ \bibinfo {pages} {123016} (\bibinfo {year}
  {2010})}\BibitemShut {NoStop}%
\bibitem [{\citenamefont {Yagi}\ and\ \citenamefont
  {Yunes}(2013)}]{YagiYunes2013ILoveQ}%
  \BibitemOpen
  \bibfield  {author} {\bibinfo {author} {\bibfnamefont {K.}~\bibnamefont
  {Yagi}}\ and\ \bibinfo {author} {\bibfnamefont {N.}~\bibnamefont {Yunes}},\
  }\bibfield  {title} {\bibinfo {title} {{I-Love-Q}: Unexpected universal
  relations for neutron stars and quark stars},\ }\href
  {https://doi.org/10.1126/science.1236462} {\bibfield  {journal} {\bibinfo
  {journal} {Science}\ }\textbf {\bibinfo {volume} {341}},\ \bibinfo {pages}
  {365} (\bibinfo {year} {2013})}\BibitemShut {NoStop}%
\bibitem [{\citenamefont {Hinderer}\ \emph {et~al.}(2016)\citenamefont
  {Hinderer}, \citenamefont {Taracchini}, \citenamefont {Foucart},
  \citenamefont {Buonanno}, \citenamefont {Steinhoff}, \citenamefont {Duez},
  \citenamefont {Kidder}, \citenamefont {Pfeiffer}, \citenamefont {Scheel},
  \citenamefont {Szilagyi}, \citenamefont {Hotokezaka}, \citenamefont
  {Kyutoku}, \citenamefont {Shibata},\ and\ \citenamefont
  {Carpenter}}]{Hinderer2016DynamicTides}%
  \BibitemOpen
  \bibfield  {author} {\bibinfo {author} {\bibfnamefont {T.}~\bibnamefont
  {Hinderer}}, \bibinfo {author} {\bibfnamefont {A.}~\bibnamefont
  {Taracchini}}, \bibinfo {author} {\bibfnamefont {F.}~\bibnamefont {Foucart}},
  \bibinfo {author} {\bibfnamefont {A.}~\bibnamefont {Buonanno}}, \bibinfo
  {author} {\bibfnamefont {J.}~\bibnamefont {Steinhoff}}, \bibinfo {author}
  {\bibfnamefont {M.}~\bibnamefont {Duez}}, \bibinfo {author} {\bibfnamefont
  {L.~E.}\ \bibnamefont {Kidder}}, \bibinfo {author} {\bibfnamefont {H.~P.}\
  \bibnamefont {Pfeiffer}}, \bibinfo {author} {\bibfnamefont {M.~A.}\
  \bibnamefont {Scheel}}, \bibinfo {author} {\bibfnamefont {B.}~\bibnamefont
  {Szilagyi}}, \bibinfo {author} {\bibfnamefont {K.}~\bibnamefont
  {Hotokezaka}}, \bibinfo {author} {\bibfnamefont {K.}~\bibnamefont {Kyutoku}},
  \bibinfo {author} {\bibfnamefont {M.}~\bibnamefont {Shibata}},\ and\ \bibinfo
  {author} {\bibfnamefont {C.~W.}\ \bibnamefont {Carpenter}},\ }\bibfield
  {title} {\bibinfo {title} {Effects of neutron-star dynamic tides on
  gravitational waveforms within the effective-one-body approach},\ }\href
  {https://doi.org/10.1103/PhysRevLett.116.181101} {\bibfield  {journal}
  {\bibinfo  {journal} {Physical Review Letters}\ }\textbf {\bibinfo {volume}
  {116}},\ \bibinfo {pages} {181101} (\bibinfo {year} {2016})}\BibitemShut
  {NoStop}%
\bibitem [{\citenamefont {Steinhoff}\ \emph {et~al.}(2016)\citenamefont
  {Steinhoff}, \citenamefont {Hinderer}, \citenamefont {Buonanno},\ and\
  \citenamefont {Taracchini}}]{Steinhoff2016DynamicTides}%
  \BibitemOpen
  \bibfield  {author} {\bibinfo {author} {\bibfnamefont {J.}~\bibnamefont
  {Steinhoff}}, \bibinfo {author} {\bibfnamefont {T.}~\bibnamefont {Hinderer}},
  \bibinfo {author} {\bibfnamefont {A.}~\bibnamefont {Buonanno}},\ and\
  \bibinfo {author} {\bibfnamefont {A.}~\bibnamefont {Taracchini}},\ }\bibfield
   {title} {\bibinfo {title} {Dynamical tides in general relativity: Effective
  action and effective-one-body hamiltonian},\ }\href
  {https://doi.org/10.1103/PhysRevD.94.104028} {\bibfield  {journal} {\bibinfo
  {journal} {Physical Review D}\ }\textbf {\bibinfo {volume} {94}},\ \bibinfo
  {pages} {104028} (\bibinfo {year} {2016})}\BibitemShut {NoStop}%
\bibitem [{\citenamefont {Pratten}\ \emph {et~al.}(2022)\citenamefont
  {Pratten}, \citenamefont {Schmidt},\ and\ \citenamefont
  {Williams}}]{Pratten2022DynamicTidesEOS}%
  \BibitemOpen
  \bibfield  {author} {\bibinfo {author} {\bibfnamefont {G.}~\bibnamefont
  {Pratten}}, \bibinfo {author} {\bibfnamefont {P.}~\bibnamefont {Schmidt}},\
  and\ \bibinfo {author} {\bibfnamefont {N.}~\bibnamefont {Williams}},\
  }\bibfield  {title} {\bibinfo {title} {Impact of dynamical tides on the
  reconstruction of the neutron star equation of state},\ }\href
  {https://doi.org/10.1103/PhysRevLett.129.081102} {\bibfield  {journal}
  {\bibinfo  {journal} {Physical Review Letters}\ }\textbf {\bibinfo {volume}
  {129}},\ \bibinfo {pages} {081102} (\bibinfo {year} {2022})}\BibitemShut
  {NoStop}%
\bibitem [{\citenamefont {Kantor}\ and\ \citenamefont
  {Gusakov}(2014)}]{KantorGusakov2014Gmodes}%
  \BibitemOpen
  \bibfield  {author} {\bibinfo {author} {\bibfnamefont {E.~M.}\ \bibnamefont
  {Kantor}}\ and\ \bibinfo {author} {\bibfnamefont {M.~E.}\ \bibnamefont
  {Gusakov}},\ }\bibfield  {title} {\bibinfo {title} {Composition
  temperature-dependent g modes in superfluid neutron stars},\ }\href
  {https://doi.org/10.1093/mnrasl/slu061} {\bibfield  {journal} {\bibinfo
  {journal} {Monthly Notices of the Royal Astronomical Society: Letters}\
  }\textbf {\bibinfo {volume} {442}},\ \bibinfo {pages} {L90} (\bibinfo {year}
  {2014})}\BibitemShut {NoStop}%
\bibitem [{\citenamefont {Passamonti}\ \emph {et~al.}(2016)\citenamefont
  {Passamonti}, \citenamefont {Andersson},\ and\ \citenamefont
  {Ho}}]{PassamontiAnderssonHo2016Buoyancy}%
  \BibitemOpen
  \bibfield  {author} {\bibinfo {author} {\bibfnamefont {A.}~\bibnamefont
  {Passamonti}}, \bibinfo {author} {\bibfnamefont {N.}~\bibnamefont
  {Andersson}},\ and\ \bibinfo {author} {\bibfnamefont {W.~C.~G.}\ \bibnamefont
  {Ho}},\ }\bibfield  {title} {\bibinfo {title} {Buoyancy and g-modes in young
  superfluid neutron stars},\ }\href {https://doi.org/10.1093/mnras/stv2361}
  {\bibfield  {journal} {\bibinfo  {journal} {Monthly Notices of the Royal
  Astronomical Society}\ }\textbf {\bibinfo {volume} {455}},\ \bibinfo {pages}
  {1489} (\bibinfo {year} {2016})}\BibitemShut {NoStop}%
\bibitem [{\citenamefont {Constantinou}\ \emph {et~al.}(2021)\citenamefont
  {Constantinou}, \citenamefont {Han}, \citenamefont {Jaikumar},\ and\
  \citenamefont {Prakash}}]{Constantinou2021GmodesCrossover}%
  \BibitemOpen
  \bibfield  {author} {\bibinfo {author} {\bibfnamefont {C.}~\bibnamefont
  {Constantinou}}, \bibinfo {author} {\bibfnamefont {S.}~\bibnamefont {Han}},
  \bibinfo {author} {\bibfnamefont {P.}~\bibnamefont {Jaikumar}},\ and\
  \bibinfo {author} {\bibfnamefont {M.}~\bibnamefont {Prakash}},\ }\bibfield
  {title} {\bibinfo {title} {g-modes of neutron stars with hadron-to-quark
  crossover transitions},\ }\href {https://doi.org/10.1103/PhysRevD.104.123032}
  {\bibfield  {journal} {\bibinfo  {journal} {Physical Review D}\ }\textbf
  {\bibinfo {volume} {104}},\ \bibinfo {pages} {123032} (\bibinfo {year}
  {2021})}\BibitemShut {NoStop}%
\bibitem [{\citenamefont {Jaikumar}\ \emph {et~al.}(2021)\citenamefont
  {Jaikumar}, \citenamefont {Semposki}, \citenamefont {Prakash},\ and\
  \citenamefont {Constantinou}}]{Jaikumar2021HybridGmodes}%
  \BibitemOpen
  \bibfield  {author} {\bibinfo {author} {\bibfnamefont {P.}~\bibnamefont
  {Jaikumar}}, \bibinfo {author} {\bibfnamefont {A.}~\bibnamefont {Semposki}},
  \bibinfo {author} {\bibfnamefont {M.}~\bibnamefont {Prakash}},\ and\ \bibinfo
  {author} {\bibfnamefont {C.}~\bibnamefont {Constantinou}},\ }\bibfield
  {title} {\bibinfo {title} {g-mode oscillations in hybrid stars: A tale of two
  sounds},\ }\href {https://doi.org/10.1103/PhysRevD.103.123009} {\bibfield
  {journal} {\bibinfo  {journal} {Physical Review D}\ }\textbf {\bibinfo
  {volume} {103}},\ \bibinfo {pages} {123009} (\bibinfo {year}
  {2021})}\BibitemShut {NoStop}%
\bibitem [{\citenamefont {Counsell}\ \emph {et~al.}(2025)\citenamefont
  {Counsell}, \citenamefont {Gittins}, \citenamefont {Andersson},\ and\
  \citenamefont {Tews}}]{Counsell2025InterfaceModes}%
  \BibitemOpen
  \bibfield  {author} {\bibinfo {author} {\bibfnamefont {A.~R.}\ \bibnamefont
  {Counsell}}, \bibinfo {author} {\bibfnamefont {F.}~\bibnamefont {Gittins}},
  \bibinfo {author} {\bibfnamefont {N.}~\bibnamefont {Andersson}},\ and\
  \bibinfo {author} {\bibfnamefont {I.}~\bibnamefont {Tews}},\ }\bibfield
  {title} {\bibinfo {title} {Interface modes in inspiralling neutron stars: A
  gravitational-wave probe of first-order phase transitions},\ }\href
  {https://doi.org/10.1103/PhysRevLett.135.081402} {\bibfield  {journal}
  {\bibinfo  {journal} {Physical Review Letters}\ }\textbf {\bibinfo {volume}
  {135}},\ \bibinfo {pages} {081402} (\bibinfo {year} {2025})}\BibitemShut
  {NoStop}%
\bibitem [{\citenamefont {Passamonti}\ \emph {et~al.}(2021)\citenamefont
  {Passamonti}, \citenamefont {Andersson},\ and\ \citenamefont
  {Pnigouras}}]{passamonti2021crust}%
  \BibitemOpen
  \bibfield  {author} {\bibinfo {author} {\bibfnamefont {A.}~\bibnamefont
  {Passamonti}}, \bibinfo {author} {\bibfnamefont {N.}~\bibnamefont
  {Andersson}},\ and\ \bibinfo {author} {\bibfnamefont {P.}~\bibnamefont
  {Pnigouras}},\ }\bibfield  {title} {\bibinfo {title} {Dynamical tides in
  neutron stars: The impact of the crust},\ }\href
  {https://doi.org/10.1093/mnras/stab870} {\bibfield  {journal} {\bibinfo
  {journal} {Monthly Notices of the Royal Astronomical Society}\ }\textbf
  {\bibinfo {volume} {504}},\ \bibinfo {pages} {1273} (\bibinfo {year}
  {2021})},\ \Eprint {https://arxiv.org/abs/2012.09637} {arXiv:2012.09637
  [astro-ph.HE]} \BibitemShut {NoStop}%
\bibitem [{\citenamefont {Lai}(1994)}]{Lai1994ResonantTides}%
  \BibitemOpen
  \bibfield  {author} {\bibinfo {author} {\bibfnamefont {D.}~\bibnamefont
  {Lai}},\ }\bibfield  {title} {\bibinfo {title} {Resonant oscillations and
  tidal heating in coalescing binary neutron stars},\ }\href
  {https://doi.org/10.1093/mnras/270.3.611} {\bibfield  {journal} {\bibinfo
  {journal} {Monthly Notices of the Royal Astronomical Society}\ }\textbf
  {\bibinfo {volume} {270}},\ \bibinfo {pages} {611} (\bibinfo {year}
  {1994})}\BibitemShut {NoStop}%
\bibitem [{\citenamefont {Reisenegger}\ and\ \citenamefont
  {Goldreich}(1994)}]{ReiseneggerGoldreich1994Modes}%
  \BibitemOpen
  \bibfield  {author} {\bibinfo {author} {\bibfnamefont {A.}~\bibnamefont
  {Reisenegger}}\ and\ \bibinfo {author} {\bibfnamefont {P.}~\bibnamefont
  {Goldreich}},\ }\bibfield  {title} {\bibinfo {title} {Excitation of neutron
  star normal modes during binary inspiral},\ }\href
  {https://doi.org/10.1086/174105} {\bibfield  {journal} {\bibinfo  {journal}
  {The Astrophysical Journal}\ }\textbf {\bibinfo {volume} {426}},\ \bibinfo
  {pages} {688} (\bibinfo {year} {1994})}\BibitemShut {NoStop}%
\bibitem [{\citenamefont {Kokkotas}\ and\ \citenamefont
  {Sch{\"a}fer}(1995)}]{KokkotasSchafer1995TidalResonant}%
  \BibitemOpen
  \bibfield  {author} {\bibinfo {author} {\bibfnamefont {K.~D.}\ \bibnamefont
  {Kokkotas}}\ and\ \bibinfo {author} {\bibfnamefont {G.}~\bibnamefont
  {Sch{\"a}fer}},\ }\bibfield  {title} {\bibinfo {title} {Tidal and
  tidal-resonant effects in coalescing binaries},\ }\href
  {https://doi.org/10.1093/mnras/275.2.301} {\bibfield  {journal} {\bibinfo
  {journal} {Monthly Notices of the Royal Astronomical Society}\ }\textbf
  {\bibinfo {volume} {275}},\ \bibinfo {pages} {301} (\bibinfo {year}
  {1995})}\BibitemShut {NoStop}%
\bibitem [{\citenamefont {Balachandran}\ and\ \citenamefont
  {Flanagan}(2007)}]{BalachandranFlanagan2007Detectability}%
  \BibitemOpen
  \bibfield  {author} {\bibinfo {author} {\bibfnamefont {P.}~\bibnamefont
  {Balachandran}}\ and\ \bibinfo {author} {\bibfnamefont {{\'E}.~{\'E}.}\
  \bibnamefont {Flanagan}},\ }\bibfield  {title} {\bibinfo {title}
  {Detectability of mode resonances in coalescing neutron star binaries},\
  }\href {https://doi.org/10.48550/arXiv.gr-qc/0701076} {\bibfield  {journal}
  {\bibinfo  {journal} {arXiv e-prints}\ ,\ \bibinfo {pages} {arXiv:gr}}
  (\bibinfo {year} {2007})},\ \Eprint {https://arxiv.org/abs/gr-qc/0701076}
  {arXiv:gr-qc/0701076 [gr-qc]} \BibitemShut {NoStop}%
\bibitem [{\citenamefont {Ho}\ and\ \citenamefont
  {Andersson}(2023)}]{HoAndersson2023DynamicalTides}%
  \BibitemOpen
  \bibfield  {author} {\bibinfo {author} {\bibfnamefont {W.~C.~G.}\
  \bibnamefont {Ho}}\ and\ \bibinfo {author} {\bibfnamefont {N.}~\bibnamefont
  {Andersson}},\ }\bibfield  {title} {\bibinfo {title} {New dynamical tide
  constraints from current and future gravitational wave detections of
  inspiralling neutron stars},\ }\href
  {https://doi.org/10.1103/PhysRevD.108.043003} {\bibfield  {journal} {\bibinfo
   {journal} {Physical Review D}\ }\textbf {\bibinfo {volume} {108}},\ \bibinfo
  {pages} {043003} (\bibinfo {year} {2023})},\ \Eprint
  {https://arxiv.org/abs/2307.10721} {arXiv:2307.10721 [gr-qc]} \BibitemShut
  {NoStop}%
\bibitem [{\citenamefont {Revilla-Pe{\~n}a}\ \emph {et~al.}(2026)\citenamefont
  {Revilla-Pe{\~n}a}, \citenamefont {Bondarescu}, \citenamefont {Lundgren},\
  and\ \citenamefont {Miralda-Escud{\'e}}}]{RevillaPena2026Detectability}%
  \BibitemOpen
  \bibfield  {author} {\bibinfo {author} {\bibfnamefont {A.}~\bibnamefont
  {Revilla-Pe{\~n}a}}, \bibinfo {author} {\bibfnamefont {R.}~\bibnamefont
  {Bondarescu}}, \bibinfo {author} {\bibfnamefont {A.~P.}\ \bibnamefont
  {Lundgren}},\ and\ \bibinfo {author} {\bibfnamefont {J.}~\bibnamefont
  {Miralda-Escud{\'e}}},\ }\bibfield  {title} {\bibinfo {title} {Rethinking
  resonance detectability during binary neutron star inspiral: Accurate
  mismatch computations for low-lying dynamical tides},\ }\href@noop {}
  {\bibfield  {journal} {\bibinfo  {journal} {arXiv e-prints}\ ,\ \bibinfo
  {pages} {arXiv:2601.21086}} (\bibinfo {year} {2026})},\ \Eprint
  {https://arxiv.org/abs/2601.21086} {arXiv:2601.21086 [gr-qc]} \BibitemShut
  {NoStop}%
\bibitem [{\citenamefont {Gittins}\ \emph {et~al.}(2026)\citenamefont
  {Gittins}, \citenamefont {Narola}, \citenamefont {Wouters}, \citenamefont
  {Pang}, \citenamefont {Hinderer},\ and\ \citenamefont {Van
  Den~Broeck}}]{Gittins2026DetectingResonances}%
  \BibitemOpen
  \bibfield  {author} {\bibinfo {author} {\bibfnamefont {F.}~\bibnamefont
  {Gittins}}, \bibinfo {author} {\bibfnamefont {H.}~\bibnamefont {Narola}},
  \bibinfo {author} {\bibfnamefont {T.}~\bibnamefont {Wouters}}, \bibinfo
  {author} {\bibfnamefont {P.~T.~H.}\ \bibnamefont {Pang}}, \bibinfo {author}
  {\bibfnamefont {T.}~\bibnamefont {Hinderer}},\ and\ \bibinfo {author}
  {\bibfnamefont {C.}~\bibnamefont {Van Den~Broeck}},\ }\bibfield  {title}
  {\bibinfo {title} {Detecting tidal resonances in binary neutron stars},\
  }\href {https://doi.org/10.48550/arXiv.2606.06376} {\bibfield  {journal}
  {\bibinfo  {journal} {arXiv e-prints}\ ,\ \bibinfo {pages}
  {arXiv:2606.06376}} (\bibinfo {year} {2026})},\ \Eprint
  {https://arxiv.org/abs/2606.06376} {arXiv:2606.06376 [gr-qc]} \BibitemShut
  {NoStop}%
\bibitem [{\citenamefont {Vick}\ and\ \citenamefont
  {Lai}(2019)}]{VickLai2019EccentricTides}%
  \BibitemOpen
  \bibfield  {author} {\bibinfo {author} {\bibfnamefont {M.}~\bibnamefont
  {Vick}}\ and\ \bibinfo {author} {\bibfnamefont {D.}~\bibnamefont {Lai}},\
  }\bibfield  {title} {\bibinfo {title} {Tidal effects in eccentric coalescing
  neutron star binaries},\ }\href {https://doi.org/10.1103/PhysRevD.100.063001}
  {\bibfield  {journal} {\bibinfo  {journal} {Physical Review D}\ }\textbf
  {\bibinfo {volume} {100}},\ \bibinfo {pages} {063001} (\bibinfo {year}
  {2019})}\BibitemShut {NoStop}%
\bibitem [{\citenamefont {Tak{\'a}tsy}\ \emph {et~al.}(2024)\citenamefont
  {Tak{\'a}tsy}, \citenamefont {Kocsis},\ and\ \citenamefont
  {Kov{\'a}cs}}]{Takatsy2024EccentricDynamicTides}%
  \BibitemOpen
  \bibfield  {author} {\bibinfo {author} {\bibfnamefont {J.}~\bibnamefont
  {Tak{\'a}tsy}}, \bibinfo {author} {\bibfnamefont {B.}~\bibnamefont
  {Kocsis}},\ and\ \bibinfo {author} {\bibfnamefont {P.}~\bibnamefont
  {Kov{\'a}cs}},\ }\bibfield  {title} {\bibinfo {title} {Observability of
  dynamical tides in merging eccentric neutron star binaries},\ }\href
  {https://doi.org/10.1103/PhysRevD.110.103043} {\bibfield  {journal} {\bibinfo
   {journal} {Physical Review D}\ }\textbf {\bibinfo {volume} {110}},\ \bibinfo
  {pages} {103043} (\bibinfo {year} {2024})},\ \Eprint
  {https://arxiv.org/abs/2407.17560} {arXiv:2407.17560 [astro-ph.HE]}
  \BibitemShut {NoStop}%
\bibitem [{\citenamefont {Tak{\'a}tsy}\ \emph {et~al.}(2026)\citenamefont
  {Tak{\'a}tsy}, \citenamefont {Zwick}, \citenamefont {Saini},\ and\
  \citenamefont {Samsing}}]{takatsy2026gmode}%
  \BibitemOpen
  \bibfield  {author} {\bibinfo {author} {\bibfnamefont {J.}~\bibnamefont
  {Tak{\'a}tsy}}, \bibinfo {author} {\bibfnamefont {L.}~\bibnamefont {Zwick}},
  \bibinfo {author} {\bibfnamefont {P.}~\bibnamefont {Saini}},\ and\ \bibinfo
  {author} {\bibfnamefont {J.}~\bibnamefont {Samsing}},\ }\bibfield  {title}
  {\bibinfo {title} {Orbital eccentricity can make neutron star g-mode
  resonances observable with current gravitational-wave detectors},\ }\href
  {https://doi.org/10.1103/PhysRevD.113.103047} {\bibfield  {journal} {\bibinfo
   {journal} {Phys. Rev. D}\ }\textbf {\bibinfo {volume} {113}},\ \bibinfo
  {pages} {103047} (\bibinfo {year} {2026})},\ \Eprint
  {https://arxiv.org/abs/2602.15111} {arXiv:2602.15111 [astro-ph.HE]}
  \BibitemShut {NoStop}%
\bibitem [{\citenamefont {Dietrich}\ \emph {et~al.}(2021)\citenamefont
  {Dietrich}, \citenamefont {Hinderer},\ and\ \citenamefont
  {Samajdar}}]{DietrichHindererSamajdar2021BNSReview}%
  \BibitemOpen
  \bibfield  {author} {\bibinfo {author} {\bibfnamefont {T.}~\bibnamefont
  {Dietrich}}, \bibinfo {author} {\bibfnamefont {T.}~\bibnamefont {Hinderer}},\
  and\ \bibinfo {author} {\bibfnamefont {A.}~\bibnamefont {Samajdar}},\
  }\bibfield  {title} {\bibinfo {title} {Interpreting binary neutron star
  mergers: Describing the binary neutron star dynamics, modelling gravitational
  waveforms, and analyzing detections},\ }\href
  {https://doi.org/10.1007/s10714-020-02751-6} {\bibfield  {journal} {\bibinfo
  {journal} {General Relativity and Gravitation}\ }\textbf {\bibinfo {volume}
  {53}},\ \bibinfo {pages} {27} (\bibinfo {year} {2021})}\BibitemShut {NoStop}%
\bibitem [{\citenamefont {Peters}\ and\ \citenamefont
  {Mathews}(1963)}]{Peters1963}%
  \BibitemOpen
  \bibfield  {author} {\bibinfo {author} {\bibfnamefont {P.~C.}\ \bibnamefont
  {Peters}}\ and\ \bibinfo {author} {\bibfnamefont {J.}~\bibnamefont
  {Mathews}},\ }\bibfield  {title} {\bibinfo {title} {Gravitational radiation
  from point masses in a keplerian orbit},\ }\href
  {https://doi.org/10.1103/PhysRev.131.435} {\bibfield  {journal} {\bibinfo
  {journal} {Physical Review}\ }\textbf {\bibinfo {volume} {131}},\ \bibinfo
  {pages} {435} (\bibinfo {year} {1963})}\BibitemShut {NoStop}%
\bibitem [{\citenamefont {Peters}(1964)}]{Peters1964}%
  \BibitemOpen
  \bibfield  {author} {\bibinfo {author} {\bibfnamefont {P.~C.}\ \bibnamefont
  {Peters}},\ }\bibfield  {title} {\bibinfo {title} {Gravitational radiation
  and the motion of two point masses},\ }\href
  {https://doi.org/10.1103/PhysRev.136.B1224} {\bibfield  {journal} {\bibinfo
  {journal} {Physical Review}\ }\textbf {\bibinfo {volume} {136}},\ \bibinfo
  {pages} {B1224} (\bibinfo {year} {1964})}\BibitemShut {NoStop}%
\bibitem [{\citenamefont {Gold}\ \emph {et~al.}(2012)\citenamefont {Gold},
  \citenamefont {Bernuzzi}, \citenamefont {Thierfelder}, \citenamefont
  {Br{\"u}gmann},\ and\ \citenamefont {Pretorius}}]{Gold2012EccentricBNS}%
  \BibitemOpen
  \bibfield  {author} {\bibinfo {author} {\bibfnamefont {R.}~\bibnamefont
  {Gold}}, \bibinfo {author} {\bibfnamefont {S.}~\bibnamefont {Bernuzzi}},
  \bibinfo {author} {\bibfnamefont {M.}~\bibnamefont {Thierfelder}}, \bibinfo
  {author} {\bibfnamefont {B.}~\bibnamefont {Br{\"u}gmann}},\ and\ \bibinfo
  {author} {\bibfnamefont {F.}~\bibnamefont {Pretorius}},\ }\bibfield  {title}
  {\bibinfo {title} {Eccentric binary neutron star mergers},\ }\href
  {https://doi.org/10.1103/PhysRevD.86.121501} {\bibfield  {journal} {\bibinfo
  {journal} {Physical Review D}\ }\textbf {\bibinfo {volume} {86}},\ \bibinfo
  {pages} {121501} (\bibinfo {year} {2012})}\BibitemShut {NoStop}%
\bibitem [{\citenamefont {East}\ \emph {et~al.}(2012)\citenamefont {East},
  \citenamefont {Pretorius},\ and\ \citenamefont
  {Stephens}}]{East2012DynamicalCapture}%
  \BibitemOpen
  \bibfield  {author} {\bibinfo {author} {\bibfnamefont {W.~E.}\ \bibnamefont
  {East}}, \bibinfo {author} {\bibfnamefont {F.}~\bibnamefont {Pretorius}},\
  and\ \bibinfo {author} {\bibfnamefont {B.~C.}\ \bibnamefont {Stephens}},\
  }\bibfield  {title} {\bibinfo {title} {Dynamical capture binary neutron star
  mergers},\ }\href {https://doi.org/10.1088/2041-8205/760/1/L4} {\bibfield
  {journal} {\bibinfo  {journal} {The Astrophysical Journal Letters}\ }\textbf
  {\bibinfo {volume} {760}},\ \bibinfo {pages} {L4} (\bibinfo {year}
  {2012})}\BibitemShut {NoStop}%
\bibitem [{\citenamefont {Paschalidis}\ \emph {et~al.}(2015)\citenamefont
  {Paschalidis}, \citenamefont {East}, \citenamefont {Pretorius},\ and\
  \citenamefont {Shapiro}}]{Paschalidis2015OneArm}%
  \BibitemOpen
  \bibfield  {author} {\bibinfo {author} {\bibfnamefont {V.}~\bibnamefont
  {Paschalidis}}, \bibinfo {author} {\bibfnamefont {W.~E.}\ \bibnamefont
  {East}}, \bibinfo {author} {\bibfnamefont {F.}~\bibnamefont {Pretorius}},\
  and\ \bibinfo {author} {\bibfnamefont {S.~L.}\ \bibnamefont {Shapiro}},\
  }\bibfield  {title} {\bibinfo {title} {One-arm spiral instability in
  hypermassive neutron stars formed by dynamical-capture binary neutron star
  mergers},\ }\href {https://doi.org/10.1103/PhysRevD.92.121502} {\bibfield
  {journal} {\bibinfo  {journal} {Physical Review D}\ }\textbf {\bibinfo
  {volume} {92}},\ \bibinfo {pages} {121502} (\bibinfo {year}
  {2015})}\BibitemShut {NoStop}%
\bibitem [{\citenamefont {Morras}\ \emph {et~al.}(2026)\citenamefont {Morras},
  \citenamefont {Pratten},\ and\ \citenamefont
  {Schmidt}}]{Morras2026EccentricNSBH}%
  \BibitemOpen
  \bibfield  {author} {\bibinfo {author} {\bibfnamefont {G.}~\bibnamefont
  {Morras}}, \bibinfo {author} {\bibfnamefont {G.}~\bibnamefont {Pratten}},\
  and\ \bibinfo {author} {\bibfnamefont {P.}~\bibnamefont {Schmidt}},\
  }\bibfield  {title} {\bibinfo {title} {Orbital eccentricity in a neutron
  star--black hole merger},\ }\href {https://doi.org/10.3847/2041-8213/ae474c}
  {\bibfield  {journal} {\bibinfo  {journal} {The Astrophysical Journal
  Letters}\ }\textbf {\bibinfo {volume} {1000}},\ \bibinfo {pages} {L2}
  (\bibinfo {year} {2026})},\ \Eprint {https://arxiv.org/abs/2503.15393}
  {arXiv:2503.15393 [astro-ph.HE]} \BibitemShut {NoStop}%
\bibitem [{\citenamefont {Clarke}\ \emph {et~al.}(2026)\citenamefont {Clarke},
  \citenamefont {Romero-Shaw}, \citenamefont {Hoy}, \citenamefont {Stegmann},
  \citenamefont {Lasky},\ and\ \citenamefont {Thrane}}]{Clarke2026GW200105}%
  \BibitemOpen
  \bibfield  {author} {\bibinfo {author} {\bibfnamefont {T.~A.}\ \bibnamefont
  {Clarke}}, \bibinfo {author} {\bibfnamefont {I.~M.}\ \bibnamefont
  {Romero-Shaw}}, \bibinfo {author} {\bibfnamefont {C.}~\bibnamefont {Hoy}},
  \bibinfo {author} {\bibfnamefont {J.}~\bibnamefont {Stegmann}}, \bibinfo
  {author} {\bibfnamefont {P.~D.}\ \bibnamefont {Lasky}},\ and\ \bibinfo
  {author} {\bibfnamefont {E.}~\bibnamefont {Thrane}},\ }\bibfield  {title}
  {\bibinfo {title} {A universal framework to identify eccentric binary
  mergers: {GW200105} case study},\ }\href
  {https://doi.org/10.48550/arXiv.2605.18742} {\bibfield  {journal} {\bibinfo
  {journal} {arXiv e-prints}\ ,\ \bibinfo {pages} {arXiv:2605.18742}} (\bibinfo
  {year} {2026})},\ \Eprint {https://arxiv.org/abs/2605.18742}
  {arXiv:2605.18742 [astro-ph.HE]} \BibitemShut {NoStop}%
\bibitem [{\citenamefont {Nitz}\ \emph {et~al.}(2020)\citenamefont {Nitz},
  \citenamefont {Lenon},\ and\ \citenamefont {Brown}}]{nitz2019eccbns}%
  \BibitemOpen
  \bibfield  {author} {\bibinfo {author} {\bibfnamefont {A.~H.}\ \bibnamefont
  {Nitz}}, \bibinfo {author} {\bibfnamefont {A.}~\bibnamefont {Lenon}},\ and\
  \bibinfo {author} {\bibfnamefont {D.~A.}\ \bibnamefont {Brown}},\ }\bibfield
  {title} {\bibinfo {title} {Search for eccentric binary neutron star mergers
  in the first and second observing runs of advanced ligo},\ }\href
  {https://doi.org/10.3847/1538-4357/ab6611} {\bibfield  {journal} {\bibinfo
  {journal} {The Astrophysical Journal}\ }\textbf {\bibinfo {volume} {890}},\
  \bibinfo {pages} {1} (\bibinfo {year} {2020})},\ \Eprint
  {https://arxiv.org/abs/1912.05464} {arXiv:1912.05464 [astro-ph.HE]}
  \BibitemShut {NoStop}%
\bibitem [{\citenamefont {Lenon}\ \emph {et~al.}(2020)\citenamefont {Lenon},
  \citenamefont {Nitz},\ and\ \citenamefont {Brown}}]{lenon2020ecc}%
  \BibitemOpen
  \bibfield  {author} {\bibinfo {author} {\bibfnamefont {A.~K.}\ \bibnamefont
  {Lenon}}, \bibinfo {author} {\bibfnamefont {A.~H.}\ \bibnamefont {Nitz}},\
  and\ \bibinfo {author} {\bibfnamefont {D.~A.}\ \bibnamefont {Brown}},\
  }\bibfield  {title} {\bibinfo {title} {Measuring the eccentricity of gw170817
  and gw190425},\ }\href {https://doi.org/10.1093/mnras/staa2120} {\bibfield
  {journal} {\bibinfo  {journal} {Monthly Notices of the Royal Astronomical
  Society}\ }\textbf {\bibinfo {volume} {497}},\ \bibinfo {pages} {1966}
  (\bibinfo {year} {2020})},\ \Eprint {https://arxiv.org/abs/2005.14146}
  {arXiv:2005.14146 [astro-ph.HE]} \BibitemShut {NoStop}%
\bibitem [{\citenamefont {Lenon}\ \emph {et~al.}(2021)\citenamefont {Lenon},
  \citenamefont {Brown},\ and\ \citenamefont {Nitz}}]{Lenon2021CEeccentric}%
  \BibitemOpen
  \bibfield  {author} {\bibinfo {author} {\bibfnamefont {A.~K.}\ \bibnamefont
  {Lenon}}, \bibinfo {author} {\bibfnamefont {D.~A.}\ \bibnamefont {Brown}},\
  and\ \bibinfo {author} {\bibfnamefont {A.~H.}\ \bibnamefont {Nitz}},\
  }\bibfield  {title} {\bibinfo {title} {Eccentric binary neutron star search
  prospects for {Cosmic Explorer}},\ }\href
  {https://doi.org/10.1103/PhysRevD.104.063011} {\bibfield  {journal} {\bibinfo
   {journal} {Physical Review D}\ }\textbf {\bibinfo {volume} {104}},\ \bibinfo
  {pages} {063011} (\bibinfo {year} {2021})}\BibitemShut {NoStop}%
\bibitem [{\citenamefont {Dutta~Roy}\ and\ \citenamefont
  {Saini}(2024)}]{DuttaRoySaini2024UnmodeledEccentricity}%
  \BibitemOpen
  \bibfield  {author} {\bibinfo {author} {\bibfnamefont {P.}~\bibnamefont
  {Dutta~Roy}}\ and\ \bibinfo {author} {\bibfnamefont {P.}~\bibnamefont
  {Saini}},\ }\bibfield  {title} {\bibinfo {title} {Impact of unmodeled
  eccentricity on the tidal deformability measurement and implications for
  gravitational wave physics inference},\ }\href
  {https://doi.org/10.1103/PhysRevD.110.024002} {\bibfield  {journal} {\bibinfo
   {journal} {Physical Review D}\ }\textbf {\bibinfo {volume} {110}},\ \bibinfo
  {pages} {024002} (\bibinfo {year} {2024})}\BibitemShut {NoStop}%
\bibitem [{\citenamefont {Kacanja}\ \emph {et~al.}(2025)\citenamefont
  {Kacanja}, \citenamefont {Soni},\ and\ \citenamefont
  {Nitz}}]{KacanjaSoniNitz2025EccentricitySignatures}%
  \BibitemOpen
  \bibfield  {author} {\bibinfo {author} {\bibfnamefont {K.}~\bibnamefont
  {Kacanja}}, \bibinfo {author} {\bibfnamefont {K.}~\bibnamefont {Soni}},\ and\
  \bibinfo {author} {\bibfnamefont {A.~H.}\ \bibnamefont {Nitz}},\ }\bibfield
  {title} {\bibinfo {title} {Eccentricity signatures in ligo-virgo-kagra's bns
  and nsbh binaries},\ }\href@noop {} {\bibfield  {journal} {\bibinfo
  {journal} {arXiv e-prints}\ ,\ \bibinfo {pages} {arXiv:2508.00179}} (\bibinfo
  {year} {2025})},\ \Eprint {https://arxiv.org/abs/2508.00179}
  {arXiv:2508.00179 [gr-qc]} \BibitemShut {NoStop}%
\bibitem [{\citenamefont {Punturo}\ \emph {et~al.}(2010)\citenamefont
  {Punturo}, \citenamefont {Abernathy}, \citenamefont {Acernese}, \citenamefont
  {Allen} \emph {et~al.}}]{Punturo2010ET}%
  \BibitemOpen
  \bibfield  {author} {\bibinfo {author} {\bibfnamefont {M.}~\bibnamefont
  {Punturo}}, \bibinfo {author} {\bibfnamefont {M.}~\bibnamefont {Abernathy}},
  \bibinfo {author} {\bibfnamefont {F.}~\bibnamefont {Acernese}}, \bibinfo
  {author} {\bibfnamefont {B.}~\bibnamefont {Allen}}, \emph {et~al.},\
  }\bibfield  {title} {\bibinfo {title} {The einstein telescope: A
  third-generation gravitational wave observatory},\ }\href
  {https://doi.org/10.1088/0264-9381/27/19/194002} {\bibfield  {journal}
  {\bibinfo  {journal} {Classical and Quantum Gravity}\ }\textbf {\bibinfo
  {volume} {27}},\ \bibinfo {pages} {194002} (\bibinfo {year}
  {2010})}\BibitemShut {NoStop}%
\bibitem [{\citenamefont {Hild}\ \emph {et~al.}(2011)\citenamefont {Hild},
  \citenamefont {Abernathy}, \citenamefont {Acernese}, \citenamefont
  {Amaro-Seoane} \emph {et~al.}}]{Hild2011ET}%
  \BibitemOpen
  \bibfield  {author} {\bibinfo {author} {\bibfnamefont {S.}~\bibnamefont
  {Hild}}, \bibinfo {author} {\bibfnamefont {M.}~\bibnamefont {Abernathy}},
  \bibinfo {author} {\bibfnamefont {F.}~\bibnamefont {Acernese}}, \bibinfo
  {author} {\bibfnamefont {P.}~\bibnamefont {Amaro-Seoane}}, \emph {et~al.},\
  }\bibfield  {title} {\bibinfo {title} {Sensitivity studies for
  third-generation gravitational wave observatories},\ }\href
  {https://doi.org/10.1088/0264-9381/28/9/094013} {\bibfield  {journal}
  {\bibinfo  {journal} {Classical and Quantum Gravity}\ }\textbf {\bibinfo
  {volume} {28}},\ \bibinfo {pages} {094013} (\bibinfo {year}
  {2011})}\BibitemShut {NoStop}%
\bibitem [{\citenamefont {Maggiore}\ \emph {et~al.}(2019)\citenamefont
  {Maggiore}, \citenamefont {{van den Broeck}}, \citenamefont {Bartolo},
  \citenamefont {Belgacem}, \citenamefont {Bertacca} \emph
  {et~al.}}]{maggiore2020et}%
  \BibitemOpen
  \bibfield  {author} {\bibinfo {author} {\bibfnamefont {M.}~\bibnamefont
  {Maggiore}}, \bibinfo {author} {\bibfnamefont {C.}~\bibnamefont {{van den
  Broeck}}}, \bibinfo {author} {\bibfnamefont {N.}~\bibnamefont {Bartolo}},
  \bibinfo {author} {\bibfnamefont {E.}~\bibnamefont {Belgacem}}, \bibinfo
  {author} {\bibfnamefont {D.}~\bibnamefont {Bertacca}}, \emph {et~al.},\
  }\bibfield  {title} {\bibinfo {title} {Science case for the einstein
  telescope},\ }\bibfield  {journal} {\bibinfo  {journal} {arXiv preprint
  arXiv:1912.02622}\ }\href {https://doi.org/10.48550/arXiv.1912.02622}
  {10.48550/arXiv.1912.02622} (\bibinfo {year} {2019}),\ \Eprint
  {https://arxiv.org/abs/1912.02622} {arXiv:1912.02622 [astro-ph.CO]}
  \BibitemShut {NoStop}%
\bibitem [{\citenamefont {Reitze}\ \emph {et~al.}(2019)\citenamefont {Reitze}
  \emph {et~al.}}]{Reitze2019CosmicExplorer}%
  \BibitemOpen
  \bibfield  {author} {\bibinfo {author} {\bibfnamefont {D.}~\bibnamefont
  {Reitze}} \emph {et~al.},\ }\bibfield  {title} {\bibinfo {title} {Cosmic
  explorer: The u.s. contribution to gravitational-wave astronomy beyond
  ligo},\ }\href {https://doi.org/10.48550/arXiv.1907.04833} {\bibfield  {journal} {\bibinfo  {journal} {Bulletin
  of the American Astronomical Society}\ }\textbf {\bibinfo {volume} {51}},\
  \bibinfo {pages} {35} (\bibinfo {year} {2019})},\ \Eprint
  {https://arxiv.org/abs/1907.04833} {arXiv:1907.04833 [astro-ph.IM]}
  \BibitemShut {NoStop}%
\bibitem [{\citenamefont {Evans}\ \emph {et~al.}(2021)\citenamefont {Evans},
  \citenamefont {Adhikari}, \citenamefont {Afle}, \citenamefont {Ballmer} \emph
  {et~al.}}]{evans2021ce}%
  \BibitemOpen
  \bibfield  {author} {\bibinfo {author} {\bibfnamefont {M.}~\bibnamefont
  {Evans}}, \bibinfo {author} {\bibfnamefont {R.~X.}\ \bibnamefont {Adhikari}},
  \bibinfo {author} {\bibfnamefont {C.}~\bibnamefont {Afle}}, \bibinfo {author}
  {\bibfnamefont {S.}~\bibnamefont {Ballmer}}, \emph {et~al.},\ }\bibfield
  {title} {\bibinfo {title} {A horizon study for cosmic explorer: Science,
  observatories, and community},\ }\bibfield  {journal} {\bibinfo  {journal}
  {arXiv preprint arXiv:2109.09882}\ }\href
  {https://doi.org/10.48550/arXiv.2109.09882} {10.48550/arXiv.2109.09882}
  (\bibinfo {year} {2021}),\ \Eprint {https://arxiv.org/abs/2109.09882}
  {arXiv:2109.09882 [astro-ph.IM]} \BibitemShut {NoStop}%
\bibitem [{\citenamefont {Hall}(2022)}]{Hall2022CosmicExplorer}%
  \BibitemOpen
  \bibfield  {author} {\bibinfo {author} {\bibfnamefont {E.~D.}\ \bibnamefont
  {Hall}},\ }\bibfield  {title} {\bibinfo {title} {Cosmic explorer: A
  next-generation ground-based gravitational-wave observatory},\ }\href
  {https://doi.org/10.3390/galaxies10040090} {\bibfield  {journal} {\bibinfo
  {journal} {Galaxies}\ }\textbf {\bibinfo {volume} {10}},\ \bibinfo {pages}
  {90} (\bibinfo {year} {2022})}\BibitemShut {NoStop}%
\bibitem [{\citenamefont {Allen}\ \emph {et~al.}(2012)\citenamefont {Allen},
  \citenamefont {Anderson}, \citenamefont {Brady}, \citenamefont {Brown},\ and\
  \citenamefont {Creighton}}]{Allen2012FindChirp}%
  \BibitemOpen
  \bibfield  {author} {\bibinfo {author} {\bibfnamefont {B.}~\bibnamefont
  {Allen}}, \bibinfo {author} {\bibfnamefont {W.~G.}\ \bibnamefont {Anderson}},
  \bibinfo {author} {\bibfnamefont {P.~R.}\ \bibnamefont {Brady}}, \bibinfo
  {author} {\bibfnamefont {D.~A.}\ \bibnamefont {Brown}},\ and\ \bibinfo
  {author} {\bibfnamefont {J.~D.~E.}\ \bibnamefont {Creighton}},\ }\bibfield
  {title} {\bibinfo {title} {{FINDCHIRP}: An algorithm for detection of
  gravitational waves from inspiraling compact binaries},\ }\href
  {https://doi.org/10.1103/PhysRevD.85.122006} {\bibfield  {journal} {\bibinfo
  {journal} {Physical Review D}\ }\textbf {\bibinfo {volume} {85}},\ \bibinfo
  {pages} {122006} (\bibinfo {year} {2012})}\BibitemShut {NoStop}%
\bibitem [{\citenamefont {Veitch}\ \emph {et~al.}(2015)\citenamefont {Veitch},
  \citenamefont {Raymond}, \citenamefont {Farr}, \citenamefont {Farr},
  \citenamefont {Graff}, \citenamefont {Vitale}, \citenamefont {Aylott},
  \citenamefont {Blackburn}, \citenamefont {Christensen}, \citenamefont
  {Coughlin}, \citenamefont {Del~Pozzo}, \citenamefont {Feroz}, \citenamefont
  {Gair}, \citenamefont {Haster}, \citenamefont {Kalogera}, \citenamefont
  {Littenberg}, \citenamefont {Mandel}, \citenamefont {O'Shaughnessy},
  \citenamefont {Pitkin}, \citenamefont {Rodriguez}, \citenamefont {R{\"o}ver},
  \citenamefont {Sidery}, \citenamefont {Smith}, \citenamefont {Van Der~Sluys},
  \citenamefont {Vecchio}, \citenamefont {Vousden},\ and\ \citenamefont
  {Wade}}]{Veitch2015LALInference}%
  \BibitemOpen
  \bibfield  {author} {\bibinfo {author} {\bibfnamefont {J.}~\bibnamefont
  {Veitch}}, \bibinfo {author} {\bibfnamefont {V.}~\bibnamefont {Raymond}},
  \bibinfo {author} {\bibfnamefont {B.}~\bibnamefont {Farr}}, \bibinfo {author}
  {\bibfnamefont {W.}~\bibnamefont {Farr}}, \bibinfo {author} {\bibfnamefont
  {P.}~\bibnamefont {Graff}}, \bibinfo {author} {\bibfnamefont
  {S.}~\bibnamefont {Vitale}}, \bibinfo {author} {\bibfnamefont
  {B.}~\bibnamefont {Aylott}}, \bibinfo {author} {\bibfnamefont
  {K.}~\bibnamefont {Blackburn}}, \bibinfo {author} {\bibfnamefont
  {N.}~\bibnamefont {Christensen}}, \bibinfo {author} {\bibfnamefont
  {M.}~\bibnamefont {Coughlin}}, \bibinfo {author} {\bibfnamefont
  {W.}~\bibnamefont {Del~Pozzo}}, \bibinfo {author} {\bibfnamefont
  {F.}~\bibnamefont {Feroz}}, \bibinfo {author} {\bibfnamefont
  {J.}~\bibnamefont {Gair}}, \bibinfo {author} {\bibfnamefont {C.-J.}\
  \bibnamefont {Haster}}, \bibinfo {author} {\bibfnamefont {V.}~\bibnamefont
  {Kalogera}}, \bibinfo {author} {\bibfnamefont {T.}~\bibnamefont
  {Littenberg}}, \bibinfo {author} {\bibfnamefont {I.}~\bibnamefont {Mandel}},
  \bibinfo {author} {\bibfnamefont {R.}~\bibnamefont {O'Shaughnessy}}, \bibinfo
  {author} {\bibfnamefont {M.}~\bibnamefont {Pitkin}}, \bibinfo {author}
  {\bibfnamefont {C.}~\bibnamefont {Rodriguez}}, \bibinfo {author}
  {\bibfnamefont {C.}~\bibnamefont {R{\"o}ver}}, \bibinfo {author}
  {\bibfnamefont {T.}~\bibnamefont {Sidery}}, \bibinfo {author} {\bibfnamefont
  {R.}~\bibnamefont {Smith}}, \bibinfo {author} {\bibfnamefont
  {M.}~\bibnamefont {Van Der~Sluys}}, \bibinfo {author} {\bibfnamefont
  {A.}~\bibnamefont {Vecchio}}, \bibinfo {author} {\bibfnamefont
  {W.}~\bibnamefont {Vousden}},\ and\ \bibinfo {author} {\bibfnamefont
  {L.}~\bibnamefont {Wade}},\ }\bibfield  {title} {\bibinfo {title} {Parameter
  estimation for compact binaries with ground-based gravitational-wave
  observations using the {LALInference} software library},\ }\href
  {https://doi.org/10.1103/PhysRevD.91.042003} {\bibfield  {journal} {\bibinfo
  {journal} {Physical Review D}\ }\textbf {\bibinfo {volume} {91}},\ \bibinfo
  {pages} {042003} (\bibinfo {year} {2015})}\BibitemShut {NoStop}%
\bibitem [{\citenamefont {Ashton}\ \emph {et~al.}(2019)\citenamefont {Ashton},
  \citenamefont {H{\"u}bner}, \citenamefont {Lasky}, \citenamefont {Talbot},
  \citenamefont {Ackley}, \citenamefont {Biscoveanu}, \citenamefont {Chu},
  \citenamefont {Divakarla}, \citenamefont {Easter}, \citenamefont {Goncharov},
  \citenamefont {Hernandez~Vivanco}, \citenamefont {Harms}, \citenamefont
  {Lower}, \citenamefont {Meadors}, \citenamefont {Melchor}, \citenamefont
  {Payne}, \citenamefont {Pitkin}, \citenamefont {Powell}, \citenamefont
  {Sarin}, \citenamefont {Smith},\ and\ \citenamefont
  {Thrane}}]{Ashton2019Bilby}%
  \BibitemOpen
  \bibfield  {author} {\bibinfo {author} {\bibfnamefont {G.}~\bibnamefont
  {Ashton}}, \bibinfo {author} {\bibfnamefont {M.}~\bibnamefont {H{\"u}bner}},
  \bibinfo {author} {\bibfnamefont {P.~D.}\ \bibnamefont {Lasky}}, \bibinfo
  {author} {\bibfnamefont {C.}~\bibnamefont {Talbot}}, \bibinfo {author}
  {\bibfnamefont {K.}~\bibnamefont {Ackley}}, \bibinfo {author} {\bibfnamefont
  {S.}~\bibnamefont {Biscoveanu}}, \bibinfo {author} {\bibfnamefont
  {Q.}~\bibnamefont {Chu}}, \bibinfo {author} {\bibfnamefont {A.}~\bibnamefont
  {Divakarla}}, \bibinfo {author} {\bibfnamefont {P.~J.}\ \bibnamefont
  {Easter}}, \bibinfo {author} {\bibfnamefont {B.}~\bibnamefont {Goncharov}},
  \bibinfo {author} {\bibfnamefont {F.}~\bibnamefont {Hernandez~Vivanco}},
  \bibinfo {author} {\bibfnamefont {J.}~\bibnamefont {Harms}}, \bibinfo
  {author} {\bibfnamefont {M.~E.}\ \bibnamefont {Lower}}, \bibinfo {author}
  {\bibfnamefont {G.~D.}\ \bibnamefont {Meadors}}, \bibinfo {author}
  {\bibfnamefont {D.}~\bibnamefont {Melchor}}, \bibinfo {author} {\bibfnamefont
  {E.}~\bibnamefont {Payne}}, \bibinfo {author} {\bibfnamefont {M.~D.}\
  \bibnamefont {Pitkin}}, \bibinfo {author} {\bibfnamefont {J.}~\bibnamefont
  {Powell}}, \bibinfo {author} {\bibfnamefont {N.}~\bibnamefont {Sarin}},
  \bibinfo {author} {\bibfnamefont {R.~J.~E.}\ \bibnamefont {Smith}},\ and\
  \bibinfo {author} {\bibfnamefont {E.}~\bibnamefont {Thrane}},\ }\bibfield
  {title} {\bibinfo {title} {{Bilby}: A user-friendly bayesian inference
  library for gravitational-wave astronomy},\ }\href
  {https://doi.org/10.3847/1538-4365/ab06fc} {\bibfield  {journal} {\bibinfo
  {journal} {The Astrophysical Journal Supplement Series}\ }\textbf {\bibinfo
  {volume} {241}},\ \bibinfo {pages} {27} (\bibinfo {year} {2019})}\BibitemShut
  {NoStop}%
\bibitem [{\citenamefont {George}\ and\ \citenamefont
  {Huerta}(2018)}]{george2017deeplearn}%
  \BibitemOpen
  \bibfield  {author} {\bibinfo {author} {\bibfnamefont {D.}~\bibnamefont
  {George}}\ and\ \bibinfo {author} {\bibfnamefont {E.~A.}\ \bibnamefont
  {Huerta}},\ }\bibfield  {title} {\bibinfo {title} {Deep learning for
  real-time gravitational wave detection and parameter estimation: Results with
  advanced ligo data},\ }\href {https://doi.org/10.1016/j.physletb.2017.12.053}
  {\bibfield  {journal} {\bibinfo  {journal} {Physics Letters B}\ }\textbf
  {\bibinfo {volume} {778}},\ \bibinfo {pages} {64} (\bibinfo {year} {2018})},\
  \Eprint {https://arxiv.org/abs/1711.03121} {arXiv:1711.03121 [gr-qc]}
  \BibitemShut {NoStop}%
\bibitem [{\citenamefont {Sun}\ \emph {et~al.}(2024)\citenamefont {Sun},
  \citenamefont {Li}, \citenamefont {Cao},\ and\ \citenamefont
  {Liu}}]{Sun2024DeepLearningDECIGO}%
  \BibitemOpen
  \bibfield  {author} {\bibinfo {author} {\bibfnamefont {M.}~\bibnamefont
  {Sun}}, \bibinfo {author} {\bibfnamefont {J.}~\bibnamefont {Li}}, \bibinfo
  {author} {\bibfnamefont {S.}~\bibnamefont {Cao}},\ and\ \bibinfo {author}
  {\bibfnamefont {X.}~\bibnamefont {Liu}},\ }\bibfield  {title} {\bibinfo
  {title} {Deep learning forecasts of cosmic acceleration parameters from
  {DECi}-hertz interferometer gravitational-wave observatory},\ }\href
  {https://doi.org/10.1051/0004-6361/202347221} {\bibfield  {journal} {\bibinfo
   {journal} {Astronomy \& Astrophysics}\ }\textbf {\bibinfo {volume} {682}},\
  \bibinfo {pages} {A177} (\bibinfo {year} {2024})}\BibitemShut {NoStop}%
\bibitem [{\citenamefont {Gabbard}\ \emph {et~al.}(2018)\citenamefont
  {Gabbard}, \citenamefont {Williams}, \citenamefont {Hayes},\ and\
  \citenamefont {Messenger}}]{Gabbard2018DeepNetworks}%
  \BibitemOpen
  \bibfield  {author} {\bibinfo {author} {\bibfnamefont {H.}~\bibnamefont
  {Gabbard}}, \bibinfo {author} {\bibfnamefont {M.}~\bibnamefont {Williams}},
  \bibinfo {author} {\bibfnamefont {F.}~\bibnamefont {Hayes}},\ and\ \bibinfo
  {author} {\bibfnamefont {C.}~\bibnamefont {Messenger}},\ }\bibfield  {title}
  {\bibinfo {title} {Matching matched filtering with deep networks for
  gravitational-wave astronomy},\ }\href
  {https://doi.org/10.1103/PhysRevLett.120.141103} {\bibfield  {journal}
  {\bibinfo  {journal} {Physical Review Letters}\ }\textbf {\bibinfo {volume}
  {120}},\ \bibinfo {pages} {141103} (\bibinfo {year} {2018})}\BibitemShut
  {NoStop}%
\bibitem [{\citenamefont {Sun}\ \emph {et~al.}(2025)\citenamefont {Sun},
  \citenamefont {Wu}, \citenamefont {Li}, \citenamefont {{McCane}},
  \citenamefont {Yang}, \citenamefont {Ma}, \citenamefont {Wang},\ and\
  \citenamefont {Zhang}}]{Sun2025ConditionalAutoencoderBNS}%
  \BibitemOpen
  \bibfield  {author} {\bibinfo {author} {\bibfnamefont {M.}~\bibnamefont
  {Sun}}, \bibinfo {author} {\bibfnamefont {J.}~\bibnamefont {Wu}}, \bibinfo
  {author} {\bibfnamefont {J.}~\bibnamefont {Li}}, \bibinfo {author}
  {\bibfnamefont {B.}~\bibnamefont {{McCane}}}, \bibinfo {author}
  {\bibfnamefont {N.}~\bibnamefont {Yang}}, \bibinfo {author} {\bibfnamefont
  {X.}~\bibnamefont {Ma}}, \bibinfo {author} {\bibfnamefont {B.}~\bibnamefont
  {Wang}},\ and\ \bibinfo {author} {\bibfnamefont {M.}~\bibnamefont {Zhang}},\
  }\bibfield  {title} {\bibinfo {title} {Conditional autoencoder for generating
  binary neutron star waveforms with tidal and precession effects},\ }\href
  {https://doi.org/10.1103/kmlw-y7yw} {\bibfield  {journal} {\bibinfo
  {journal} {Physical Review D}\ }\textbf {\bibinfo {volume} {112}},\ \bibinfo
  {pages} {084016} (\bibinfo {year} {2025})}\BibitemShut {NoStop}%
\bibitem [{\citenamefont {Cuoco}\ \emph {et~al.}(2020)\citenamefont {Cuoco},
  \citenamefont {Powell}, \citenamefont {Cavagli{\`a}}, \citenamefont {Ackley}
  \emph {et~al.}}]{Cuoco2020MLReview}%
  \BibitemOpen
  \bibfield  {author} {\bibinfo {author} {\bibfnamefont {E.}~\bibnamefont
  {Cuoco}}, \bibinfo {author} {\bibfnamefont {J.}~\bibnamefont {Powell}},
  \bibinfo {author} {\bibfnamefont {M.}~\bibnamefont {Cavagli{\`a}}}, \bibinfo
  {author} {\bibfnamefont {K.}~\bibnamefont {Ackley}}, \emph {et~al.},\
  }\bibfield  {title} {\bibinfo {title} {Enhancing gravitational-wave science
  with machine learning},\ }\href {https://doi.org/10.1088/2632-2153/abb93a}
  {\bibfield  {journal} {\bibinfo  {journal} {Machine Learning: Science and
  Technology}\ }\textbf {\bibinfo {volume} {2}},\ \bibinfo {pages} {011002}
  (\bibinfo {year} {2020})}\BibitemShut {NoStop}%
\bibitem [{\citenamefont {Sun}\ and\ \citenamefont
  {Li}(2023)}]{Sun2023DECIGOIMBBH}%
  \BibitemOpen
  \bibfield  {author} {\bibinfo {author} {\bibfnamefont {M.}~\bibnamefont
  {Sun}}\ and\ \bibinfo {author} {\bibfnamefont {J.}~\bibnamefont {Li}},\
  }\bibfield  {title} {\bibinfo {title} {Parameter estimation for
  intermediate-mass binary black holes through gravitational waves observed by
  {DECIGO}},\ }\href@noop {} {\bibfield  {journal} {\bibinfo  {journal} {arXiv
  e-prints}\ ,\ \bibinfo {pages} {arXiv:2312.07834}} (\bibinfo {year}
  {2023})},\ \Eprint {https://arxiv.org/abs/2312.07834} {arXiv:2312.07834
  [gr-qc]} \BibitemShut {NoStop}%
\bibitem [{\citenamefont {Chua}\ and\ \citenamefont
  {Vallisneri}(2020)}]{ChuaVallisneri2020Posteriors}%
  \BibitemOpen
  \bibfield  {author} {\bibinfo {author} {\bibfnamefont {A.~J.~K.}\
  \bibnamefont {Chua}}\ and\ \bibinfo {author} {\bibfnamefont {M.}~\bibnamefont
  {Vallisneri}},\ }\bibfield  {title} {\bibinfo {title} {Learning bayesian
  posteriors with neural networks for gravitational-wave inference},\ }\href
  {https://doi.org/10.1103/PhysRevLett.124.041102} {\bibfield  {journal}
  {\bibinfo  {journal} {Physical Review Letters}\ }\textbf {\bibinfo {volume}
  {124}},\ \bibinfo {pages} {041102} (\bibinfo {year} {2020})}\BibitemShut
  {NoStop}%
\bibitem [{\citenamefont {Sun}\ \emph {et~al.}(2026)\citenamefont {Sun},
  \citenamefont {Wu}, \citenamefont {Hu}, \citenamefont {Li}, \citenamefont
  {Yang}, \citenamefont {Ma}, \citenamefont {Wang}, \citenamefont {Zhang},\
  and\ \citenamefont {Zhong}}]{Sun2025LensedDMHalos}%
  \BibitemOpen
  \bibfield  {author} {\bibinfo {author} {\bibfnamefont {M.}~\bibnamefont
  {Sun}}, \bibinfo {author} {\bibfnamefont {J.}~\bibnamefont {Wu}}, \bibinfo
  {author} {\bibfnamefont {Q.}~\bibnamefont {Hu}}, \bibinfo {author}
  {\bibfnamefont {J.}~\bibnamefont {Li}}, \bibinfo {author} {\bibfnamefont
  {N.}~\bibnamefont {Yang}}, \bibinfo {author} {\bibfnamefont {X.}~\bibnamefont
  {Ma}}, \bibinfo {author} {\bibfnamefont {B.}~\bibnamefont {Wang}}, \bibinfo
  {author} {\bibfnamefont {M.}~\bibnamefont {Zhang}},\ and\ \bibinfo {author}
  {\bibfnamefont {Y.}~\bibnamefont {Zhong}},\ }\bibfield  {title} {\bibinfo
  {title} {Detection of multiband lensed gravitational waves from dark matter
  halos with deep learning},\ }\href {https://doi.org/10.1103/p8jc-kgp2}
  {\bibfield  {journal} {\bibinfo  {journal} {Physical Review D}\ }\textbf
  {\bibinfo {volume} {113}},\ \bibinfo {pages} {103004} (\bibinfo {year}
  {2026})},\ \Eprint {https://arxiv.org/abs/2511.09107} {arXiv:2511.09107
  [astro-ph.IM]} \BibitemShut {NoStop}%
\bibitem [{\citenamefont {Dax}\ \emph {et~al.}(2021)\citenamefont {Dax},
  \citenamefont {Green}, \citenamefont {Gair}, \citenamefont {Macke},
  \citenamefont {Buonanno},\ and\ \citenamefont {Sch{\"o}lkopf}}]{Dax2021NPE}%
  \BibitemOpen
  \bibfield  {author} {\bibinfo {author} {\bibfnamefont {M.}~\bibnamefont
  {Dax}}, \bibinfo {author} {\bibfnamefont {S.~R.}\ \bibnamefont {Green}},
  \bibinfo {author} {\bibfnamefont {J.}~\bibnamefont {Gair}}, \bibinfo {author}
  {\bibfnamefont {J.~H.}\ \bibnamefont {Macke}}, \bibinfo {author}
  {\bibfnamefont {A.}~\bibnamefont {Buonanno}},\ and\ \bibinfo {author}
  {\bibfnamefont {B.}~\bibnamefont {Sch{\"o}lkopf}},\ }\bibfield  {title}
  {\bibinfo {title} {Real-time gravitational wave science with neural posterior
  estimation},\ }\href {https://doi.org/10.1103/PhysRevLett.127.241103}
  {\bibfield  {journal} {\bibinfo  {journal} {Physical Review Letters}\
  }\textbf {\bibinfo {volume} {127}},\ \bibinfo {pages} {241103} (\bibinfo
  {year} {2021})}\BibitemShut {NoStop}%
\bibitem [{\citenamefont {Flanagan}\ and\ \citenamefont
  {Hinderer}(2008)}]{FlanaganHinderer2008TidalLove}%
  \BibitemOpen
  \bibfield  {author} {\bibinfo {author} {\bibfnamefont {{\'E}.~{\'E}.}\
  \bibnamefont {Flanagan}}\ and\ \bibinfo {author} {\bibfnamefont
  {T.}~\bibnamefont {Hinderer}},\ }\bibfield  {title} {\bibinfo {title}
  {Constraining neutron-star tidal love numbers with gravitational-wave
  detectors},\ }\href {https://doi.org/10.1103/PhysRevD.77.021502} {\bibfield
  {journal} {\bibinfo  {journal} {Phys. Rev. D}\ }\textbf {\bibinfo {volume}
  {77}},\ \bibinfo {pages} {021502} (\bibinfo {year} {2008})},\ \Eprint
  {https://arxiv.org/abs/0709.1915} {arXiv:0709.1915 [astro-ph]} \BibitemShut
  {NoStop}%
\bibitem [{\citenamefont {Flanagan}\ and\ \citenamefont
  {Racine}(2007)}]{FlanaganRacine2007Rossby}%
  \BibitemOpen
  \bibfield  {author} {\bibinfo {author} {\bibfnamefont {{\'E}.~{\'E}.}\
  \bibnamefont {Flanagan}}\ and\ \bibinfo {author} {\bibfnamefont
  {{\'E}.}~\bibnamefont {Racine}},\ }\bibfield  {title} {\bibinfo {title}
  {Gravitomagnetic resonant excitation of rossby modes in coalescing neutron
  star binaries},\ }\href {https://doi.org/10.1103/PhysRevD.75.044001}
  {\bibfield  {journal} {\bibinfo  {journal} {Physical Review D}\ }\textbf
  {\bibinfo {volume} {75}},\ \bibinfo {pages} {044001} (\bibinfo {year}
  {2007})}\BibitemShut {NoStop}%
\bibitem [{\citenamefont {Nitz}\ \emph {et~al.}(2024)\citenamefont {Nitz},
  \citenamefont {Harry}, \citenamefont {Brown}, \citenamefont {Biwer},
  \citenamefont {Willis}, \citenamefont {Canton}, \citenamefont {Capano},
  \citenamefont {Dent}, \citenamefont {Pekowsky}, \citenamefont {Davies},
  \citenamefont {De}, \citenamefont {Cabero}, \citenamefont {Wu}, \citenamefont
  {Williamson}, \citenamefont {Machenschalk}, \citenamefont {Macleod},
  \citenamefont {Pannarale}, \citenamefont {Kumar}, \citenamefont {Reyes},
  \citenamefont {dfinstad}, \citenamefont {Kumar}, \citenamefont {T{\'a}pai},
  \citenamefont {Singer}, \citenamefont {Kumar}, \citenamefont {veronica
  villa}, \citenamefont {maxtrevor}, \citenamefont {Gadre}, \citenamefont
  {Khan}, \citenamefont {Fairhurst},\ and\ \citenamefont
  {Tolley}}]{alex_nitz_2024_10473621}%
  \BibitemOpen
  \bibfield  {author} {\bibinfo {author} {\bibfnamefont {A.}~\bibnamefont
  {Nitz}}, \bibinfo {author} {\bibfnamefont {I.}~\bibnamefont {Harry}},
  \bibinfo {author} {\bibfnamefont {D.}~\bibnamefont {Brown}}, \bibinfo
  {author} {\bibfnamefont {C.~M.}\ \bibnamefont {Biwer}}, \bibinfo {author}
  {\bibfnamefont {J.}~\bibnamefont {Willis}}, \bibinfo {author} {\bibfnamefont
  {T.~D.}\ \bibnamefont {Canton}}, \bibinfo {author} {\bibfnamefont
  {C.}~\bibnamefont {Capano}}, \bibinfo {author} {\bibfnamefont
  {T.}~\bibnamefont {Dent}}, \bibinfo {author} {\bibfnamefont {L.}~\bibnamefont
  {Pekowsky}}, \bibinfo {author} {\bibfnamefont {G.~S.~C.}\ \bibnamefont
  {Davies}}, \bibinfo {author} {\bibfnamefont {S.}~\bibnamefont {De}}, \bibinfo
  {author} {\bibfnamefont {M.}~\bibnamefont {Cabero}}, \bibinfo {author}
  {\bibfnamefont {S.}~\bibnamefont {Wu}}, \bibinfo {author} {\bibfnamefont
  {A.~R.}\ \bibnamefont {Williamson}}, \bibinfo {author} {\bibfnamefont
  {B.}~\bibnamefont {Machenschalk}}, \bibinfo {author} {\bibfnamefont
  {D.}~\bibnamefont {Macleod}}, \bibinfo {author} {\bibfnamefont
  {F.}~\bibnamefont {Pannarale}}, \bibinfo {author} {\bibfnamefont
  {P.}~\bibnamefont {Kumar}}, \bibinfo {author} {\bibfnamefont
  {S.}~\bibnamefont {Reyes}}, \bibinfo {author} {\bibnamefont {dfinstad}},
  \bibinfo {author} {\bibfnamefont {S.}~\bibnamefont {Kumar}}, \bibinfo
  {author} {\bibfnamefont {M.}~\bibnamefont {T{\'a}pai}}, \bibinfo {author}
  {\bibfnamefont {L.}~\bibnamefont {Singer}}, \bibinfo {author} {\bibfnamefont
  {P.}~\bibnamefont {Kumar}}, \bibinfo {author} {\bibnamefont {veronica
  villa}}, \bibinfo {author} {\bibnamefont {maxtrevor}}, \bibinfo {author}
  {\bibfnamefont {B.~U.~V.}\ \bibnamefont {Gadre}}, \bibinfo {author}
  {\bibfnamefont {S.}~\bibnamefont {Khan}}, \bibinfo {author} {\bibfnamefont
  {S.}~\bibnamefont {Fairhurst}},\ and\ \bibinfo {author} {\bibfnamefont
  {A.}~\bibnamefont {Tolley}},\ }\href
  {https://doi.org/10.5281/zenodo.10473621} {\bibinfo {title} {gwastro/pycbc:
  v2.3.3 release of pycbc}} (\bibinfo {year} {2024})\BibitemShut {NoStop}%
\bibitem [{\citenamefont {Zhao}\ \emph {et~al.}(2011)\citenamefont {Zhao},
  \citenamefont {Van Den~Broeck}, \citenamefont {Baskaran},\ and\ \citenamefont
  {Li}}]{Zhao2011ETCosmology}%
  \BibitemOpen
  \bibfield  {author} {\bibinfo {author} {\bibfnamefont {W.}~\bibnamefont
  {Zhao}}, \bibinfo {author} {\bibfnamefont {C.}~\bibnamefont {Van
  Den~Broeck}}, \bibinfo {author} {\bibfnamefont {D.}~\bibnamefont
  {Baskaran}},\ and\ \bibinfo {author} {\bibfnamefont {T.~G.~F.}\ \bibnamefont
  {Li}},\ }\bibfield  {title} {\bibinfo {title} {Determination of dark energy
  by the {Einstein Telescope}: Comparing with {CMB}, {BAO}, and {SNIa}
  observations},\ }\href@noop {} {\bibfield  {journal} {\bibinfo  {journal}
  {Physical Review D}\ }\textbf {\bibinfo {volume} {83}},\ \bibinfo {pages}
  {023005} (\bibinfo {year} {2011})}\BibitemShut {NoStop}%
\bibitem [{\citenamefont {Schneider}\ \emph {et~al.}(2001)\citenamefont
  {Schneider}, \citenamefont {Ferrari}, \citenamefont {Matarrese},\ and\
  \citenamefont {Portegies~Zwart}}]{Schneider2001LowFrequency}%
  \BibitemOpen
  \bibfield  {author} {\bibinfo {author} {\bibfnamefont {R.}~\bibnamefont
  {Schneider}}, \bibinfo {author} {\bibfnamefont {V.}~\bibnamefont {Ferrari}},
  \bibinfo {author} {\bibfnamefont {S.}~\bibnamefont {Matarrese}},\ and\
  \bibinfo {author} {\bibfnamefont {S.~F.}\ \bibnamefont {Portegies~Zwart}},\
  }\bibfield  {title} {\bibinfo {title} {Low-frequency gravitational waves from
  cosmological compact binaries},\ }\href@noop {} {\bibfield  {journal}
  {\bibinfo  {journal} {Monthly Notices of the Royal Astronomical Society}\
  }\textbf {\bibinfo {volume} {324}},\ \bibinfo {pages} {797} (\bibinfo {year}
  {2001})}\BibitemShut {NoStop}%
\bibitem [{\citenamefont {Cutler}\ and\ \citenamefont
  {Holz}(2009)}]{CutlerHolz2009Cosmology}%
  \BibitemOpen
  \bibfield  {author} {\bibinfo {author} {\bibfnamefont {C.}~\bibnamefont
  {Cutler}}\ and\ \bibinfo {author} {\bibfnamefont {D.~E.}\ \bibnamefont
  {Holz}},\ }\bibfield  {title} {\bibinfo {title} {Ultrahigh precision
  cosmology from gravitational waves},\ }\href@noop {} {\bibfield  {journal}
  {\bibinfo  {journal} {Physical Review D}\ }\textbf {\bibinfo {volume} {80}},\
  \bibinfo {pages} {104009} (\bibinfo {year} {2009})}\BibitemShut {NoStop}%
\bibitem [{\citenamefont {Cai}\ and\ \citenamefont
  {Yang}(2017)}]{CaiYang2017ETCosmology}%
  \BibitemOpen
  \bibfield  {author} {\bibinfo {author} {\bibfnamefont {R.-G.}\ \bibnamefont
  {Cai}}\ and\ \bibinfo {author} {\bibfnamefont {T.}~\bibnamefont {Yang}},\
  }\bibfield  {title} {\bibinfo {title} {Estimating cosmological parameters by
  the simulated data of gravitational waves from the {Einstein Telescope}},\
  }\href@noop {} {\bibfield  {journal} {\bibinfo  {journal} {Physical Review
  D}\ }\textbf {\bibinfo {volume} {95}},\ \bibinfo {pages} {044024} (\bibinfo
  {year} {2017})}\BibitemShut {NoStop}%
\bibitem [{\citenamefont {Albanesi}\ \emph {et~al.}(2025)\citenamefont
  {Albanesi}, \citenamefont {Gamba}, \citenamefont {Bernuzzi}, \citenamefont
  {Fontbut{\'e}}, \citenamefont {Gonzalez},\ and\ \citenamefont
  {Nagar}}]{albanesi2025effective}%
  \BibitemOpen
  \bibfield  {author} {\bibinfo {author} {\bibfnamefont {S.}~\bibnamefont
  {Albanesi}}, \bibinfo {author} {\bibfnamefont {R.}~\bibnamefont {Gamba}},
  \bibinfo {author} {\bibfnamefont {S.}~\bibnamefont {Bernuzzi}}, \bibinfo
  {author} {\bibfnamefont {J.}~\bibnamefont {Fontbut{\'e}}}, \bibinfo {author}
  {\bibfnamefont {A.}~\bibnamefont {Gonzalez}},\ and\ \bibinfo {author}
  {\bibfnamefont {A.}~\bibnamefont {Nagar}},\ }\bibfield  {title} {\bibinfo
  {title} {Effective-one-body modeling for generic compact binaries with
  arbitrary orbits},\ }\href@noop {} {\bibfield  {journal} {\bibinfo  {journal}
  {Physical Review D}\ }\textbf {\bibinfo {volume} {112}},\ \bibinfo {pages}
  {L121503} (\bibinfo {year} {2025})}\BibitemShut {NoStop}%
\bibitem [{\citenamefont {Smith}\ \emph {et~al.}(2021)\citenamefont {Smith},
  \citenamefont {Borhanian}, \citenamefont {Sathyaprakash}, \citenamefont
  {Hernandez~Vivanco}, \citenamefont {Field}, \citenamefont {Lasky},
  \citenamefont {Mandel}, \citenamefont {Morisaki}, \citenamefont {Ottaway},
  \citenamefont {Slagmolen}, \citenamefont {Thrane}, \citenamefont
  {T{\"o}yr{\"a}},\ and\ \citenamefont
  {Vitale}}]{Smith2021ThirdGenerationInference}%
  \BibitemOpen
  \bibfield  {author} {\bibinfo {author} {\bibfnamefont {R.}~\bibnamefont
  {Smith}}, \bibinfo {author} {\bibfnamefont {S.}~\bibnamefont {Borhanian}},
  \bibinfo {author} {\bibfnamefont {B.}~\bibnamefont {Sathyaprakash}}, \bibinfo
  {author} {\bibfnamefont {F.}~\bibnamefont {Hernandez~Vivanco}}, \bibinfo
  {author} {\bibfnamefont {S.~E.}\ \bibnamefont {Field}}, \bibinfo {author}
  {\bibfnamefont {P.}~\bibnamefont {Lasky}}, \bibinfo {author} {\bibfnamefont
  {I.}~\bibnamefont {Mandel}}, \bibinfo {author} {\bibfnamefont
  {S.}~\bibnamefont {Morisaki}}, \bibinfo {author} {\bibfnamefont
  {D.}~\bibnamefont {Ottaway}}, \bibinfo {author} {\bibfnamefont
  {B.}~\bibnamefont {Slagmolen}}, \bibinfo {author} {\bibfnamefont
  {E.}~\bibnamefont {Thrane}}, \bibinfo {author} {\bibfnamefont
  {D.}~\bibnamefont {T{\"o}yr{\"a}}},\ and\ \bibinfo {author} {\bibfnamefont
  {S.}~\bibnamefont {Vitale}},\ }\bibfield  {title} {\bibinfo {title} {Bayesian
  inference for gravitational waves from binary neutron star mergers in
  third-generation observatories},\ }\href
  {https://doi.org/10.1103/PhysRevLett.127.081102} {\bibfield  {journal}
  {\bibinfo  {journal} {Physical Review Letters}\ }\textbf {\bibinfo {volume}
  {127}},\ \bibinfo {pages} {081102} (\bibinfo {year} {2021})},\ \Eprint
  {https://arxiv.org/abs/2103.12274} {arXiv:2103.12274} \BibitemShut {NoStop}%
\bibitem [{\citenamefont {Goodfellow}\ \emph {et~al.}(2016)\citenamefont
  {Goodfellow}, \citenamefont {Bengio},\ and\ \citenamefont
  {Courville}}]{Goodfellow2016DeepLearning}%
  \BibitemOpen
  \bibfield  {author} {\bibinfo {author} {\bibfnamefont {I.}~\bibnamefont
  {Goodfellow}}, \bibinfo {author} {\bibfnamefont {Y.}~\bibnamefont {Bengio}},\
  and\ \bibinfo {author} {\bibfnamefont {A.}~\bibnamefont {Courville}},\ }\href
  {https://www.deeplearningbook.org/} {\emph {\bibinfo {title} {Deep
  Learning}}}\ (\bibinfo  {publisher} {MIT Press},\ \bibinfo {year}
  {2016})\BibitemShut {NoStop}%
\bibitem [{\citenamefont {He}\ \emph {et~al.}(2016)\citenamefont {He},
  \citenamefont {Zhang}, \citenamefont {Ren},\ and\ \citenamefont
  {Sun}}]{He2016DeepResidualLearning}%
  \BibitemOpen
  \bibfield  {author} {\bibinfo {author} {\bibfnamefont {K.}~\bibnamefont
  {He}}, \bibinfo {author} {\bibfnamefont {X.}~\bibnamefont {Zhang}}, \bibinfo
  {author} {\bibfnamefont {S.}~\bibnamefont {Ren}},\ and\ \bibinfo {author}
  {\bibfnamefont {J.}~\bibnamefont {Sun}},\ }\href@noop {} {\bibinfo {title}
  {Deep residual learning for image recognition}} (\bibinfo {year} {2016}),\
  \Eprint {https://arxiv.org/abs/1512.03385} {arXiv:1512.03385} \BibitemShut
  {NoStop}%
\bibitem [{\citenamefont {{Kimi Team}}(2026)}]{Kimi2026AttentionResiduals}%
  \BibitemOpen
  \bibfield  {author} {\bibinfo {author} {\bibnamefont {{Kimi Team}}},\
  }\href@noop {} {\bibinfo {title} {Attention residuals}} (\bibinfo {year}
  {2026}),\ \Eprint {https://arxiv.org/abs/2603.15031} {arXiv:2603.15031}
  \BibitemShut {NoStop}%
\end{thebibliography}
\end{document}